\documentclass[12pt]{article}
\usepackage[english]{babel}
\usepackage[utf8]{inputenc}
\usepackage{microtype}

\usepackage[T1]{fontenc}
\usepackage[toc,page]{appendix}
\usepackage{lmodern}
\usepackage[top=1in, bottom=1in, left=1in, right=1in, letterpaper]{geometry}
\usepackage{amsmath}
\usepackage{amssymb}
\usepackage{amsthm}
\usepackage{amsfonts}
\usepackage{latexsym}
\usepackage{amscd}
\usepackage{euscript}
\usepackage{mathrsfs}
\usepackage{mathtools}
\usepackage{graphicx, xcolor}
\definecolor{darkred}{RGB}{144,0,0}
\definecolor{darkblue}{RGB}{0,0,144}
\usepackage[colorlinks=true, linkcolor=darkred, citecolor=darkblue]{hyperref}
\usepackage{subcaption}
\usepackage{textcomp}
\usepackage{booktabs}
\usepackage{array}
\usepackage{bm}
\usepackage{accents}
\newcolumntype{L}[1]{>{\raggedright\let\newline\\\arraybackslash\hspace{0pt}}m{#1}}
\newcolumntype{C}[1]{>{\centering\let\newline\\\arraybackslash\hspace{0pt}}m{#1}}
\newcolumntype{R}[1]{>{\raggedleft\let\newline\\\arraybackslash\hspace{0pt}}m{#1}}
\usepackage[ruled,vlined,noresetcount,linesnumbered]{algorithm2e}
\makeatletter
\renewcommand{\underbar}[1]{\underaccent{\bar}{#1}}
\makeatother
\renewcommand{\k}{K}
\newcommand{\uk}{\k_0}
\newcommand{\ok}{\k_1}
\newcommand{\ook}{\bar{\k}_1}
\newcommand{\uj}{{j}_0}
\newcommand{\oj}{{j}_1}

\newcommand{\of}{\bar{f}}
\newcommand{\uf}{\underbar{f}}

\newcommand{\ukk}{k_0}
\newcommand{\okk}{k_1}

\newcommand{\x}{X}
\newcommand{\xv}{x}
\renewcommand{\i}{\mathrm{i}}

\renewcommand{\bf}{B}
\newcommand{\ubf}{\underbar{B}}
\newcommand{\obf}{\bar{B}}

\newcommand{\ci}[1]{C_{{i#1}}}

\newcommand{\y}{Y}
\newcommand{\yi}[1]{Y_{{i}}(#1)}
\newcommand{\yis}[1]{Y_{{i}}^*(#1)}
\newcommand{\ys}[1]{Y^*(#1)}
\newcommand{\yii}{Y_{{i}}}
\newcommand{\yiis}{Y_{{i}}^*}

\newcommand{\yv}{y}
\newcommand{\dyv}[1]{\dif_{\yv}}
\newcommand{\uy}{\underbar {\mathcal{Y}}_0}
\newcommand{\oy}{\bar {\mathcal{Y}}_0}

\newcommand{\knot}{t}

\renewcommand{\oe}{\dot{\k}}

\newcommand{\vis}[1]{V_{{i}}^*(#1)}

\newcommand{\rem}{\mathcal{R}}

\newcommand{\con}{c}

\newcommand{\tc}{\ell}

\newcommand{\ib}{\psi}
\newcommand{\kap}{\kappa}       
\newcommand{\co}{\delta}

\newcommand{\conv}{R}
\newcommand{\nn}{N}

\newcommand{\quant}{Q}

\newcommand{\dif}{D}

\newcommand{\leP}{\lesssim_{\P}}
\newcommand{\convp}{\overset{\P}{\to}}
\newcommand{\convd}{\Rightarrow}

\newcommand{\vv}{\yv}
\newcommand{\ubars}{a}
\newcommand{\wv}{w}
\newcommand{\sv}{w}
\renewcommand{\v}{\y}

\renewcommand{\t}{T}
\newcommand{\s}{W}
\newcommand{\bars}{b}
\newcommand{\rad}{\rho}
\newcommand{\ti}{T_i}
\newcommand{\qua}{\tau}

\newcommand{\qb}{g^{(\Q)}}
\newcommand{\qc}{g^{(\Q)}}
\newcommand{\cv}[1]{\operatorname{cv}_{{#1}}}
\newcommand{\appftna}[1]{f}
\newcommand{\appftnb}[1]{g_{{#1}}}
\newcommand{\ueta}{\underbar{\eta}}
\newcommand{\oeta}{\bar{\eta}}
\newcommand{\utheta}{\underbar{\theta}}
\newcommand{\otheta}{\bar{\theta}}

\newcommand{\OA}{the Appendix}

\newcommand{\maxb}{\bar{b}}

\renewcommand{\lg}{
  \mathrel{
    \vcenter{\offinterlineskip
      \ialign{##\cr$\leqslant$\cr\noalign{\kern-1.5pt}$\geqslant$\cr}
    }
  }
}

\newcommand{\gl}{%
  \mathrel{%
    \vcenter{\offinterlineskip
      \ialign{##\cr$\geqslant$\cr\noalign{\kern-1.5pt}$\leqslant$\cr}%
    }%
  }%
}
\newcommand{\argmax}[1]{\operatornamewithlimits{arg\hspace{0.1em} max}_{{#1}}\hspace{0.1em} }
\newcommand{\argmin}[1]{\operatornamewithlimits{arg\hspace{0.1em} min}_{{#1}}\hspace{0.1em} }

\usepackage[shortlabels]{enumitem}
\usepackage[retainorgcmds]{IEEEtrantools}
\usepackage{tabularx}
\usepackage{array}
\usepackage{caption}
\usepackage{xcolor}
\usepackage{float}
\usepackage{tcolorbox}
\usepackage{tabularx}
\usepackage{threeparttable}
\usepackage{multirow}
\usepackage{setspace}
\usepackage{lscape, pdflscape}
\usepackage{bbm}
\usepackage{natbib}
\usepackage{bibentry}
\setcitestyle{authoryear, open={(},close={)}}
\renewcommand{\citep}[2]{\citeauthor{#1} (\citeyear{#1}, {#2})}
\newcommand{\citeayfull}[1]{\citeauthor*{{#1}}, \citeyear{{#1}}}
\newcommand{\citeay}[1]{\citeauthor{{#1}}, \citeyear{{#1}}}

\def\oo{\infty}
\def\0oo{[0,\oo)}

\def\C{\mathbb{C}}
\def\E{E}
\def\N{\mathbb{N}}
\def\P{P}
\def\Q{{P_{T}}}
\def\R{\mathbb{R}}
\def\S{\mathcal S}

\def\Z{\mathbb{Z}}
\def\CA{A}
\def\CB{\mathcal{B}}
\def\CC{C}

\def\CF{\mathcal{F}}

\def\CH{\mathcal{H}}

\def\CJ{\mathcal{J}}

\def\CP{\mathcal{P}}

\def\CS{\mathcal{S}}

\def\CY{\mathcal{Y}}

\def\1{\mathbbm{1}}
\def\var{\opw{var}}

\def\lm{\left[ \begin{matrix}}
\def\rm{\end{matrix}\right]}

\def\cds{\cdots}

\def\te{\text}
\def\op{\operatornamewithlimits}
\def\opw{\operatorname}

\def\what{\hat}

\def\sgn{\op{sign}}
\def\re{\operatornamewithlimits{Re}}

\def\tbf{\textbf}

\def\OP{O_{P}}
\def\oP{o_{P}}

\def\ub{\underbrace}

\newcommand{\I}[1]{ \mathbbm{1}\{  {#1}  \}}

\newcommand{\tr}[1]{\op{tr}\left({#1}\right)}

\newcommand{\diag}{\op{diag}}
\newcommand{\VV}[1]{{\left\vert\kern-0.25ex\left\vert\kern-0.25ex\left\vert #1 
    \right\vert\kern-0.25ex\right\vert\kern-0.25ex\right\vert}}

\theoremstyle{definition} \newtheorem{define}{Definition}
\theoremstyle{plain} \newtheorem*{assumption*}{\assumptionletter}
\makeatletter
\providecommand{\assumptionletter}{}

\newenvironment{assumption}[1]
 {%
  \renewcommand{\assumptionletter}{Assumption #1}%
  \begin{assumption*}%
  \protected@edef\@currentlabel{#1}%
 }
 {%
  \end{assumption*}
 }

\theoremstyle{plain}
\newtheorem{thm}{\protect\theoremname}
\newtheorem{thmA}{\protect\theoremname}
\numberwithin{thmA}{section}

\newtheorem{lemA}{Lemma}
\numberwithin{lemA}{section}
\newtheorem{propA}{Proposition}
\numberwithin{propA}{section}

\newtheorem*{asm*}{Assumption}
\newtheorem{asmA}{Assumption}
\numberwithin{asmA}{section}
\providecommand{\theoremname}{Theorem}
\newtheorem{prop}{Proposition}
\theoremstyle{definition}
\newtheorem{remark}{Remark}

\makeatother

\usepackage{etoolbox}
\makeatletter
\patchcmd{\@maketitle}
  {\ifx\@empty\@dedicatory}
  {\ifx\@empty\@date \else {\vskip3ex \centering\footnotesize\@date\par\vskip1ex}\fi
   \ifx\@empty\@dedicatory}
  {}{}
\patchcmd{\@adminfootnotes}
  {\ifx\@empty\@date\else \@footnotetext{\@setdate}\fi}
  {}{}{}
\makeatother

\usepackage{bookmark}

\begin{document}
\title{Identification and Inference in General Bunching Designs}
\author{Myunghyun Song
\thanks{
Department of Economics, Columbia University, \texttt{ms6347@columbia.edu}.
I thank my advisor Sokbae (Simon) Lee for his gracious advice and continuous support throughout this project.
I also thank Serena Ng, Jushan Bai, Bernard Salanié, Lihua Lei, Mikkel Plagborg-Møller, and Soonwoo Kwon for their helpful comments.
Seminar participants at Columbia and the KBER Summer Institute at Seoul National University have provided invaluable feedback.
I am responsible for any and all errors.
}}
\date{This version: \today.}




\maketitle
\begin{abstract}
This paper develops an econometric framework and tools for the identification and inference of a structural parameter in general bunching designs.
We present point and partial identification results, which generalize previous approaches in the literature.
The key assumption for point identification is the analyticity of the counterfactual density, which defines a broader class of distributions than many commonly used parametric families.
In the partial identification approach, the analyticity condition is relaxed and various inequality restrictions can be incorporated.
Both of our identification approaches allow for observed covariates in the model, which has previously been permitted only in limited ways.
These covariates allow us to account for observable factors that influence decisions regarding the running variable.
We provide a suite of counterfactual estimation and inference methods, termed the generalized polynomial strategy.
Our method restores the merits of the original polynomial strategy proposed by \citet{chettyAdjustmentCostsFirm2011} while addressing several weaknesses in the widespread practice.
The efficacy of the proposed method is demonstrated compared to the polynomial estimator in a series of Monte Carlo studies within the augmented isoelastic model.
We revisit the data used in \citet{saezTaxpayersBunchKink2010} and find substantially different results relative to those from the polynomial strategy.
\end{abstract}

    
\newpage
\section{Introduction}

The bunching method utilizes discontinuities in an incentive schedule to elicit behavioral responses from individuals.
It is facilitated by a choice variable, also known as a running variable, manipulated by individuals to maximize their payoffs.
Governments, or other authoritative entities, establish the incentive schedule applied to these individuals, which change discretely at specific cutoffs based on the chosen running variable.
Such policies create varying incentives for the selection of the running variable across these cutoffs, leading to noticeable bunching -- manifesting as either excessive or missing mass in the distribution.
Initiated by \citet{saezTaxpayersBunchKink2010}, who observed bunching in the U.S. taxable income distribution and used it to measure the elasticity of taxable income, the bunching method has been further grounded in subsequent works (\citeayfull{chettyAdjustmentCostsFirm2011}; \citeayfull{klevenUsingNotchesUncover2013}, etc).
Since then, it has found widespread applications across various economic studies: e.g., for the study of the effects of minimum wages on employment (\citeayfull{cengizEffectMinimumWages2019}; \citeayfull{derenoncourtMinimumWagesRacial2020}), the impact of tax base on corporate tax evasion in developing countries (\citeayfull{bestProductionRevenueEfficiency2015}), 
the impact of taxation and fiscal stimulus in the U.K. housing market
(\citeayfull{bestHousingMarketResponses2018}),
the elasticity of intertemporal substitution for U.K. households (\citeayfull{bestEstimatingElasticityIntertemporal2020}),
regulatory costs faced by publicly listed firms (\citeayfull{ewensRegulatoryCostsBeing2024}), 
and detection of score/data manipulation around cutoffs (\citeayfull{foremnyGhostCitizensUsing2017}; \citeayfull{deeCausesConsequencesTest2019}; \citeayfull{ghanemCensoredMaximumLikelihood2020}) among many others.

This paper develops a formal econometric framework and tools for the identification and inference of a structural parameter in the bunching design.
Our focus in this paper is on the kink design, in which a concave payoff function exhibits kinks at specific cutoffs.
Another commonly used bunching design, known as the notch design, applies to cases where the incentive schedule features discontinuities, referred to as notches.
Discussions on how the notch design could fare in our framework are deferred to \OA.

Building on a potential outcome framework similar to that of \citet{goffTreatmentEffectsBunching2024}, we lay out an analytic framework for structural models in the bunching design.
We analyze two dimensions of the counterfactual, where each counterfactual policy—referred to as the pre- and post-kink policies—has an incentive schedule that varies smoothly across the entire range of a choice variable.
The counterfactual choices are defined as hypothetical choices an individual would make if subjected to either of these counterfactual policies.
The prekink and postkink counterfactual choices are denoted by $\ys{0}$ and $\ys{1}$, respectively.

Identification in the bunching design hinges on two key ingredients: the counterfactual distribution of $\ys{0}$ and the structural model of individuals' responses to policy changes.
Our identification approach requires less stringent conditions on these ingredients than in existing methods.
First, we explore the use of an analytic model for the counterfactual distribution for bunching identification.
As noted by \citet*{blomquistBunchingIdentificationTaxable2021}, identification in bunching designs relies on implicit shape restrictions on the counterfactual distribution.
Unlike many commonly used parametric approaches, our method is not restricted to an ex-ante finite-dimensional family of distributions, but it leverages a dense family of analytic functions.
Under the analyticity condition, the counterfactual density and its derivatives can be stably approximated by polynomials of increasing orders.
This allows us to achieve two seemingly conflicting goals: flexibly fitting the observed distribution while uniquely extrapolating the counterfactual into the unobserved region.


The idea of leveraging the analyticity assumption on the counterfactual distribution for bunching identification was explored in the earliest version of \citet{pollinger2024kinks}, which developed identification and consistent estimation in an extended isoelastic model with intensive and extensive margin responses.
Our approach is distinguished from it in two notable ways.
First, we address both point and partial identification in a general class of structural models.
Second, following the identification step, we focus on inference of a structural parameter, which requires a different set of tools and assumptions compared to estimation.

We allow for covariates included in the structural model to account for observable factors that influence individual decisions regarding the choice variable.
By contrast, a common feature of structural models in bunching design is that the choice variable is determined by a scalar unobserved heterogeneity, whose distribution is directly linked to the structural parameter.
As noted in \citet*{bestEstimatingElasticityIntertemporal2020}, which examines the elasticity of intertemporal substitution in dynamic consumption models, the inclusion of additional variables or parameters can significantly increase the complexity of the analysis.
We address these challenges within a unified framework for structural modeling, identification, and subsequent procedures.

Previous approaches have permitted covariates only in limited ways.
When certain covariates are continuously distributed, they need to be grouped into a small number of discrete categories.
Moreover, such an approach requires partitioning the sample based on these categories, which could lead to aggregation bias or loss of power.
This strategy has been employed in \citet{saezTaxpayersBunchKink2010} and \citet{bestEstimatingElasticityIntertemporal2020} among many others.
Our approach integrates covariates into the model without requiring such coarseness.

We propose a partial identification approach that incorporates prior information on the counterfactual distribution.
This method is characterized by encapsulating all potential instances of the counterfactual distribution function between the upper and lower envelope functions that are piecewise analytic, which leads to a set of inequality restrictions akin to first-order stochastic dominance.
Under such a condition, the method can provide sharp identification bounds for a scalar structural parameter.
This method applies to a wider range of counterfactual distributions than the first approach and accommodates various shape restrictions discussed in the literature.
For instance, in the context of the isoelastic model, our result yields the elasticity bounds that are identical to those established in \citet{blomquistBunchingIdentificationTaxable2021} and \citet*{bertanhaBetterBunchingNicer2023} under the corresponding assumptions.

The polynomial strategy, initially proposed by \citet{chettyAdjustmentCostsFirm2011}, is a widespread approach to estimating the unobserved counterfactual distribution.
It proceeds by fitting a flexible polynomial to the observed distribution excluding a narrow window surrounding the cutoff.
The fitted polynomial provides the initial estimate for the unobserved counterfactual density inside the window.\footnote{\citet{chettyAdjustmentCostsFirm2011} requires an iterative procedure to ensure that the integral constraint holds for the estimated counterfactual density.
If excess bunching is small, this iterative procedure has no major impacts.}
\citet{chettyAdjustmentCostsFirm2011} used the difference between the observed mass and the estimated counterfactual mass within the window as an estimate for the excess bunching fraction.
The estimated excess bunching fraction $\what{B}$, after being normalized by the counterfactual density $\what{f}$ at the cutoff, identifies the average policy response via the small kink approximation
$
{\what{B}}/{\what{f}} \approx \E[\ys{0}-\ys{1}|\ys{0}=K].
$
The structural model then identifies the average policy response as a function of the structural parameter $\theta$, represented by a function $h$ such that $\E[\ys{0}-\ys{1}|\ys{0}=K] = h(\theta)$.
Such a restriction can justify using minimum distance estimation for $\theta$.

We highlight some weaknesses in the traditional polynomial estimation and propose an alternative scheme, termed the generalized polynomial strategy.
First, the conventional polynomial estimation works under the assumption that $\ys{1}$ and $\ys{0}$ have proportional densities above the cutoff.
However, this assumption is often incompatible with the commonly held structural assumption that both counterfactual choices are related but have different geneses.
To address this issue, we propose a valid counterfactual adjustment procedure, referred to as the counterfactual correction.
Consistency of the estimation and inference based on the proposed method is theoretically established and substantiated by numerical evidence.

\begin{figure}[hbtp!]
    \centering
    \caption{\textbf{Estimation results using different adjustment methods.}
        The left panel displays the observed income histogram, and the right panel compares the estimated counterfactual densities, each based on the traditional proportional adjustment method and the proposed counterfactual correction.
        The income data are shifted upward by 20000 to avoid negative values.
        The dashed lines in each panel demarcate the excluded window around the cutoff $\k=20000$.
        The bin width is set to $1000$.
        A tenth-degree polynomial is used to fit the counterfactual density for each method.
    }
    \label{fig:prop-and-counterfactual-corrections}
    \hspace{-2em}
    \begin{subfigure}[b]{0.46\linewidth} 
        \caption{Observed income distribution}
        \includegraphics[scale=0.4]{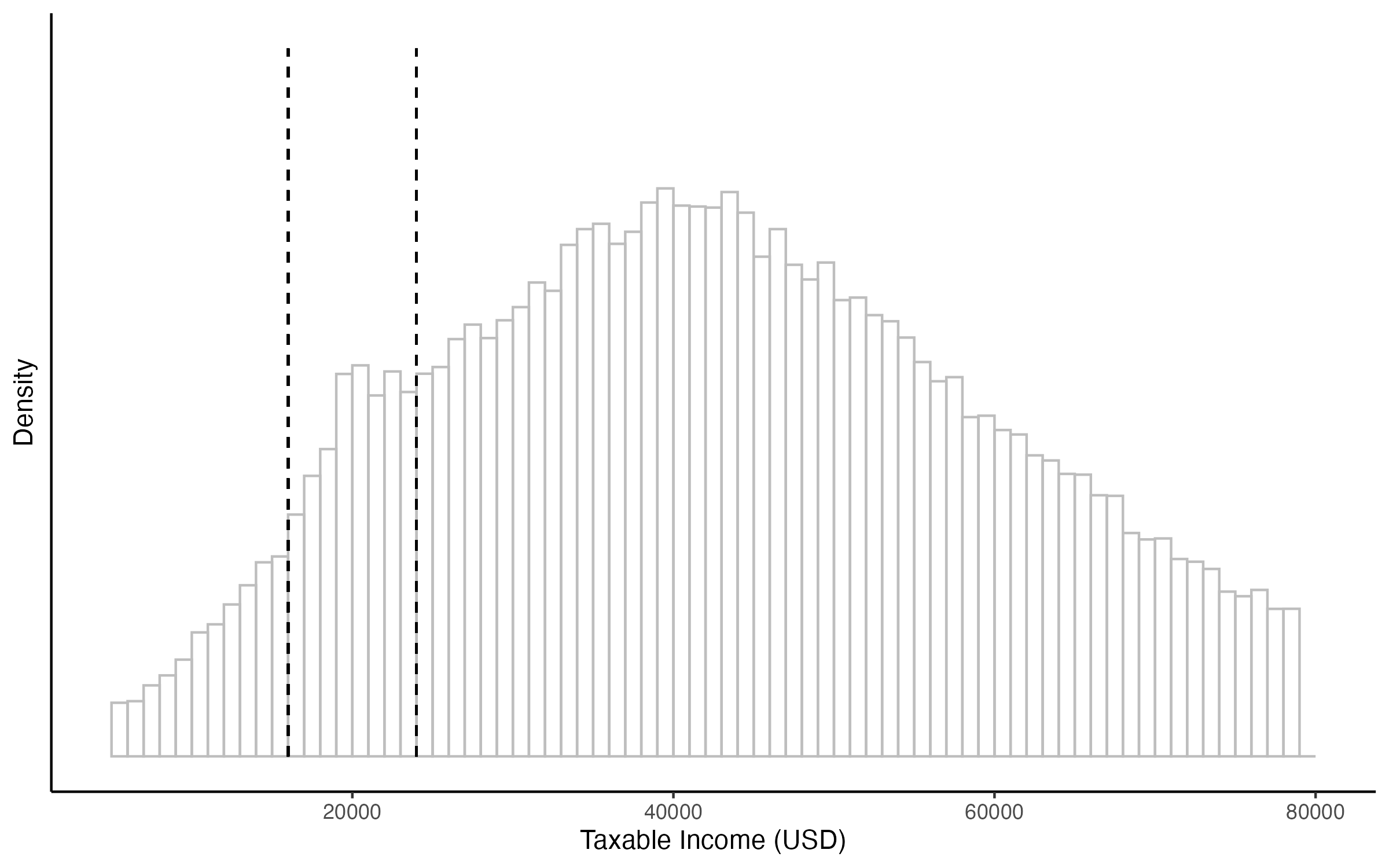}
        \label{fig:prop-correction}
    \end{subfigure}
    \hspace{2em}
    \begin{subfigure}[b]{0.46\linewidth}
        \caption{Estimated counterfactual densities}
        \includegraphics[scale=0.4]{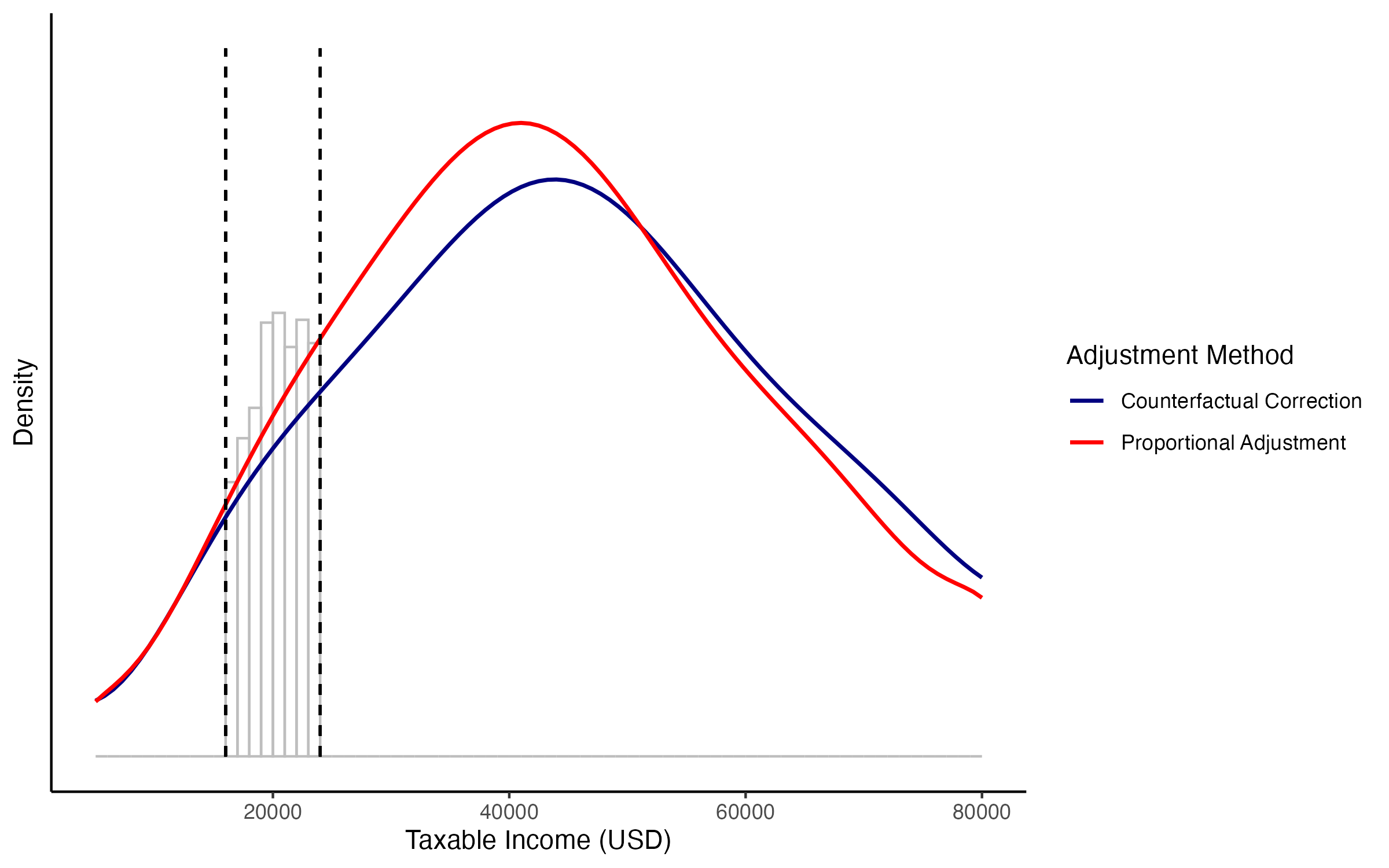}
        \label{fig:c-correction}
    \end{subfigure}
    
\end{figure}

The choice of the counterfactual adjustment procedure may have significant implications in practice.
Figure~\ref{fig:prop-and-counterfactual-corrections}(b) displays the counterfactual estimation results based on the U.S. tax records for married filers, used in \citet{saezTaxpayersBunchKink2010}.
The solid red curve in the right panel represents the estimated counterfactual density using the proportional adjustment by \citet{chettyAdjustmentCostsFirm2011}, while the solid blue curve shows the estimated counterfactual density based on the counterfactual correction procedure.
In this application, our method hinges on the structural assumption that $\ys{0} = (0.84)^{-\theta} \ys{1}$, where $\theta$ represents the taxable income elasticity and $0.84$ reflects the discrete change in the net-of-tax rate at the income cutoff.
It proceeds by applying this transformation to all $\ys{1}$ observed to the right of the window,\footnote{Our counterfactual correction applies to each hypothesized value of $\theta$. To draw Figure~\ref{fig:prop-and-counterfactual-corrections}(b), we used the value of $\theta = 0.4$, which is drawn from a preliminary analysis as a plausible value.} in order to revert $\ys{1}$ to $\ys{0}$, followed by fitting a polynomial while excluding the data within the window.
Applying this procedure produces a noticeably distinct counterfactual density compared to the one estimated using the proportional adjustment method.
Numerically, $(\what B, \what f)$ are estimated as $(1.29\times 10^{-2},1.19\times 10^{-4})$ for the proportional adjustment method, and $(1.60\times 10^{-2},1.07\times 10^{-4})$ for the counterfactual correction method.
Based on the small kink approximation, these estimates correspond to elasticities of $\what\theta = 0.30$ and $\what\theta = 0.41$, respectively.\footnote{Both estimates rely on the small kink approximation for fair comparison. Our estimates for $\theta$ remain stable across various polynomial degrees ranging from $5$ to $11$, up to the second decimal point.
However, the estimates based on the proportional adjustment exhibit much higher sensitivity, decreasing from $0.44$ to $0.30$ as the degree ranges from $7$ to $10$.}
This discrepancy could have substantial implications for policy making.
According to \citet{saezUsingElasticitiesDerive2001}, these elasticities translate into the optimal top rates of $0.63$ and $0.55$, assuming a common calibration of other parameters.\footnote{See footnote~\ref{ftnote:saez} for details.}

Secondly, identification via the small kink approximation, which underpins many existing approaches, only reflects the first-order term in an infinite series expansion of the bunching moment.
This potentially leads to a significant bias from omitting higher-order terms.
For instance, incorporating the second-order term that accounts for the slope of the counterfactual density, as shown in Figure~\ref{fig:prop-and-counterfactual-corrections}(b), corrects the overstatement of our estimate for $\theta$ by approximately 10\%.
For consistent estimation and correct inference, it is necessary to consider errors from finite-order approximation.

We close this gap by increasing the approximation order and incorporating omitted higher-order terms.
Each of these terms can be consistently estimated from the sample based on their closed-form expression.
We propose a sieve M-estimator based on a convex risk function, which can be solved easily using off-the-shelf numerical tools.
We establish statistical guarantees of our estimator contingent on selecting appropriate tuning parameters, based on the theory of sieve derivative estimation and analytic function approximation.

Turning to the bunching inference problem, we focus on testing the hypothesis $H_0 : \theta_0 = \theta$.
Based on our estimation strategy, we propose a t-test for a scalar parameter.
To address multiple parameters, a Wald test can be developed using various bunching moments generated by choices of the weighting function.
A valid confidence set can be constructed by inverting the proposed test.
Our test adopts bias-aware critical values that account for potential approximation bias, whose upper bound can be calibrated using the data.
The growth rates of the sieve and approximation orders are specified to ensure the asymptotic validity of counterfactual estimation and inference.

We assess the efficacy of the proposed inferential method compared to the polynomial estimator through a series of simulation studies.
The findings suggest that the proposed test demonstrates good size and power properties and addresses bias in the polynomial estimator.
The relevance of our method to real-world data is demonstrated through an empirical application revisiting the U.S. tax data originally used in \citet{saezTaxpayersBunchKink2010}.

The paper is organized as follows.
Section~\ref{sect:econometric-framework} introduces the econometric framework.
Section~\ref{sect:identification-results} presents our main results on point and partial identification.
Section~\ref{sect:confidence-region} develops a counterfactual estimation and inference method.
Sections~\ref{sect:monte-carlo} and \ref{sect:empirical} illustrate the effectiveness of our method through Monte Carlo experiments and the empirical application.
Finally, Section~\ref{sect:conclusion} concludes the paper.

\subsection{Notation}
Let $\|A\| = \sup_{x\in\mathbb{R}^n,x\ne 0}\|Ax\|/\|x\|$ denote the spectral norm of a matrix $A \in \mathbb{R}^{m\times n}$.
We denote by $A^-$ the pseudo-inverse of a square matrix $A \in \mathbb{R}^{m\times m}$.
For a function $f:U\to\mathbb{R}$, where $U \subseteq \R^n$ is open, its mixed partial of order $\alpha = (\alpha_1, \alpha_2,\cds, \alpha_n) \in \mathbb{Z}_+^n$ is denoted by $\dif_{x_1}^ {\alpha_1}\dif_{x_2}^{\alpha_2}\cdots \dif_{x_n}^ {\alpha_n} f = \frac{\partial^{|\alpha|} f}{\partial x_1^{\alpha_1}\partial x_2^{\alpha_2}\cdots \partial x_n ^{\alpha_n}}$ for $|\alpha| = \sum_{j=1}^n \alpha_j$.
Here, $\mathbb{Z}_+ = \{ 0,1,2,\ldots \}$.




Denote $(\Omega, \mathcal{F})$ as the underlying measurable space, and $\P$ as a probability measure on it.
We denote $\E_\P[.]$ as the expectation computed with respect to $\P$.
For a random variable $X: \Omega \to \mathbb{R}$, let $F^{(\P)}_X(x) = \P(X \le x)$ denote the cumulative distribution function (CDF) of $X$.
When $F^{(\P)}_X$ is absolutely continuous, we denote $f^{(\P)}_X(x) = dF^{(\P)}_X(x)/dx$ as its probability density function (PDF).
Likewise, we denote $F^{(\P)}_{X|Y}(x, y)$ as the conditional CDF of $X$ given $Y=y$ and $f^{(\P)}_{X|Y}(x , y)$ as the corresponding conditional PDF.
For each $\tau \in [0,1]$, we denote $\quant^{(\P)}_{X|Y}(\tau, y) = \inf \{ x \in \mathbb{R}\cup\{ \pm \infty \}: F^{(\P)}_{X|Y}(x | y) \ge \tau \}$ as the conditional $\tau$ quantile of $X$ given $Y = y$.
We suppress $\P$ in these functions when no confusion likely arises regarding the underlying measure.
The indicator function, denoted by $\I{.}$, takes the value $1$ if the event in the brackets holds and $0$ otherwise.
The $(1-\alpha)$ quantile of the chi-squared distribution with $q$ degrees of freedom is denoted by $\chi^2_q(1-\alpha)$ for $\alpha \in (0,1)$.

For sequences of nonnegative real numbers $(a_n)_{n \in \N}$ and $(b_n)_{n \in \N}$, we write $a_n \lesssim b_n$ if there exists an absolute constant $C > 0$ such that $a_n \le C b_n$ for all $n \in \N$.
For a sequence of random variables $(X_n)_{n \in \N}$ and a random variable $X$, we write $X_n \convp X$ if $X_n$ converges to $X$ in probability, and $X_n \convd X$ if the law of $X_n$ converges weakly to that of $X$.
If $(Y_n)_{n \in \N}$ is another sequence of random variables, we write $X_n = \OP(Y_n)$ or $X_n \leP Y_n$ if $(|X_n/Y_n|)_{n \in \N}$ is uniformly tight, and $X_n = \oP(Y_n)$ if $|X_n/Y_n| \convp 0$.
We adopt the convention $0/0 = 0$.

\section{Econometric Framework}

\label{sect:econometric-framework}
This section prepares an econometric framework for the analysis in the subsequent sections.
We first present the isoelastic model as an illustrative example.
We then expand on this model by formulating general conditions for the structural model to exhibit bunching under the compound policy to which individuals are actually subject.
We discuss regularity conditions required for the structural model and the data, along with our treatment of optimization frictions.
An extended framework for the notch design is relegated to \OA.

\subsection{Isoelastic Model}
\label{subsect:isoelastic-model}
The isoelastic model refers to a model of consumption and income choices, where the individual preferences are represented by the utility function of the form
\begin{equation*}
U(c, \yv, \eta_i, \theta_0) := c - \frac{\eta_i}{1 + 1/\theta_0} \left( \frac{\yv}{\eta_i}\right)^{1 + 1/\theta_0}.
\end{equation*}
Here, $\eta_i > 0$ represents a unit-specific, unobserved preference parameter and $\theta_0 > 0$ is a structural parameter representing the taxable income elasticity to the net-of-tax rate.

In the isoelastic model, unit $i$ is assumed to select a bundle $(\ci{}^*, \yiis)$ that maximizes their utility subject to the income tax schedule specified by the government.
Here, we focus on the piecewise linear tax schedule that applies to many countries including the U.S.

A piecewise linear tax system feature multiple tax brackets that are defined by values of the chosen $\y$.
At the cutoffs between adjacent brackets, the marginal tax rates change discretely, while tax liabilities vary continuously.
As demonstrated in \citet{bertanhaBetterBunchingNicer2023}, we can focus on local analysis with only two brackets and a single cutoff.
Specifically, in the global analysis, the local analysis can be applied sequentially to each pair of adjacent brackets, and the results can then be aggregated, leveraging the structure of the kink problem.

The payoff and the budget constraint faced by unit $i$ in this setup can be written as
\begin{align}
\label{eq:util-maximization}
(\ci{}^*,\yiis) &= \argmax{c > 0, \yv > 0} U(c,\yv, \eta_i, \theta_0) \\
 \text{s.t.}  \ \  c &= A_0(\yv) \I{\yv \le \k} + A_1(\yv) \I{\yv > \k}. \nonumber
\end{align}
Here, $A_0(\yv)$ and $A_1(\yv)$ represent the after-tax income defined by the \textit{linear} tax schedules applied to the lower and upper tax brackets, respectively.
We define  $\tau_0 < \tau_1$  as the marginal tax rates for each bracket and $I$ as the intercept at the cutoff, denoted by $\k$.
Then, we can write
$
A_d(\yv) = (1-\tau_d) (\yv - \k) + I
$
for each $d \in \{ 0,1 \}$.
This implies that the actual after-tax income corresponding to the taxable income $\yv$ is given by
$$
A(\yv) = (1-\tau_0)(\yv-\k) \I{\yv \le \k} + (1-\tau_1)(\yv-\k) \I{\yv > \k} + I,
$$
which is a piecewise linear function of $\yv$ with a negative kink at $\yv = \k$.


\subsubsection{Counterfactual choices}
For each $d \in \{ 0,1 \}$, we define the counterfactual choice $\yis{d}$ as the potential income that unit $i$ would have chosen if they had been subject to the linear tax schedule $c = A_d(\yv)$.
I.e., $\yis{d}$ is defined as the maximizer of the following counterfactual payoff function $U_d$:
\begin{align*}
   \yis{d} &:= \argmax{\yv > 0} U_d(\yv,\eta_i, \theta_0),\ \ \text{where}\\
  & U_d(\yv, \eta_i, \theta) = (1 - \tau_d) \cdot (\yv -\k) + I - \frac{\eta_i}{1 + 1/\theta_0}\left( \frac{\yv}{\eta_i}\right)^{1 + 1/\theta}.
\end{align*}
Here, we have concentrated out $c$ from the original utility function by substituting in the budget constraint $c = A_d(\yv)$.
This indirect utility representation allows us to focus on the choice of $\y$.
By the first-order condition and the strict concavity of $U_d$, this implies the counterfactual choice
\begin{equation*}
    \yis{d} = (1-\tau_d)^{\theta_0} \eta_i, \quad d \in \{ 0,1 \}.
\end{equation*}

\subsubsection{Actual choice}
Using the counterfactual payoff functions defined above, we can formulate the actual income choice in \eqref{eq:util-maximization} as the solution to the following unconstrained problem:
\begin{equation*}
    \yiis = \argmax{\yv > 0} U(\yv,\eta_i,\theta_0),
\end{equation*}
where the actual payoff $U$ under the piecewise linear tax schedule is given by
\begin{equation*}
U(\yv,\eta_i,\theta_0) = U_0(\yv, \eta_i,\theta_0) \I{\yv \le \k} + U_1(\yv,\eta_i,\theta_0)\I{\yv > \k}.
\end{equation*}
This compound payoff function reflects the kink in the actual income tax schedule.
Since $\dif_\yv U_0(\yv,\eta,\theta_0)|_{\yv=\k} > \dif_\yv U_1(\yv,\eta,\theta_0)|_{\yv = \k}$, the actual payoff function also exhibits a negative kink at $\k$:
\begin{align*}
\dif_\yv  U_0(\k,\eta,\theta_0) = \dif_\yv  U(\yv,\eta,\theta_0) |_{\yv = \k-}  > \ \dif_\yv  U(\yv,\eta,\theta_0) |_{\yv = \k+} =  \dif_\yv U_1(\k,\eta,\theta_0).
\end{align*}
The actual payoff function is strictly concave in $\yv$ and smooth at all points except the cutoff.

As demonstrated in \citet{saezTaxpayersBunchKink2010}, the actual income choice can be solved as follows:
\begin{align}
\label{eq:solution-isoelastic}
\yiis = \begin{cases}
 (1-\tau_0)^{\theta_0} \eta_i & \text{if }\eta_i \in (0, \underbar \eta) \\
 \k & \text{if } \eta_i \in [\underbar \eta, \bar \eta]\\
 (1-\tau_1)^{\theta_0} \eta_i & \text{if } \eta_i \in (\bar \eta, \infty)
\end{cases}.
\end{align}
Here, $[\underbar \eta ,\bar \eta]$ is given by
$
\left[ \k (1-\tau_0)^{-\theta_0}, \k (1-\tau_1)^{-\theta_0}\right],
$
which is a function of the unknown $\theta_0$ and other known parameters.
Such an interval is referred to as the bunching interval, as it characterizes the range of the bunching subpopulation in the entire distribution.


\subsection{Utility Model}
We expand beyond the isoelastic model and explore bunching in a broader context.
We consider a situation where two policies, indexed by $0$ and $1$, present smooth but different incentive schedules that are defined across the entire range of $\yv$.
The distinct incentive schedules are embedded in the counterfactual payoff functions $U_d$, which represent the hypothetical utility that would result under policy $d \in \{ 0,1 \}$.

The counterfactual choice $\yis{d}$ is defined as the maximizer of $U_d$ with respect to the argument $\yv$.
For brevity, we consider a scalar running variable $\y \in \mathcal{Y}$, where $\mathcal{Y}$ is a compact choice interval in $\mathbb{R}$.
We assume there is a single cutoff $\k$ in the actual policy that lies in the interior of $\mathcal{Y}$.
In the actual policy, whether $\y$ exceeds $\k$ determines the treatment under each policy.
The incentive schedule for policy $0$ is applied when $\y \le \k$, and the incentive schedule for policy $1$ is applied otherwise.
Polices $0$ and $1$ are also referred to as the prekink and postkink policies, respectively.

We allow for observed covariates in the counterfactual payoff function, denoted by a vector $\x \in \mathcal{X} \subseteq \R^{d_x}$.
This set of variables represents observably heterogeneous factors involved in the decision-making process of the running variable.
In the tax context, $\x$ may include demographic characteristics, occupation, additional income sources, etc., depending on the modeling details.
As before, $\eta \in \mathcal{H} \subseteq \R$ represents a scalar unobserved preference heterogeneity.
The counterfactual payoff is then summarized as a function $U_d(\yv,\xv,\eta)$.

Now, we make the following assumptions about the counterfactual payoff functions.
\begin{assumption}{1}
\label{asm:kinked-payoff}
For each $d \in \{ 0,1 \}$ and $(\xv,\eta) \in \mathcal{X}\times\mathcal{H}$, the counterfactual payoff $U_d(\yv,\xv, \eta)$ is a strictly concave function in $\yv \in \mathcal{Y}$.
Moreover, it holds that

\begin{align}
\label{eq:single-crossing}
U_0(\yv,\xv,\eta)  -   U_1(\yv,\xv,\eta) \begin{cases}
 \le 0   &\quad \text{if}\quad \yv < \k   \\
 =0 &\quad \text{if}\quad \yv = \k   \\
 \ge 0  &\quad \text{if}\quad \yv > \k
\end{cases}.
\end{align}
for all $\yv \in \mathcal{Y}$ and $(\xv,\eta)\in \mathcal{X}\times\mathcal{H}$.

\end{assumption}

Condition~\eqref{eq:single-crossing} requires that those choosing $\y < \k$ find the postkink policy more beneficial than the prekink policy, and the converse is also true.
The two policies offer the same payoff at the cutoff, which ensures the continuity of the compound payoff function.
Such situations can be commonly found in many progressive tax systems and regulatory policies.

Condition~\eqref{eq:single-crossing} introduces bunching in this general setup.
Assuming differentiability, Condition~\eqref{eq:single-crossing} implies that
$
\dif_{\yv}  U_0(\yv,\xv,\eta)|_{\yv = \k} \ge \dif_{\yv}  U_1(\yv,\xv,\eta)|_{\yv = \k}.
$
If the inequality is strict, the compound payoff function, which can be written as 
$$
U(\yv,\xv,\eta) = U_0(\yv,\xv,\eta)\I{\yv \le \k} + U_1(\yv,\xv,\eta) \I{\yv > \k},
$$
displays a negative kink at the cutoff.
Consequently, the distribution of the actually chosen $\y^*$ exhibits a positive point mass at $\k$ if there is a continuum of $\eta$ that satisfies
\begin{equation*}
     \dif_{\yv}  U_1(\yv,\xv,\eta)|_{\yv = \k} \le 0 \le \dif_{\yv}  U_0(\yv,\xv,\eta)|_{\yv = \k}.
\end{equation*}
Assumption~\ref{asm:kinked-payoff} can be straightforwardly modified to allow for payoff maximization with equality constraints or multiple choice variables.
Equality constraints can be integrated with the payoff function using the Lagrangian or direct substitution, as in the case of the isoelastic model.

The next proposition provides a simple connection between the actual and the counterfactual choices under Assumption~\ref{asm:kinked-payoff}.

\begin{prop}
\label{prop:bunching-characterization}
Let Assumption~\ref{asm:kinked-payoff} hold. Then, it holds
\begin{equation*}
\yiis = \begin{cases}
    \yis{0}  &  \text{if} \quad \yis{0} < \k\\
    \yis{1}  &  \text{if} \quad \yis{1} > \k\\
    \k       &  \text{if} \quad \yis{1} \le \k \le \yis{0}
\end{cases}.
\end{equation*}
The three events correspond to $\{ \yiis < \k \}$, $\{ \yiis > \k \}$, and $\{ \yiis = \k \}$, respectively.
As a result, they are mutually exclusive and comprehensive.
\end{prop}
\citep{goffTreatmentEffectsBunching2024}{Lemma~1} derived the same relationship in the payoff maximization problem under piecewise smooth budget constraints.
Proposition~\ref{prop:bunching-characterization} offers two useful insights into the distribution of $\y^*$.
First, the segments of $\y^*$ to the left and right of the cutoff reveal the distributions of the prekink and postkink counterfactual choices, respectively.
Second, the bunching mass is the intersection of the right-censored portion of the prekink counterfactual distribution and the left-censored portion of the postkink counterfactual distribution.


\subsection{Structural Equations}
\label{subsect:structural-model}
Assume the structural equations for $\ys{d}$ specified as follows:
\begin{align}
\label{eq:structural-equations}
\yis{d} = m(d, \x_i, \eta_i, \theta_0),\ \ d \in \left\{ 0,1\right\}.  
\end{align}
The structural parameter is denoted by $\theta_0 \in \Theta \subseteq \R^{d_\theta}$.
We assume that there exists a utility model satisfying Assumption~\ref{asm:kinked-payoff}, from which the structural equations \eqref{eq:structural-equations} can be derived.
For that utility model to satisfy Condition~\eqref{eq:single-crossing}, the structural equations must satisfy
\begin{equation}
    \label{eq:revealed-preference-restriction}
m(0, \xv, \eta, \theta) \le \k \ \ \implies \ \  m(1, \xv, \eta, \theta) \le \k, \quad \forall \theta \in \Theta,
\end{equation}
in light of Proposition~\ref{prop:bunching-characterization}.
Conversely, if the structural equations meet Condition~\eqref{eq:revealed-preference-restriction}, they are compatible with a utility model satisfying Condition~\eqref{eq:single-crossing}.

To proceed with our analysis, we require the following assumptions about the structural model.

\begin{assumption}{2}
\label{asm:invertible-structural-equations}
For each $d \in \{ 0,1 \}$, the following hold.
\begin{enumerate}[leftmargin = 0.05\linewidth]

\item[(i)] 
For each $\xv \in \mathcal{X}$ and $\theta \in \Theta$, $\yv = m(d,\xv, \eta,\theta)$ is strictly increasing in $\eta$ with the inverse denoted by $\eta = m^{-1}(d,\xv,\yv,\theta)$.

\item [(ii)]
For each $\theta \in \Theta$, $(\xv,\eta) \mapsto m(d,\xv,\eta,\theta)$ and $(\xv,\yv) \mapsto m^{-1}(d,\xv,\yv,\theta)$ are both continuous.

\end{enumerate}
\end{assumption}

Assumption~\ref{asm:invertible-structural-equations} imposes that a scalar unobservable heterogeneity $\eta$ determines the common quantiles of both counterfactual choices conditional on $\x_i = \xv$.
This assumption is standard in the literature on structural estimation  (\citeay{matzkinNonparametricEstimationNonadditive2003}; \citeauthor*{blundellIndividualCounterfactualsMultidimensional2017}, \citeyear{blundellIndividualCounterfactualsMultidimensional2017}, etc).
A similar rank invariance condition can be found in \citet{chernozhukovIVModelQuantile2005} in the context of IV quantile estimation.
We may allow for an extension to cases where $\yv$ is a vector containing a running variable $\yv_1$ and $\eta$ is of the same dimension as $\yv$, although such an extension lies outside the scope of this paper.

A sufficient condition for the utility model to satisfy Assumption~\ref{asm:invertible-structural-equations}(i) is the following: for each $d \in \{ 0,1 \}$ and $\xv \in \mathcal{X}$, $(\yv,\eta)\mapsto U_d(\yv,\xv,\eta)$ has single crossing differences in $\eta$.\footnote{If $U_d(\yv,\xv,\eta)-U_d(\yv',\xv,\eta) \ge 0 $, then $U_d(\yv,\xv,\eta')-U_d(\yv',\xv,\eta') \ge 0$ for all $\eta' \ge \eta$.
}
This condition is implied by a stronger condition of increasing differences: ${\partial^2 U_d(\yv,\xv,\eta)}/{\partial \yv \partial \eta} \ge 0$ for each $d$ and $\xv$.
Assumption~\ref{asm:invertible-structural-equations}(ii) is a standard regularity assumption ensuring continuous responses of individuals.

An important implication of Assumption~\ref{asm:invertible-structural-equations}(i) is the existence of the \textit{reversion}, defined as
\begin{equation}
\label{eq:def-reversion}
    \conv( \yv, \xv, \theta) \equiv m(0, \xv, m^{-1}(1, \yv, \xv,\theta),\theta).
\end{equation}
The reversion function couples one counterfactual choice with the other choice via the identities $\yis{0} = \conv(\yis{1},\x_i,\theta)$ and $\yis{1} = \conv^{-1}(\yis{0},\x_i,\theta)$.
Here, $\conv^{-1}(\yv,\xv,\theta)$ represents the inverse of $\conv(\yv,\xv,\theta)$ with respect to the argument $\yv$, which is strictly increasing by Assumption~\ref{asm:invertible-structural-equations}.
Applying the reversion on both sides of $\ys{1} \le \k$, the bunching condition can be written equivalently as
\begin{equation*}
\I{\yis{1}\le \k \le \yis{0}} = \I{\k \le \yis{0} \le \conv(\k,\x_i,\theta_0)}.
\end{equation*}
Here, $\conv(\k,\xv,\theta_0)-\k \ge 0$ represents the propensity to bunch for those with $\x_i = \xv$.
This implies that the bunching moment is determined by the marginal distribution of $\ys{0}$ and the conditional distribution of $\conv(\k, \x, \theta_0)$ given $\ys{0}$, which highlights a reduction in dimensionality from $\x$.

\subsection{Optimization Frictions}
\label{subsect:optimization-frictions}

The spread of genuine bunchers around the cutoff is referred to as fuzzy bunching in this paper.
The lack of tight bunching poses challenges to the bunching method, as the true bunching status is obscured by those incidentally located around the cutoff without the intention of bunching.

Fuzzy bunching has typically been attributed to optimization frictions in the literature.
Various sources of these frictions, such as adjustment costs, inattention, and other factors, have been extensively discussed in the literature for their roles in dampening the observed elasticity
(e.g., \citeay{chettyAdjustmentCostsFirm2011}; \citeay{klevenUsingNotchesUncover2013}).
This paper chooses to treat these frictions as measurement errors without specifying their nature.
A key assumption is the existence of a narrow window encapsulating the range of measurement errors.
This is specified in the following assumptions along with those on the data-generating process.
\begin{assumption}{3}
\label{asm:data-optimization-errors}
Let $[\uk, \ok] \subseteq \mathcal{Y}$ be a known interval containing $\k$.
For each $d \in \{ 0,1 \}$, $\mathcal{Y}_d = [\underbar{\mathcal{Y}}_d, \bar{\mathcal{Y}}_d]$ contains $\k_d$ in its interior.
Moreover, the following hold.
\begin{enumerate}[leftmargin = 0.05\linewidth]

\item [(i)] If $\yiis = \k$, then $\yii \in [\uk, \ok]$. Moreover, $\yii = \yiis$ holds if $\yii \notin [\uk, \ok]$.

\item [(ii)] The data consist of i.i.d. draws $\{ (\yii, \x_i) : i=1,\ldots, n \}$ following the structural equations.

\item[(iii)] 
For each $d \in \{ 0,1 \}$, the counterfactual choice $\yis{d}$ has a bounded and positive density $f_{\ys{d}}$ on $\mathcal{Y}_d$ such that $\con_{1} \le f_{\ys{d}}(\yv) \le \con_{2}$ for all $\yv \in \mathcal{Y}_d$ and for some $0 <\con_{1} \le \con_{2}$.

\end{enumerate}
\end{assumption}

We assume that the data are i.i.d.
A form of dependence may also be accommodated, although such extensions are beyond the scope of this paper.
The counterfactual distribution must be continuous according to Assumption~\ref{asm:data-optimization-errors}(iii).
This assumption is standard in the literature and implies no point mass observed in the absence of a kink.

Assumption~\ref{asm:data-optimization-errors}(i) requires that observed choices outside the window equal the desired choices.
This assumption is crucial for us to infer the counterfactual distribution from the observed distribution.
This assumption could be economically justified, e.g., in a model where bounded adjustment costs are incurred conditional on one's decision to bunch, as in \citet{chettyAdjustmentCostsFirm2011}.

\citet{blomquistBunchingIdentificationTaxable2021} have adopted the same assumption about the optimization error window.
This assumption is also important in the polynomial strategy, which is a widespread practice in the bunching method.
Practically, the choice of the window is often guided by visual inspection, which can be validated through some robustness checks.

There are alternative approaches to addressing optimization errors.
In the notch literature, it is relatively common to model optimization frictions as a proportion of individuals who do not respond to the notches (\citeay{klevenUsingNotchesUncover2013}; \citeay{bestEstimatingElasticityIntertemporal2020}, etc).
Recently, \citet*{anagolDiffuseBunchingFrictions2024} proposed a different model-based approach, where each unit $i$ receives a sparse set of offers $\{ \yiis + \varepsilon_1,\ldots,\yiis + \varepsilon_M \}$ and locates at $\yii$ that maximizes their payoff among these offers.
The offers are randomly drawn in the vicinity of the desired choice following a certain distribution, represented by $\varepsilon_1,\ldots,\varepsilon_M \overset{iid}\sim F_\varepsilon$.
Their procedure is operationalized without specifying a priori range of optimization errors like ours.
However, it requires researchers' insights into $M$ and the distribution of the offers.
Their treatment of optimization frictions can seemingly be integrated with our method based on the analytic model, which could be a fruitful avenue for future research.

\section{Identification}
\label{sect:identification-results}

This section presents our bunching identification results.
We first propose our semiparametric approach based on the analyticity condition of the counterfactual functions.
It is demonstrated that the structural parameter can be point-identified as an implicit function of the bunching moment under the analyticity condition and mild regularity assumptions.

Our second approach relaxes the analyticity condition on the counterfactual density used in the first approach and considers instead (piecewise) analytic envelope functions bounding the counterfactual distribution.
This approach accommodates a range of counterfactual scenarios, albeit at the expense of resulting in potentially multiple identified values of $\theta$.

\subsection{Point Identification}
\label{subsect:point-identification}
\citet{blomquistBunchingIdentificationTaxable2021} demonstrated that without imposing further restrictions on the unobservable counterfactual density, the elasticity is unrestricted by the observables in the isoelastic model.
Such identification failure occurs similarly in the general bunching design, intuitively because the bunching moment depends on latent component of the counterfactual distribution, which is independent of both the structural parameter and the observables.

This implies that the econometrician must narrow a model of the counterfactual density to reconstruct the unobservable counterfactual from the observables.
We first discuss the required shape restrictions on the counterfactual density that allow point identification.
Let $\mathcal{F}$ denote a model of $f_{\ys{0}}: \CY_0 \to \mathbb{R}_+$.
We assume that if $f\in \mathcal{F}$, $\tilde{f} \in \mathcal{F}$, and $f(\yv) = \tilde{f}(\yv)$ for all $\yv < \uk$, then $f \equiv \tilde{f}$ on $\CY_0$.
If $\mathcal{F}$ satisfies this condition, we say that the model possesses the identity property.
The identity property for the model of the counterfactual CDF was previously proposed by \citep{bertanhaBetterBunchingNicer2023}{Assumption~1}.

Various models enjoy the identity property: many parametric classes such as exponential families with smooth sufficient statistics, their discrete and continuous mixtures, finite-degree polynomials, and most crucially for our analysis, the analytic class in light of the identity theorem from complex analysis (\citeauthor{steinComplexAnalysis2010}, \citeyear{steinComplexAnalysis2010}).
Generally speaking, there is a trade-off between a model's flexibility to fit the data and its capability of allowing unique extrapolation.
For instance, many nonparametric classes, such as $\CC^k$ (where $k \in \Z_+$), Hölder spaces, or Sobolev spaces, offer nearly full flexibility in data fitting.
However, at the expense of this freedom, they cannot provide a unique extrapolation into the unobservable region, highlighting the absence of the identity property.
In this paper, we investigate the use of the analytic function class for bunching identification in a general structural model, as a semiparametric approach to balancing these conflicting goals.
Similar ideas can be found in the econometric literature. 
Since its earliest version in 2020, \citet{pollinger2024kinks} has explored the usefulness of the analyticity assumption for identifying the intensive and extensive margin elasticities in an extended isoelastic model.
With similar motivation but in a different context, \citet{iariaRealAnalyticDiscrete2024} examined the role of real analyticity in identification, extrapolation, and numerical implementation in discrete choice models.

Now, we describe our first approach to point identification.
For simplicity, we focus on the case where $\theta_0$ is a scalar parameter.
However, it is possible to address a vector of structural parameters by developing multiple bunching moments.
Throughout this paper, we follow the common convention of defining the status quo counterfactual choice as $\ys{0}$ rather than $\ys{1}$.
Nevertheless, the roles of $\ys{0}$ and $\ys{1}$ can be interchanged by treating $\{ (-\yii, \x_i) \}_{i=1}^n$ as a new dataset with the corresponding adjustments to the cutoff and the optimization error window.

Let $\ti = \t(\x_i)$ be a bounded, nonnegative function of the covariates such that $\E[\ti] > 0$.
Given $\ti$, we define a new probability measure as $\Q(A) = \E[\1_A \ti]/\E[\ti]$.
This offers a measure of bunching that may be different from the unweighted fraction, based on the weighting function $\t(.)$.
For various choices of $\t(.)$, these measures can provide a set of moments that can be used as identifying restrictions for $\theta_0$.
If $\ti$ is set to a constant 1, $\Q$ reduces to $\P$, which is sufficient for identifying a scalar parameter.

For each $\qua \in (0,1)$ and $\theta \in \Theta$, we define
\begin{IEEEeqnarray*}{rll}
	& f^{(\Q)}_{\ys{0}}(\yv) &:= \  f_{\ys{0}}(\yv) \frac{\E[\ti |\yis{0} = \yv]}{\E[\ti]} \equiv \dif_\yv\Q(\ys{0} \le \yv),\\
	& \qb(\qua, \theta, \yv)\  &:=\ \quant^{(\Q)}_{\conv( \ok ,\x,\theta)|\ys{0}}(\qua | \yv),
\end{IEEEeqnarray*}
as the counterfactual density of $\ys{0}$ and the conditional quantile function of $\conv(\ok,\x,\theta)$ given $\ys{0}$, both measured with respect to $\Q$.
Here, $\conv(\yv,\xv,\theta)$ denotes the reversion as defined in \eqref{eq:def-reversion}.
We may interpret $\conv(\yv,\xv,\theta)$ as the predicted value of $\ys{0}$ in response to the shift from policy $1$ to $0$, given that the unit was at $\ys{1} = \yv$ and $\x = \xv$.
In this sense, $\conv(\k,\x,\theta) - \k$ quantifies the predicted response to a potential repeal of the kink for individuals located at the right edge of the bunching population.

A function $f:[\ubars,\bars] \to \mathbb{R}$ is said to be analytic if, for every $\yv \in [\ubars, \bars]$, there is an open neighborhood around $\yv$ on which $f$ agrees with its Taylor expansion at $\yv$.
Moreover, it is well-known that an analytic function $f:[\ubars,\bars]\to\mathbb{R}$ can be extended to a complex-analytic function $\tilde{f}:U\to\mathbb{C}$ over an open planar set $U \subseteq \mathbb{C}$ containing $[\ubars,\bars]$.
Our identification theory leverages rich results in the theory of complex analysis applicable to such analytic continuations.

We now formally define the analyticity condition of the counterfactual density and conditional quantile functions.

\begin{define}
\label{def:analyticity-condition}
Let $\appftna{\qua}:[\ubars, \bars] \to \R$ be an analytic function and $\{ \appftnb{\qua}: [\ubars, \bars] \to\R; \qua \in (0,1)\}$ be a collection of analytic functions.
Then, $\appftna{\qua}$ and $(\appftnb{\qua})_{\qua\in (0,1)}$ satisfy the analyticity condition on $[\ubars, \bars]$ for smoothness constants $(\rad,\beta,\delta) \in \mathbb{R}_+^3$, if the following are true:\footnote{
    $\Re(z) = x$ denotes the real part of a complex number $z =x+\i y$, and $|z|=\sqrt{x^2+y^2}$ denotes the modulus.
    }
\begin{enumerate}[leftmargin = 0.05\linewidth]

    \item [(i)] $\int_{\ubars}^{\bars}\int_0^{1} |\appftna{\qua}(\yv + \rad e^{2\pi \i x})| dx d\yv \le \beta \int_{\ubars}^{\bars} |\appftna{\qua}(\yv)| d\yv$,

    \item [(ii)] $ \sup_{x \in [0,1]}|\appftnb{\qua}(\yv + \rad e^{2\pi \i x}) - \yv | \I{\appftnb{\qua}(\yv)-\yv\ge 0} \le \delta \rad$ for all $\yv \in [\ubars, \bars]$ and $\qua \in (0,1)$,

    \item [(iii)] $ \sup_{\yv \in [\ubars,\bars], \tau \in (0,1), x \in [0,1]}|\appftnb{\qua}(\yv + \rad e^{2\pi \i x})| < \infty$.

\end{enumerate}
Here, $h(\yv + r e^{2\pi \i x}):= \sum_{j=0}^\infty \frac{1}{j!}\dif_\yv^j h(\yv) r^j e^{2\pi \i j x}$, $0\le r < R$, $x \in [0,1]$, $\yv \in [\ubars,\bars]$, represents the unique analytic continuation of an analytic function $h:[\ubars,\bars] \to \mathbb{R}$ onto the $R$-neighborhood of $[\ubars,\bars]$ in $\mathbb{C}$.
\end{define}

The smoothness constants $(\rad,\beta,\delta)$ control the variability of the functions $\appftna{\qua}$ and $(\appftnb{\qua})_{\qua \in (0,1)}$.
The constant $\rho$ is at least as large as the radii of convergence of these functions.
Given the value of $\rad$, the constants $\beta$ and $\delta$ more concretely govern variations in $\appftna{\qua}(\yv)$ and $\appftnb{\qua}(\yv)$ within the $\rad$-neighborhood of $[\ubars,\bars]$.
In Assumption~\ref{asm:identification-regularity}, we require that $\delta < 1$ to ensure that bunching does not extend to the upper limit of the support.
The intuition is that the higher $\ys{0}$ is, the fewer incentives there should be for bunching, as will be clarified in the proof of Theorem~\ref{thm:partial-identification}.
Further technical remarks are relegated to \OA.

For each $\theta \in \Theta$, let $\ook(\theta) \in [\k, \oy]$ be any constant satisfying $\Q( \conv(\ok, \x_i,\theta) > \ook(\theta),  \yii \in [\uk, \ok]  ) = 0$.
We impose the analyticity condition on the counterfactual density and the conditional quantile functions as follows.

\begin{assumption}{4}
	\label{asm:identification-regularity}
    Let $\ti = \t(\x_i) \ge 0$ satisfy $\inf_{\yv \in \CY_0}\E[\ti|\yis{0}=\yv] \ge \con_{1}$ and $\ti \le \con_{2}$ for some $0 < \con_{1} \le \con_{2}$.
    Define $\Q$ as a probability measure such that $\Q(A) = \E[\mathbbm 1_A \ti]/\E[\ti]$.
	Moreover, the following hold.
	\begin{enumerate}[leftmargin = 0.05\linewidth]
		\item [(i)] 
		$f^{(\Q)}_{\ys{0}}$ and $(\qb(\qua, \theta, .))_{\qua \in (0,1)}$ satisfy the analyticity condition (Definition~\ref{def:analyticity-condition}) on $[\uk, \ook(\theta)]$ for each $\theta \in \Theta$ with some uniform smoothness constants $(\rad,\beta,\delta)$ such that $\delta \in [0,1)$.

		\item [(ii)] $\E[\ti  \I{\uk\le \yis{0} \le \conv( \ok,\x_i,\theta)}]$ is one-to-one in $\theta \in \Theta$.

	\end{enumerate}
	
\end{assumption}

Assumption~\ref{asm:identification-regularity}(i) plays a central role in deriving our identification results.
Loosely speaking, it requires the analyticity of three functions: the counterfactual density $f_{\ys{0}}$, the conditional expectation function of $\ti$, and the conditional quantile functions of $\conv(\ok,\x_i,\theta_0)$.
Consequently, these functions must be uniformly approximated by a sequence of finite-degree polynomials with exponentially decreasing errors (\citeay{timanTheoryApproximationFunctions1994}).
Since the distribution of $\yis{0}$ is observed only in a censored manner, the analyticity of these functions is untestable in the unobservable region.
Moreover, there are no formal statistical procedures for testing the analyticity of densities or quantile functions, to the best of my knowledge, even when they are fully observable.
Thus, we suggest visually supporting this assumption by examining various bin-based estimates of these functions in the observed region.

A sufficient condition for Assumption~\ref{asm:identification-regularity}(i) is that the joint density of $(\yis{0}, \conv(\ok,\x_i,\theta))$ is analytic and bounded away from $0$ in the support.
This assumption can be further supported by the analyticity of the joint density of $(\eta_i, \x_i)$.
Such a condition would restrict $f_{\ys{0},\conv(\ok,\x,\theta)}$ to be a continuous function, as well as their derivatives.

Assumption~\ref{asm:identification-regularity}(i) may appear to restrict $\conv(\ok,\x_i,\theta)$ to be continuously distributed.
However, it can be extended to cases where $\conv(\ok,\x_i,\theta)$ has finitely many non-overlapping components in its support by conditioning on membership to each component.
This situation arises, for example, when $X_i$ is a mixture of dummy variables and continuous covariates.
In such cases, the requirement is that each conditional $\Q$-probability assigned to each component should be an analytic function of $\yv$.

Assumption~\ref{asm:identification-regularity}(ii) stipulates that the bunching moment predicted by the model must be able to differentiate the true value from the others, according to the true distribution of $(\ys{0}, \x)$.
This condition ensures global identification through the comparison of moments in Theorem~\ref{thm:point-identification}.
In the case where $\theta$ is a scalar parameter, the identification condition is fulfilled when the reversion $\conv$ is monotonic in $\theta$.

Now, we are prepared to present our point identification result.
\begin{thm}
	\label{thm:point-identification}
	\noindent Let Assumptions~\ref{asm:kinked-payoff}, \ref{asm:invertible-structural-equations}, \ref{asm:data-optimization-errors}, and \ref{asm:identification-regularity} hold.
    Then, $\theta_0 \in \Theta$ is a unique solution to the equation $\bf(\theta) = \E[\ti  \I{\yii \in [\uk, \ok]}]$, where $\bf : \Theta \to [0, \infty)$ is defined as
			\begin{align}
				\label{eq:bunching-moment-equation}
				\bf(\theta) := \sum_{j=1}^\infty \frac{1}{j!} \dif_\yv^{j-1} [ \E[\ti  (\conv( \ok,\x_i,\theta) -\uk)^j | \yis{0} = \yv] f_{\ys{0}}(\yv)]_{\yv = \uk}.
			\end{align}
		Moreover, for all $l \in \mathbb{Z}_+$ and $\theta \in \Theta$, it holds
			\begin{equation}
				\label{eq:remainder-bound}
				\left| \sum_{j=l+1}^\infty \frac{1}{j!} \dif_\yv^{j-1} [ \E[\ti  (\conv( \ok,\x_i,\theta) -\uk)^j | \yis{0} = \yv] f_{\ys{0}}(\yv)]_{\yv = \uk} \right| \le \beta \delta^l \bf(\theta).
			\end{equation}
		
		
		
		
	
\end{thm}

Our point identification result crucially rests upon Assumption~\ref{asm:identification-regularity}(i) to express a bunching moment as a series of functionals involving counterfactual quantities.
If the analyticity condition is violated, our method would identify a pseudo-true value arising from the fitted distribution within the analytic model.

To see why $\bf(\theta)$ can be inferred from the observables, we use the fact that $\yii = \yis{0}$ for all units $i$ such that $\yii < \uk$ by Proposition~\ref{prop:bunching-characterization}.
Then, we can express the $j$th summand on the right-hand side of \eqref{eq:bunching-moment-equation} as
\begin{align*}
	& \frac{1}{j!} \dif_\yv^{j-1} [ \E[\ti  (\conv( \ok,\x_i,\theta) -\uk)^j | \yis{0} = \yv] f_{\ys{0}}(\yv)]_{\yv = \uk} \\
	= &\  \frac{1}{j!} \dif_\yv^{j-1} [ \E[\ti  (\conv( \ok,\x_i,\theta) -\uk)^j | \yii = \yv] f_{\y}(\yv)]_{\yv = \uk-},	
\end{align*}
which is determined by the distribution of $(\y,\x)$.
As a result, $\theta_0$ can be identified.

The observed vector $\x$ affects the function $B(\theta)$ through the conditional moments of $\conv(\ok,\x,\theta)$ given $\ys{0}$.
Notably, it is possible to accommodate a moderately large or high-dimensional $\x$ without compromising efficiency, so long as $\conv(.,\theta)$ depends on a low-dimensional parameter $\theta$.
Previous approaches have either partitioned the sample based on a limited number of $\x$ values or simply consolidated $\x$ into its empirical average in the data, which could serve as an approximation to the aggregation in \eqref{eq:bunching-moment-equation}.
Such methods may lead to concerns about aggregation bias or information loss, which can be addressed in our approach.

Consider the model $\yis{0} = \conv(\yis{1},\x_i,\theta)$, where the observed bunching is assumed to be tight at $\k$.
Assume, however, that the correct specification is $\yis{0} = \conv(\yis{1},\x_i,\theta_i)$, where $\theta_i \in \mathbb{R}$ represents a unit-specific parameter.
Then, the pseudo-true value $\theta_*$ identified through the bunching method corresponds to the solution to the equation
\begin{align*}
\P(\k \le \yis{0} \le \conv(\k, \x_i, \theta_*)) =  \P(\k \le \yis{0} \le \conv(\k, \x_i, \theta_i)).
\end{align*}
Suppose that $f_{\ys{0}}$ is flat around the bunching interval, that is, $\dif^{j} f_{\ys{0}}(\k) \approx 0$ for all $j \in \mathbb{N}$.
By a slight extension of Theorem~\ref{thm:point-identification}, this implies that the equation above expands to
\begin{equation*}
   \E[\conv(\k, \x_i, \theta_*) - \k | \yis{0} = \k]f_{\ys{0}}(\k) \approx \E[\conv(\k, \x_i, \theta_i) - \k | \yis{0} = \k]f_{\ys{0}}(\k),
\end{equation*}
which shows that $\theta_*$ approximately satisfies
$$
\E[\conv(\k, \x_i, \theta_*)| \yis{0} = \k] \approx \E[\conv(\k, \x_i, \theta_i)| \yis{0} = \k].
$$
In this sense, when $\theta_i$ is misspecified but $\x_i$ is correctly specified, the pseudo-true value approximately identifies a weighted average of $\theta_i$:
$$
\theta_* \approx \frac{\E[ \dif_{\theta} \conv(\k, \x_i, \theta_*) \cdot \theta_i | \yis{0} = \k]}{\E[ \dif_{\theta} \conv(\k, \x_i, \theta_*)| \yis{0} = \k]},
$$
where the weights are nonnegative if $\conv(\k,\xv,\theta)$ is monotonic in $\theta$ for each $\xv \in \mathcal{X}$.
This reaffirms a similar relationship derived in \citet{klevenUsingNotchesUncover2013} in the absence of $\x$.

\paragraph*{Comparison to existing results}
There are several existing approaches to bunching identification in the literature.
The most closely related to our result is the small kink approximation, initially proposed by \citet{saezTaxpayersBunchKink2010} as a nonparametric identification method.
It says that the normalized bunching fraction approximately identifies the local average of individual responses to policy shift at the cutoff:
\begin{equation}
	\label{eq:small-kink-approximation}
	\frac{\P(\yiis = \k)}{f_{\ys{0}}(\k)} \approx \E[\yis{0} - \yis{1}|\yis{0} = \k],
\end{equation}
provided that either the policy response $\ys{0}-\ys{1}$ is small or the counterfactual density is flat around the bunching interval.
In the isoelastic model, after a further substitution $\yis{0}-\yis{1} = \theta_0 \log \left( \frac{1-\tau_0}{1-\tau_1}\right) \k$ into \eqref{eq:small-kink-approximation}, this approximately identifies the true elasticity as
\begin{equation*}
	\theta_0 \approx \frac{\P(\yiis = \k) / f_{\ys{0}}(\k)}{\k \log \left( \frac{1-\tau_0}{1-\tau_1}\right)}.
\end{equation*}
This identification method has been a predominant approach in the kink and notch design literature, with its theoretical foundation and applications further developed in subsequent works, such as \citet{chettyAdjustmentCostsFirm2011}, \citet{klevenUsingNotchesUncover2013}, \citet{blomquistBunchingIdentificationTaxable2021}, and \citet{goffTreatmentEffectsBunching2024}, among many others.

The small kink approximation relies on the first moment of the individual policy responses and the value of the counterfactual density at the cutoff to approximate the bunching fraction.
Our result complements this approximate relationship with the omitted higher-order effects of the individual policy responses and the derivatives of the counterfactual density.
Overlooking these terms may lead to significant bias in the counterfactual estimation, which can be addressed in our procedure.

\citet{bertanhaBetterBunchingNicer2023} proposed two semiparametric identification approaches to the elasticity in the isoelastic model.
The first approach is based on the mid-censored Tobit regression facilitated by the observed covariates $\x$.
It is based on the quasi-maximum likelihood estimation (QMLE) of the following model:
\begin{align*}
	y_i &= \begin{cases}
 \theta_0\log (1-\tau_0) + \eta_i & \te{ if } y_i < k\\
 k & \te{ if } y_i = k\\
 \theta_0\log (1-\tau_1) + \eta_i & \te{ if } y_i > k\\
\end{cases},
\end{align*}
where $\eta = X'\beta_0 + e$ and $e | X \sim N(0,\sigma_0^2)$ for some $\beta_0 \in \R^{d_x}$ and $\sigma_0^2 > 0$.
The conditional normality assumption enables estimation of $(\theta_0, \beta_0, \sigma_0^2)$ through the QMLE on the observable $(\yv,\x)$.
\citet{bertanhaBetterBunchingNicer2023} emphasized that the identification of $\theta_0$ does not require the full power of the conditional normality assumption.
Under a weaker condition that the unconditional distribution of $\eta$ coincides with a mixture of Gaussian distributions with mean $\x'\beta_0$ and variance $\sigma_0^2$, they demonstrated that $\theta_0$ can be identified and consistently estimated.

Their identification hinges on the assumption $F_{\eta}(h) = \E_X[\Phi( \frac{h - X'\beta_0}{\sigma_0})]$ for some pseudo-true values $(\beta_0,\sigma_0^2)$, where $\Phi$ denotes the normal CDF.
They incorporated $X$ into the model primarily to improve its fit to the counterfactual distribution.
In contrast, the our main motivation is to enhance the flexibility of the model.
Notably, the Gaussian mixture CDF $F_\eta$ defined above satisfies Assumption~\ref{asm:identification-regularity}(i).\footnote{We prove the global analyticity of $F_\eta$ and $f_\eta$ in \OA.}
Therefore, $\theta_0$ can be identified through our means without requiring additional covariates.
However, improving the estimation procedure by incorporating information provided by $\x$ would be an interesting avenue for future research.

Their second approach is based on the restriction on a conditional quantile of $\eta$.
They assume, for some $\tau \in (0,1)$ and $\beta_0$, the conditional $\tau$ quantile of $\eta$ given $X$ is given by
$
	\quant_{\eta|X}(\tau, x) = x'\beta_0.
$
This, in turn, implies that the conditional $\tau$ quantile of $y$ given $X$ assumes a parametric form similar to the mid-censored Tobit regression model.
This can be estimated using a quantile regression under a suitable rank condition on $X$.
This approach provides an alternative identification means through the conditional quantile restriction on the counterfactual distribution, instead of restricting the entire shape.
On the other hand, it does not allow for semi- or nonparametric specification of the conditional quantile function due to the rank condition.
Our approach would serve as a valuable complement in situations where researchers are uncertain about such restrictions.

\subsection{Partial Identification}
\label{subsect:partial-identification}

We relax the analyticity condition on the counterfactual density by allowing the counterfactual cumulative distribution function to be bounded by a pair of analytic envelope functions.
Specifically, we assume the following.
\begin{assumption}{5}
\label{asm:envelope-condition}
Let $\uf$ and $\of$ be envelope functions such that for all $\yv \in [\uk, \ook(\theta_0)]$, it holds
\begin{align}
\label{eq:envelope-assumption}
\Q(\uk \le \yis{0}\le \yv) & \equiv \int_{\uk}^\yv  f^{(\Q)}_{\ys{0}}(z)dz \in \left[ \int_{\uk}^\yv \uf(z)dz, \int_{\uk}^\yv  \of(z)dz\right].
\end{align}
\end{assumption}
Condition~\eqref{eq:envelope-assumption} imposes an inequality restriction similar to first-order stochastic dominance, except that $\of$ and $\uf$ are not required to be valid probability densities.
Note that \eqref{eq:envelope-assumption} is implied by
\begin{equation*}
	f^{(\Q)}_{\ys{0}}(\yv) \in [\uf(\yv), \of(\yv)], \quad \forall \yv \in [\uk, \ook(\theta_0)].
\end{equation*}
However, \eqref{eq:envelope-assumption} provides a slightly more general restriction than the direct bounds on the counterfactual density.

Our approach to partial identification, which embraces a range of counterfactual scenarios, allows for identification under less stringent assumptions at the expense of potentially resulting in a continuum of identified values.
It can accommodate several parametric and nonparametric shape restrictions in the literature, including the trapezoidal approximation (\citeay{saezTaxpayersBunchKink2010}), densities with bounded oscillations (\citeay{blomquistBunchingIdentificationTaxable2021}), bi-log-concave CDF (\citeay{goffTreatmentEffectsBunching2024}), and Lipschitz continuous densities (\citeay{bertanhaBetterBunchingNicer2023}).
We derive a set of sharp identification bounds in Theorem~\ref{thm:partial-identification} subject to the restriction \eqref{eq:envelope-assumption}.
That is, the lower bound is attained when $\uf \equiv f^{(\Q)}_{\ys{0}}$, and the same applies to the upper bound.
When applied to the isoelastic model, Theorem~\ref{thm:partial-identification} yields the same bounds as those established in \citep{blomquistBunchingIdentificationTaxable2021}{Theorem~2} and \citep{bertanhaBetterBunchingNicer2023}{Theorem~2} subject to the respective shape restrictions.
This assertion is demonstrated in \OA.
We now state the analyticity assumption on the envelope functions.

\begin{assumption}{4$'$}
\label{asm:partial-identification-regularity}
Let $\ti = \t(\x_i) \ge 0$ satisfy $\inf_{\yv \in \CY_0}\E[\ti|\yis{0}=\yv] \ge \con_{1}$ and $\ti \le \con_{2}$ for some $0 < \con_{1} \le \con_{2}$.
Moreover, $\of$ and $(\qc(\qua, \theta, .))_{\qua \in (0,1)}$ satisfy the analyticity condition on $[\uk, \ook(\theta)]$ for each $\theta \in \Theta$ with some uniform smoothness constants $(\rad,\bar{\beta}, \delta)$ such that $\delta \in [0,1)$.
Similarly, $\uf$ and $(\qc(\qua, \theta, .))_{\qua \in (0,1)}$ satisfy the analyticity condition on $[\uk, \ook(\theta)]$ for each $\theta \in \Theta$ with uniform smoothness constants $(\rad,\underbar{\beta},\delta)$.
\end{assumption}
The following theorem presents our partial identification result.

\begin{thm}
	\label{thm:partial-identification}
	Let Assumptions~\ref{asm:kinked-payoff}, \ref{asm:invertible-structural-equations}, \ref{asm:data-optimization-errors}, \ref{asm:partial-identification-regularity}, and \ref{asm:envelope-condition} hold.
	Then, it holds that $\Q({\yii \in [\uk, \ok]}) \in [\ubf(\theta_0), \obf(\theta_0)]$, where for all $\theta \in \Theta$,
		\begin{align*}
		\obf(\theta) & = \sum_{j=1}^\infty \frac{1}{j!} \dif_\yv^{j-1} [ \E_\Q[(\conv(\ok,\x_i,\theta) - \uk)^j | \yis{0} = \yv] \of(\yv) ]_{\yv = \uk},\\
		\ubf(\theta) &=  \sum_{j=1}^\infty \frac{1}{j!} \dif_\yv^{j-1} [ \E_\Q[(\conv(\ok,\x_i,\theta) - \uk)^j | \yis{0} = \yv] \uf(\yv)]_{\yv = \uk}	.
		\end{align*}
	Moreover, if $\of$ and $\uf$ are both nonnegative functions, it holds for all $l \in \mathbb{Z}_+$ and $\theta \in \Theta$,
		\begin{align*}
			\left| \sum_{j=l+1}^\infty \frac{1}{j!} \dif_\yv^{j-1} [ \E_\Q[(\conv(\ok,\x_i,\theta) - \uk)^j | \yis{0} = \yv] \of(\yv)]_{\yv = \uk} \right| \le \bar{\beta}\delta^l \obf(\theta),\\
			\left| \sum_{j=l+1}^\infty \frac{1}{j!} \dif_\yv^{j-1} [ \E_\Q[(\conv(\ok,\x_i,\theta) - \uk)^j | \yis{0} = \yv] \uf(\yv) ]_{\yv = \uk} \right| \le \underbar{\beta}\delta^l \ubf(\theta).
		\end{align*}
\end{thm}

It is possible to relax Assumption~\ref{asm:partial-identification-regularity} in Theorem~\ref{thm:partial-identification} to allow for piecewise analytic envelope functions.
For this extension, we refer to Theorem~\ref{thmA:extension-partial-iden} in \OA.
Such an extension is useful for handling piecewise linear envelope curves arising from the Lipschitz constraint (\citeay{bertanhaBetterBunchingNicer2023}).


\section{Bunching Confidence Region}
\label{sect:confidence-region}

The idea of bunching confidence region is to collect values of $\theta$ that are compatible with the observed pattern of bunching.
More specifically, the identification results in Section~\ref{sect:identification-results} enable us to draw empirically testable implications from the structural parameter, expressed as either moment equality or inequality restrictions.
We construct a test for these restrictions under the hypothesis $H_0 : \theta_0 = \theta$.

In this section, we first present our counterfactual estimation scheme, termed the generalized polynomial strategy.
We delineate our procedure based on the counterfactual correction and polynomial sieve estimation.
An asymptotically valid test is proposed with the growth conditions to guide the proper selection of tuning parameters.
Lastly, we propose a modified procedure for the inference on values in a partially-identified set derived from Theorem~\ref{thm:partial-identification}.

\subsection{Generalized Polynomial Strategy}
\label{subsect:point-estimation}
This subsection details our counterfactual estimation scheme, referred to as the generalized polynomial strategy.
We assume conditions in Theorem~\ref{thm:point-identification} for the point identification of $\theta_0$.
Our aim is to construct a confidence set $C_n$ satisfying
$
\liminf_{n\to\infty} \P(\theta_0 \in C_n) \ge 1- \alpha,
$
where $\alpha \in (0,1)$ is a prescribed coverage level.
It is well-known that such a confidence region can be constructed by inverting a test for the hypothesis $H_0: \theta_0 = \theta$, which will be our focus.

Let $\theta \in \Theta$ be a hypothesized value and assume $H_0:\theta_0 = \theta$ is correct.
Using Theorem~\ref{thm:point-identification}, we can express the observed bunching moment as the following infinite sum:
\begin{align*}
& \E[\ti \I{\uk \le \yii \le \ok }]  = \sum_{j=1}^\infty \frac{1}{j!} \dif_{\yv}^{j-1}[ \E[\ti  (\conv(\ok,\x_i,\theta)-\uk)^j |\yis{0} = \yv] f_{\ys{0}}(\yv)]_{\yv = \uk}.
\end{align*}
We test this relationship by estimating the functions $f_j(\yv) := \E[\ti  w_i^j |\yis{0} = \yv] f_{\ys{0}}(\yv)$.
Here, we denote $w_i = \conv(\ok,\x_i,\theta) - \uk \ge 0$, where the nonnegativity follows from the restriction \eqref{eq:revealed-preference-restriction} and the monotonicity of the reversion.

Our procedure comprises two steps: (i) the construction of the estimation sample and (ii) the sieve estimation of the counterfactual functions.
We present a pseudo-code in Algorithm~\ref{alg:code-test-statistic} to provide an overview of the procedure.
The construction of the estimation sample necessitates an appropriate counterfactual adjustment procedure.
We name the proposed method as the counterfactual correction.
The method crucially hinges on the existence of the reversion, $\conv(\yv, \xv, \theta)$, which is directly applied to transform the data.

Since $f_j$ is analytic on $\CY_0$, as established in Lemma~\ref{lemA:sufficient-analytic}, these functions as well as their derivatives can be globally approximated by a sequence of polynomials of increasing degrees.
This underpins our estimation scheme using the polynomial sieve.
For each $j \in \mathbb{N}$, note that $f_j$ is proportional to the density $\dif_\yv {\P}_{j}(\yis{0} \le \yv)$, where ${d{\P}_{j}}/{d \P} = \ti w_i^j/\E[\ti w_i^j]$.
Motivated by
$$
f_j \in \argmax{f > 0} \E[\ti w_i^j \log (f(\yis{0})) \I{\yis{0} \in \mathcal{S}}] - \int_{\mathcal{S}} f(\yv)d\yv
$$
for any region $\mathcal{S} \subseteq \CY_0$, we propose a sieve M-estimator for $f_j$ with mid-censored data.
By recasting the problem as the estimation of a reweighted density, we can estimate $f_j$ without binning the observations.


\SetKwInOut{KwFinOut}{Final output}

\begin{algorithm}[ht!]
\caption{Computation of the test statistic $\mathcal{T}_n(\theta)$ for $H_0: \theta_0 =\theta$.}
\label{alg:code-test-statistic}

\KwIn{Data $\{(\yii, \x_i):i=1,\ldots, n\}$, weighting function $\t(.)$, hypothesized value $\theta \in \Theta$, support $[\uy,\oy]$, excluded window $[\uk, \ok]$, polynomial order $\kap \in \N$, and approximation order $\tc \in [1, \kap]$.
}

Set $\ook(\theta) \leftarrow \max_{i:\yii \in [\uk, \ok],\ti>0} \conv(\ok,\x_i,\theta)$ and $\what B \leftarrow \frac{1}{n} \sum_{i=1}^n \ti \I{\yii \in [\uk, \ok]}$.

Set $\S \leftarrow [\uy, \uk]\cup(\ook(\theta),\oy]$.

\For{$i=1,\ldots,n$, $\yii \not\in [\uk, \ok]$}{
\If{$\yii \in [\uy , \uk)$}{
Set $(\yi{0},\x_i) \leftarrow (\yii,\x_i)$ and include it in the estimation sample.
}
\ElseIf{$\yii \in (\ok , \infty)$}{
\If{$\conv(\yii, \x_i, \theta) \in (\ook(\theta), \oy]$}
{
Set $(\yi{0},\x_i)  \leftarrow (\conv(\yii, \x_i, \theta),\x_i)$ and include it in the estimation sample.
}
}
}

\KwOut{The estimation sample $\{ (\yi{0},\x_i) : i \in \nn \}$.}

\For{$j = 1,2,\ldots, \tc$}{

    Compute $\what \gamma_{\kap j} = (\what \gamma_{\kap j, 1},\ldots, \what \gamma_{\kap j, \kap})'$ as defined in \eqref{define:estimator}.

\For{$i=1,\ldots,n$, $i \in \nn$}{
    Set $\what f_{\kap j i} \leftarrow z_{\kap i}' \what \gamma_{\kap j}$.

    Set
    $
    \what \nu_{\kap ji}= (\what \nu_{\kap ji,1},\ldots,\what \nu_{\kap ji,\kap})' \leftarrow ( \frac{1}{n} \sum_{i\in \nn} \ti w_i^j \frac{z_{\kap i} z_{\kap i}'}{\what f_{\kap j i}^2} )^{-1} \ti w_i^j \frac{z_{\kap i}}{\what f_{\kap j i}}
    $

}
}

Set $\what \mu_{n} \leftarrow \what B - \sum_{j=1}^\tc \frac{1}{j}\what \gamma_{\kap j, j}$.

\For{$i=1,\ldots, n$}{
Set
$
\what s_{n i} \leftarrow 0.
$

\If{$Y_i \in [\uk, \ok]$}{
Set 
$
\what s_{n i} \leftarrow \ti \I{\yii \in [\uk, \ok]}.
$
}
\ElseIf{$i \in \nn$}{
Set 
$
\what s_{n i} \leftarrow \sum_{j=1}^\tc \frac{1}{j}\what \nu_{\kap ji, j}.
$
}
}

Set $\what \sigma_{n}^2 \leftarrow \frac{1}{n}\sum_{i=1}^n \what s_{n i}^2$ and $\mathcal{T}_n(\theta) \leftarrow \left| {\sqrt{n} \what \mu_{n}}/{\sqrt{\what \sigma_{n}^2}} \right|$.

\KwFinOut{Test statistic $\mathcal{T}_n(\theta)$ for testing $H_0 : \theta_0 = \theta$.}

\end{algorithm}
\subsubsection{Counterfactual Correction}
This subsection presents our proposal for counterfactual adjustment that allows consistent estimation of the counterfactual quantities.
For the estimation of the counterfactual related to $\ys{0}$, we need inputs that correctly reflect the distribution of $\ys{0}$ despite censoring.
On the one hand, one can observe $f_{\ys{0}}$ to the left of the excluded window.
This enables us to extrapolate $f_{\ys{0}}$ into the unobservable region.
However, using only the data in the left segment is not the most efficient idea, as it renders counterfactual estimation highly noisy.
There are significant benefits in taking advantage of the data to the right of the window for improving counterfactual estimation.
To this end, an appropriate counterfactual adjustment procedure is necessitated since the data on the right represent $f_{\ys{1}}$, not $f_{\ys{0}}$.

The proportional adjustment is a common counterfactual adjustment scheme in the polynomial strategy in the presence of an excess bunching fraction.
In the proportional adjustment procedure, the adjustment of the form $\what f_{\ys{0}} = (1+c_j) \what f_{\ys{1}}$ is sequentially applied to $\what f_{\ys{1}}$, where the factor $c_j>0$ is updated so as to ensure the resulting estimate of $\what f_{\ys{0}}$ integrates to $1$.
However, the proportional adjustment may not align with the structural assumptions, leading to incorrect analysis.
For instance, assume that $\ys{0} = \conv(\ys{1})$.
Then, the change of variables implies that $f_{\ys{0}}(\yv) = f_{\ys{1}}(\conv^{-1}(\yv)) \frac{1}{\conv'(\conv^{-1}(\yv))}$, which in general differs from $f_{\ys{0}}(\yv) = (1+c) f_{\ys{1}}(\yv)$.

Our counterfactual correction procedure leverages the truth of the hypothesis $H_0:\theta_0 = \theta$.
The procedure can be described as follows.
Let $(\yii, \x_i)$ be a data point.
If $\yii > \ok$, we impute the missing value of $\yis{0}$ as $\yi{0} = \conv(\yii,\x_i,\theta)$ by applying the reversion.
If $\yii < \uk$, we define $\yi{0} = \yii$.
If $\yii \in [\uk, \ok]$, unit $i$ is excluded from the estimation sample.
Iterate this over all units and collect the values of $(\yi{0},\x_i)$ for the retained units.
The initial estimation sample is then defined as $\{ (\yi{0},\x_i):\yii<\uk \text{ or } \yii > \ok \}$, where
\begin{align*}
    \yi{0} := \begin{cases}
\yii & \ \text{if}\ \yii < \uk\\
\conv(\yii,\x_i,\theta) & \ \text{if}\ \yii > \ok\\
\text{NA} & \ \text{if}\ \yii \in [\uk, \ok]
\end{cases}.
\end{align*}
Assuming that $\theta_0 = \theta$ and the structural equations are correctly specified, we have $\yi{0} = \yis{0}$ for all units in the estimation sample.
However, this does not immediately imply that valid counterfactual estimation is possible based on this estimation sample.
To see this point, we note that
\begin{equation*}
    f_{\y{}(0)}(\yv) = f_{\ys{0}}(\yv) (1-\P(\yii \in [\uk, \ok]|\yis{0} =\yv)).
\end{equation*}
This equation accounts for the fact that units engaged in bunching are excluded from the estimation sample not based on their $\ys{0}$ values, but on a random selection criterion.
Since the proportion of bunchers may be nonzero for some values of $\yi{0}$ in the estimation sample, a further truncation is necessitated.

We resolve this issue by truncating $\yi{0}$ above a threshold beyond which the bunching fraction becomes $0$.
To implement this, we recall the bunching condition $\{\uk \le \yis{0} \le \conv(\ok, \x_i, \theta)\}$.
This implies that the bunching will not occur beyond
\begin{align*}
\ook(\theta) &:= \max_{i: \CB_i = 1} \conv( \ok, \x_i,  \theta),
\end{align*}
where $\CB_i = \I{\yii \in [\uk, \ok],\ti > 0}$ indicates the observed bunching status.
By construction, we have $\ook(\theta) \ge \max_{i: \CB_i=1}\yis{0}$.
Under mild conditions, $\ook(\theta)$ is a consistent estimator for an upper bound of $\ys{0}$ among bunchers.
This allows us to treat $\ook(\theta)$ as if it were a deterministic constant for the sake of counterfactual estimation.

This leads to the final estimation sample, defined as $\{ (\yi{0},\x_i): \yi{0} \in \S(\theta)\}$, where $\S(\theta) := [\uy,\uk]\cup(\ook(\theta),\oy]$ denotes the support of $\yi{0}$ in the estimation sample.
This sample will be used for the counterfactual estimation related to $\ys{0}$.
For notational convenience, the estimation sample will be denoted as $\{ (\yi{0}, \x_i)\}_{i \in \nn(\theta)}$.
Here, $\I{i \in \nn(\theta)} = \I{\yi{0} \in \S(\theta)} = \I{\yis{0} \in \S(\theta)}$ indicates unit $i$'s membership in the estimation sample.
Hereafter, we omit the dependence on $\theta$ where it is unlikely to cause confusion.

\subsubsection{Counterfactual Estimation}
\label{subsubsect:estimation-testing}
This section describes our method to estimate $f_j$ for each $j \in \N$.
Let $\mathcal{P}_k$ denote the linear span of all polynomials of degree less than $k \in \mathbb{N}$.
The polynomial basis of order $k\in\N$ is defined as
\begin{align*}
    z_k(\yv) &:= (1,\yv-\uk,(\yv-\uk)^2, \ldots, (\yv-\uk)^{k - 1})'  \in \R^{k},\\
    z_{k i} &:= z_k(\yi{0}),\quad i=1,\ldots,\nn.
\end{align*}
We denote the coefficient as $\gamma  = \gamma_k \in \R^{k}$ with $k$ suppressed for simplicity.
We define our sieve space of order $k$ for $f_j$ as
\begin{equation*}
\mathcal{F}_{k j} := \{ z_{k}'\gamma \in \mathcal{P}_k :  \sup_{\yv \in \mathcal{Y}_0} |\log(z_k(\yv)'\gamma) | \le \con_{3}j\}    
\end{equation*}
where $\con_{3}>0$ is chosen such that $f_{k j}$, the best approximation of $f_j$ in $\mathcal{F}_{k j}$ in a suitable sense, lies in the interior of $\mathcal{F}_{k j}$.
The linear growth of $|\log f_j|$ as $j$ increases is ensured by Lemma~\ref{lemA:sufficient-analytic}.
This manifests that our search space for $f_{kj}$ will be limited to functions comparable in size to $f_j$.
We will not be specific about this constant as it mainly serves as a technical and numerical device.
With some abuse of notation, we denote $\gamma \in\mathcal{F}_{kj}$ if $z_k'\gamma \in \mathcal{F}_{kj}$.

For illustration, we focus on a single bunching moment generated by a weighting function $\ti$.
If there are multiple bunching moments, they can be addressed individually in a similar fashion.
For each $k \in \N$ and $j \in \N$, define $\what \gamma_{k j} \in \mathbb{R}^k$ as
\begin{align*}
\what\gamma_{k j} := \argmax{\gamma \in \mathcal{F}_{k j} } &\ \frac{1}{n} \sum_{i = 1}^n \ti  w_i^j \log (z_{k i}'\gamma)\I{i \in \nn} \\
 &  \text{s.t.}  \ \left( \int_{\S} z_k(\yv)d\yv \right)' \gamma = \frac{1}{n} \sum_{i =1}^n \ti  w_i^j \I{i \in \nn}\nonumber.
\end{align*}
Using the Lagrangian method, $\what{\gamma}_{\kap j}$ is the solution to the following convex optimization problem:
\begin{align}
    \label{define:estimator}
\what\gamma_{k j} := \argmax{\gamma \in \mathcal{F}_{k j} }\left[ \frac{1}{n} \sum_{i \in \nn} \ti  w_i^j \log (z_{k i}'\gamma) - \left(\int_{\S} z_k(\yv)d\yv\right)' \gamma \right].
\end{align}
Assuming an interior solution, the first-order condition for $\what\gamma_{k j}$ is given by
\begin{equation}
\label{eq:first-order-cond-estimation}
\frac{1}{n} \sum_{i \in \nn} \ti  w_i^j \frac{z_{k i}}{z_{k i}'\what\gamma_{k j}} - \int_{\S} z_k(\yv)d \yv =0.
\end{equation}
This leads to our estimator for $f_j$ using the order $k$ polynomial sieve:\footnote{In practice, it is recommended to use an orthogonal polynomial basis for numerical stability, which can be obtained by applying a linear basis change.}
\begin{align*}
   & \what{f}_{k j}(\yv) = z_k(\yv)' \what \gamma_{k j},\\
   & \frac{1}{j!}\widehat{\dif_{\yv}^{j-1} f_j(\yv)}|_{\yv = \uk} = \frac{1}{j} \what \gamma_{k j,j},  
\end{align*}
where $\what \gamma_{k j,j}$ represents the $j$-th component of $\what \gamma_{k j}=(\what \gamma_{k j,1},\ldots,\what \gamma_{k j,k})'$, associated with the term $(\yv-\uk)^{j-1}$.

\subsubsection{Testing Procedure}

For each $k \in \N$ and $1 \le l \le k$, let
\begin{equation*}
\what \mu_{k l} := \what{\E}[\ti \I{\yii \in [\uk, \ok]}] - \sum_{j=1}^l \frac{1}{j}\what\gamma_{k j, j},
\end{equation*}
where $\what{\E}[\ti \I{\yii \in [\uk, \ok]}] : = \frac{1}{n} \sum_{i=1}^n \ti \I{\yii \in [\uk, \ok]}$.
Here, $k$ and $l$ represent the sieve and the approximation orders, respectively.
These tuning parameters reduce the infinite-dimensional array of coefficients to a finite-dimensional one.
They are chosen to be $k = \kap_n$ and $l = \tc_n$ based on a pre-specified schedule determined by the sample size.
We allow $\kap_n$ and $\tc_n$ to increase with $n$, where the subscript $n$ will be suppressed at times.
The growth conditions for $\kap_n$ and $\tc_n$ will be specified later.
Under suitable regularity conditions, we establish that $\what \mu_n := \what \mu_{\kap \tc} = b_{\tc} + \oP(1)$ under the null hypothesis (Lemma~\ref{lem:estimation-consistency}), where $b_{l} := \sum_{j=l+1}^\infty \frac{1}{j!} \dif_{\yv}^{j-1}[ \E[\ti  (\conv(\ok,\x_i,\theta)-\uk)^j |\yis{0} = \yv] f_{\ys{0}}(\yv)]_{\yv = \uk}$ represents the approximation error at truncation order $l$.

It remains to quantify the uncertainty associated with $\what \mu_n$.
Let $s_{k l i}$ be an influence function in the asymptotic linear expansion of $\what \mu_{k l}$, i.e., for each $k \in \N$ and $1 \le l \le k$,
$$
\sqrt{n}(\what\mu_{k l} - b_{l}) = \frac{1}{\sqrt n}\sum_{i=1}^n s_{k l i} + \oP(\sigma_{k l}).
$$
Then, we estimate $\sigma_{k l}^2 = \var(s_{k l i})$ by
$
\what \sigma_{k l}^2 = \frac{1}{n} \sum_{i=1}^n \what s_{k l i}^2,
$
where $\what s_{k l i}$ is a suitable estimate of the individual influence function that satisfies
$$
\frac{1}{n} \sum_{i=1}^n |\what{s}_{k l i}^2 - s_{k l i}^2| = \oP( \sigma^2_{k l}).
$$
For example, we can obtain $\what s_{k l i}$ by collecting the influence functions derived from the score functions of $\what{\E}[\ti \I{\yii \in [\uk,\ok]}]$ and $\frac{1}{j}\what \gamma_{k j, j}$'s.
(See Algorithm~\ref{alg:code-test-statistic}.)
This method is adopted across all our applications as well as in the proof of Theorem~\ref{thm:asymptotic-normality}.

Let $\what \sigma_n^2 := \what \sigma^2_{\kap \tc}$ denote the estimated standard error of $\what \mu_n$ and $\mathcal T_n(\theta) := \sqrt{n}|\what\mu_{n}|/\what \sigma_{n}$ be the absolute t-statistic.
To derive the null distribution of $\mathcal T_n$, we should make additional regularity assumptions on the data-generating process and the choice of the sieve and approximation orders.
We first assume that the asymptotic variances of $\what \mu_{k l}$ are uniformly bounded away from $0$, which is innocuous in most situations.
\begin{assumption}{7}
\label{asm:variance-lower-bound}
There exist $\con_{4} >0$ such that $\sigma_{k l}^2 \ge \con_{4}$ for all sufficiently large $k \in \N$ and $l \in [1, k]$ such that $l/k \to 0$.
\end{assumption}

We assume the existence of the interior solution to the sample and the corresponding population maximization program in the following.
We emphasize that this assumption is not too restrictive since $\con_{3}$ can be set large enough to ensure an interior solution.

\begin{assumption}{6} 
    \label{asm:estimator-interior-solution}
\begin{enumerate}[leftmargin = 0.05\linewidth]

\item [(i)] For each $k \in \N$ and $j \in \N$, there exists a solution $\gamma_{k j} \in \R^{k}$ in the interior of $\mathcal{F}_{k j}$ that solves
\begin{align}
    \label{eq:population-sieve-problem}
    \gamma_{k j} = \argmax{\gamma \in \mathcal{F}_{k j}}  \left[ \E[\ti w_i^j \I{\yis{0} \in \S_\infty}\log (z_{k i} '\gamma)]  -\left(\int_{\S_\infty} z_k(\yv)d\yv\right)' \gamma \right],
\end{align}
where $\CS_\infty$ denotes the probability limit of $\S$.

\item [(ii)] For each $k \in \N$ and $j \in \N$, there exists a solution $\what \gamma_{k j} \in \R^{k}$ in the interior of $\mathcal{F}_{k j}$ solving \eqref{eq:first-order-cond-estimation} with probability equal to $1$ as $n \to \infty$.

\end{enumerate}
\end{assumption}

To state the conditions for an appropriate polynomial order, we introduce a measure of the quality of the polynomial extrapolation.
For each $k \in \N$, let $H_k \in \R^{k \times k}$ and $o_k(.)$ be defined as $H_{k} = \int_{\CY_0} z_k(\yv) z_k(\yv)' d\yv$ and $o_k(\yv) = H_k^{-1/2} z_k(\yv)$, respectively.
Then, $o_k(.)$ represents the orthonormal polynomial basis of order $k$ with respect to the Lebesgue measure on $\CY_0$.
Let $0 < \chi_k \le 1$ be defined as the smallest eigenvalue of the matrix $\int_{\S} o_k(\yv) o_k(\yv)' d\yv$.
The number $1/\chi_k$ represents the worst-case ratio of the extrapolation errors within $\S^c$ to the fitted errors in $\S$, quantifying the stochastic variability of the counterfactual estimation relative to the observed fit.
We refer to this constant as the extrapolation norm.
The sequence $(1/\chi_k)_{k\in\mathbb{N}}$ increases unboundedly, at most at an exponential rate of $C^{k}$ for some $C > 1$ by Lemma~\ref{lem:Hilbert-matrix}(ii).
The magnitude of this sequence depends positively on the size of the extrapolated region relative to that of the entire support, reflecting the difficulty of extrapolation.

We need to ensure $\chi_{\kap_n}^{-1}$ does not diverge too quickly relative to the sample size, as it would result in a highly noisy extrapolation into the unobserved region despite the goodness of the observed fit.
This underscores the need to gradually expand the sieve space due to the ill-posed nature of the extrapolation problem.
The concrete conditions are presented in Assumption~\ref{asm:selection-tuning}, which guarantees consistent counterfactual estimation and correct inference.

\begin{assumption}{8}
\label{asm:selection-tuning}
$n^{\con_{5}/2} \lesssim \chi_\kap^{-1} \lesssim n^{\con_{6}/2}$,
  $\kap \lesssim \log n$, and $1 \le \tc \lesssim \log n /\log \log n$ for some $0 < \con_{5} \le  \con_{6} < 2/5$.
\end{assumption}

Notice that $n^{\con_{5}/2} \lesssim \chi_\kap^{-1} \lesssim C^\kap$ requires $\kap \gtrsim \log n$.
Thus, the sieve dimension should increase at a logarithmic rate in $n$, where the proportional factor depends on the size of the extrapolated region.

Next, we make regularity assumptions about the smoothness of the counterfactual functions, represented by $\rad$, and the lower bound of the estimated function.
The assumed conditions ensure that the errors from polynomial approximation decay sufficiently fast.

\begin{assumption}{9}
\label{asm:regularity-for-estimation}
\begin{enumerate}[leftmargin = 0.05\linewidth]

\item [(i)] Assumption~\ref{asm:identification-regularity}(i) is satisfied for $\rad>0$ such that $(1+2\tilde{\rad}+2\sqrt{\tilde{\rad}^2+\tilde{\rad}}) > (\limsup_{k\to\infty} \chi_k^{-1/k})^{1/\con_{5}}$, where $\tilde{\rad} = \rad/|\CY_0|$, $|\CY_0| = \oy-\uy$, and $\con_{5}$ is the same constant as in Assumption~\ref{asm:selection-tuning}.

\item [(ii)] $\inf_{\yv \in \CY_0} \E[\ti (\conv(\ok,\x_i,\theta_0) - \uk) |\yis{0} = \yv] \ge \con_{1} > 0$.

\end{enumerate}
\end{assumption}

Given these additional assumptions, it is possible to derive an asymptotic distribution of $\mathcal T_n$, as presented in Theorem~\ref{thm:asymptotic-normality} below.
It allows us to construct an $\alpha$-sized test for $H_0 : \theta_0 = \theta$ as 
\begin{equation*}
\phi_n(\theta) := \I{|\mathcal T_n(\theta)| \ge \cv{\alpha}(\sqrt{n}\maxb_\tc/ \what {\sigma}_n)}.
\end{equation*}
Here, $\cv{\alpha}(b)$ represents the $(1-\alpha)$ quantile of $|N(b,1)|$ and $\maxb_\tc = \beta \delta^{\tc} \E[\ti \I{\yii \in [\uk, \ok]}]$ serves as an upper bound for the approximation bias derived in Theorem~\ref{thm:point-identification}.
We may estimate $\maxb_\tc$ or another upper bound based on the sample
(see the remarks following Theorem~\ref{thm:asymptotic-normality}).
Consequently, a confidence set with an asymptotic coverage of $(1-\alpha)$ can be constructed as
\begin{equation*}
    C_n := \left\{ \theta \in \Theta : \phi_n(\theta)  = 0\right\}.
\end{equation*}

\begin{thm}
\label{thm:asymptotic-normality}
For any $\alpha \in (0,1)$, it holds
\begin{equation*}
\limsup_{n\to\infty}\sup_{\P \in \mathfrak{P}_0}\P( |\mathcal T_n(\theta)| \ge \cv{\alpha}(\sqrt{n} \maxb_\tc/ \what \sigma_{n})) \le \alpha,
\end{equation*}
where $\maxb_\tc = \beta \delta^{\tc} \E[\ti \I{\yii \in [\uk, \ok]}]$.
Here, $\mathfrak{P}_0$ denotes the set of all distributions for $(\y,\x)$ under which $H_0:\theta_0 = \theta$ holds, where each distribution satisfies Assumptions~\ref{asm:estimator-interior-solution}, \ref{asm:variance-lower-bound}, \ref{asm:selection-tuning}, and \ref{asm:regularity-for-estimation}, as well as the conditions in Theorem~\ref{thm:point-identification}(i), for the given constants $\{\con_{k}\}_{k=1}^6$.
\end{thm}

Some remarks on the choice of $\tc$ and the calibration of approximation bias are in order.
First, it could be informative to report confidence sets using critical values with $\maxb_\tc$ set to $0$.
These confidence sets should cover the pseudo-true values that are identified despite the approximation errors.
If $\tc$ is appropriately chosen and the approximation errors are small in absolute terms, these pseudo-true values can still provide meaningful information about the true parameter.
We adopt this convention in reporting our results in the Monte Carlo experiments and the empirical application.

Second, to obtain an upper bound for the approximation bias, we can estimate the smoothness constants from the data.
To implement this, we first set a grid $\mathcal{G}$ for values of $\rad$.
Next, we estimate $f^{(\Q)}_{\ys{0}}(\yv)$ and $\qb(\qua, \theta, \yv)$ along with their derivatives for $\qua$ picked from a dense grid in $(0,1)$.
The conditional quantile function can be estimated using the weighted polynomial sieve quantile regression akin to the proposed estimation procedure.
Given estimates for these functions, for each $\rad \in \mathcal{G}$, one can compute $(\beta_\rad,\delta_\rad)$ according to Definition~\ref{def:analyticity-condition}, and bound $|b_\tc| \le \min_{\rad \in \mathcal{G}} \beta_\rad \delta_\rad^{\tc} \E[\ti \I{\yii \in [\uk, \ok]}]$.
The bias can be controlled more efficiently using the upper bound provided in Lemma~\ref{lemA:sufficient-analytic}(iii), which is expressed as a Fourier coefficient related to these functions.
Such methods could further support an appropriate choice of the approximation order by evaluating the approximation quality.


\subsection{Inference on Partially Identified Values}
\label{subsect:inference-set}

This subsection outlines our proposal for inference on the values in the identified set
\begin{equation*}
    \Theta_0 = \{ \theta \in \Theta : \ubf(\theta) \le \Q(\yii \in [\uk, \ok]) \le \obf(\theta) \},
\end{equation*}
derived from Theorem~\ref{thm:partial-identification}.
As before, our aim is to construct a confidence region $C_n$ with the correct coverage
$
\liminf_{n\to\infty} \inf_{\theta \in \Theta_0}\P(\theta \in C_n) \ge 1- \alpha.
$
This can be accomplished by inverting a test for $H_0 : \theta_0 = \theta$ that satisfies $\limsup_{n\to\infty} \sup_{\theta \in \Theta_0} \E[\phi_n(\theta)] \le \alpha$.
We consider a single bunching moment weighted by $\ti$.

We assume the existence of preliminary estimators for the envelope functions and their derivatives, which permit an asymptotic linear expansion around $\yv = \uk$.
This allows for flexible incorporation of prior knowledge about  $\uf$ and $\of$.

Let $\theta\in \Theta_0$ be a hypothesized value.
We then estimate the conditional moment functions, defined as $m_j(\yv) = \E_\Q[w_i^j | \yis{0} = \yv]$.
These functions can be estimated in various ways using the weighted sieve regression.
For instance, we can define a least-squares estimator as
\begin{align*}
    \what m_{k j}(\yv) = z_k(\yv)' \what \gamma_{k j},\qquad \what \gamma_{k j} &:= \argmin{\gamma \in \mathbb{R}^k}  \frac{1}{n} \sum_{i \in \nn} \ti (w_i^j -z_{k i}' \gamma)^2,
\end{align*}
or an alternative estimator as
\begin{align*}
    \widetilde \gamma_{k j} &=\argmax{\gamma \in \mathcal{F}_{k j}} \frac{1}{n}  \sum_{i \in \nn} \ti \left\{   w_i^j \log(z_{k i}' \gamma) - z_{k i}' \gamma \right\}.
\end{align*}
Choose $\kap \in \N$ and $\tc \in [1, \kap]$ so that Assumption~\ref{asm:selection-tuning} is satisfied.
Let
\begin{align*}
    \what \mu_{1n} &=  \sum_{j=1}^\tc \frac{1}{j!} \dif_{\yv}^{j-1} [\what m_{\kap j}(\yv) \uf(\yv)]_{\yv = \uk} - \what\E[\ti \I{\yii \in [\uk, \ok]}]/\what \E[\ti],\\
    \what \mu_{2n} &=  \sum_{j=1}^\tc\frac{1}{j!} \dif_{\yv}^{j-1} [\what m_{\kap j}(\yv) \of(\yv)]_{\yv = \uk} - \what\E[\ti \I{\yii \in [\uk, \ok]}]/\what \E[\ti].
\end{align*}
The following moment inequality restrictions are tested under $H_0 : \theta_0 = \theta$.
\begin{equation*}
A\mu - \maxb := \left( \begin{matrix}
    \mu_{1} - \maxb_{1\tc}\\
    - \mu_2 - \maxb_{2 \tc}
\end{matrix} \right)\le \left( \begin{matrix}
    0 \\ 0    
\end{matrix} \right),
\end{equation*}
where $A$ is a diagonal matrix with diagonal entries $(1,-1)$, $\mu = (\mu_1,\mu_2)' = \op{plim}_{n\to\infty} (\what \mu_{1n}, \what \mu_{2n})'$, and $\maxb=(\maxb_{1\tc},\maxb_{2\tc})'$.
Here, $\maxb_{1 \tc}$ and $\maxb_{2 \tc}$ represent upper bounds for the finite approximation errors of $\ubf(\theta_0)$ and $\obf(\theta_0)$, respectively.

There is a large body of literature on inference for parameters identified by moment inequality restrictions (\citeay{romanoPracticalTwoStepMethod2014}; \citeay{andrewsInferenceLinearConditional2023}; \citeay{coxSimpleAdaptiveSizeExact2023}, etc).
Among these methods, \citet{coxSimpleAdaptiveSizeExact2023} proposed to use a chi-square test with data-adaptive critical values.
Their procedure does not require simulation of data or bootstrap, making it well-suited for our application.
Following their procedure, we define a quasi-likelihood-ratio test statistic as
\begin{equation}
\label{eq:inference-partial}
    \mathcal{T}_n(\theta) = \min_{\mu:A\mu-\maxb\le 0} n \left( \begin{matrix}
        \what\mu_{1n}(\theta)-\mu_1 \\
        -\what\mu_{2n}(\theta)+\mu_2
\end{matrix} \right)' \what V_n(\theta)^{-}\left( \begin{matrix}
        \what\mu_{1n}(\theta)-\mu_1 \\
        -\what\mu_{2n}(\theta)+\mu_2
\end{matrix} \right),
\end{equation}
where the asymptotic variance, $\what V_n(\theta)$, can be derived from the asymptotic linear expansion of $\sqrt{n}(\what \mu_{1n}(\theta),\what \mu_{2n}(\theta))'$.
The critical value is set to $\chi^2_{\what {df}(\theta)}(1-\alpha)$, where $\what {df}(\theta)$ denotes the rank of $[A \what V(\theta)^-]_{\what J}$.
Here, we denote by $[A\what V(\theta)^-]_{\what J}$ the submatrix of $A \what V(\theta)^-$ formed by the rows corresponding to the indices in $\what J$, and by $\what J$ the set of indices for the binding inequalities at the solution to \eqref{eq:inference-partial}.
For instance, suppose $\what V_n(\theta)^{-1}$ exists and only one of the inequalities is binding across all values in $\Theta$.
Then, we reject $H_0 : \theta \in \Theta_0$ when $\mathcal{T}_n(\theta) > \chi_1^2(1-\alpha)$.
The corresponding confidence region can be constructed as
$
    C_n = \{ \theta \in\Theta : \mathcal{T}_n(\theta) \le \chi^2_1(1-\alpha) \}.
$


\section{Monte Carlo Experiments}
\label{sect:monte-carlo}

To demonstrate the efficacy of the proposed method, we carry out a series of Monte Carlo experiments.
We are particularly interested in the size and power properties of our test proposed in Section~\ref{sect:confidence-region}.
We will compare the performance of our method with that of a version of the polynomial estimator to be detailed in this section.

The design of our experiments is described as follows.
We generate a random sample $\{ Y_i, X_i \}_{i=1}^n \subseteq \R\times \R$ of size $n = 100,000$ from the augmented isoelastic model.
It consists of the equations
\begin{align*}
    Y_i^*(d) &= (1-\tau_d)^{\theta_i}\eta_i,\ \ d \in \left\{ 0,1\right\},\\
    \theta_i &= \theta(\x_i)= \theta_0 + \omega_0 \x_i,\quad (\theta_0, \omega_0) \in \Theta \equiv \{(\theta,\omega) : \theta > 0, |\omega| < \theta  \},\\
    Y_i &= Y_i^* \I{Y_i^* \notin [\uk, \ok]} + \varepsilon_i \I{Y_i^* \in [\uk, \ok]},
\end{align*}
where $\eta$ possesses a positive density $f_\eta$ and $X$ is a uniform random variable on $[-1,1]$ independent of $\eta$.
The optimization errors are represented by $\varepsilon$, which assumes a triangular density in the window $[\uk, \ok]$.
The structural parameters refer to the average income elasticity $\theta_0 = \E[\theta(\x_i)]$ and the average differential $\omega_0 = \E[{\partial \theta(\x_i)}/{\partial X_i}]$.
We set $\theta_0 = 0.5$ and $\omega_0$ to either $0$ or $0.25$ in the experiment.
The case where $\omega_0 = 0$ corresponds to the standard isoelastic model, while $\omega_0 \ne 0$ suggests heterogeneous income responses across different $\x$ groups.
The policy parameters and the excluded window are set to $(\tau_0, \tau_1, \k, \uk, \ok) = (0.00,0.20,2.0,1.7,2.3)$.

\begin{figure}[ht!]
	\caption{Distributions of $\y$ and $\ys{0}$ under each DGP}
	\label{fig:monte-carlo-dist-figures}
	\begin{center}
		\begin{subfigure}[b]{0.45\linewidth} 
        \caption{DGP $1$ (Polynomial density)}
        \includegraphics[scale=0.425]{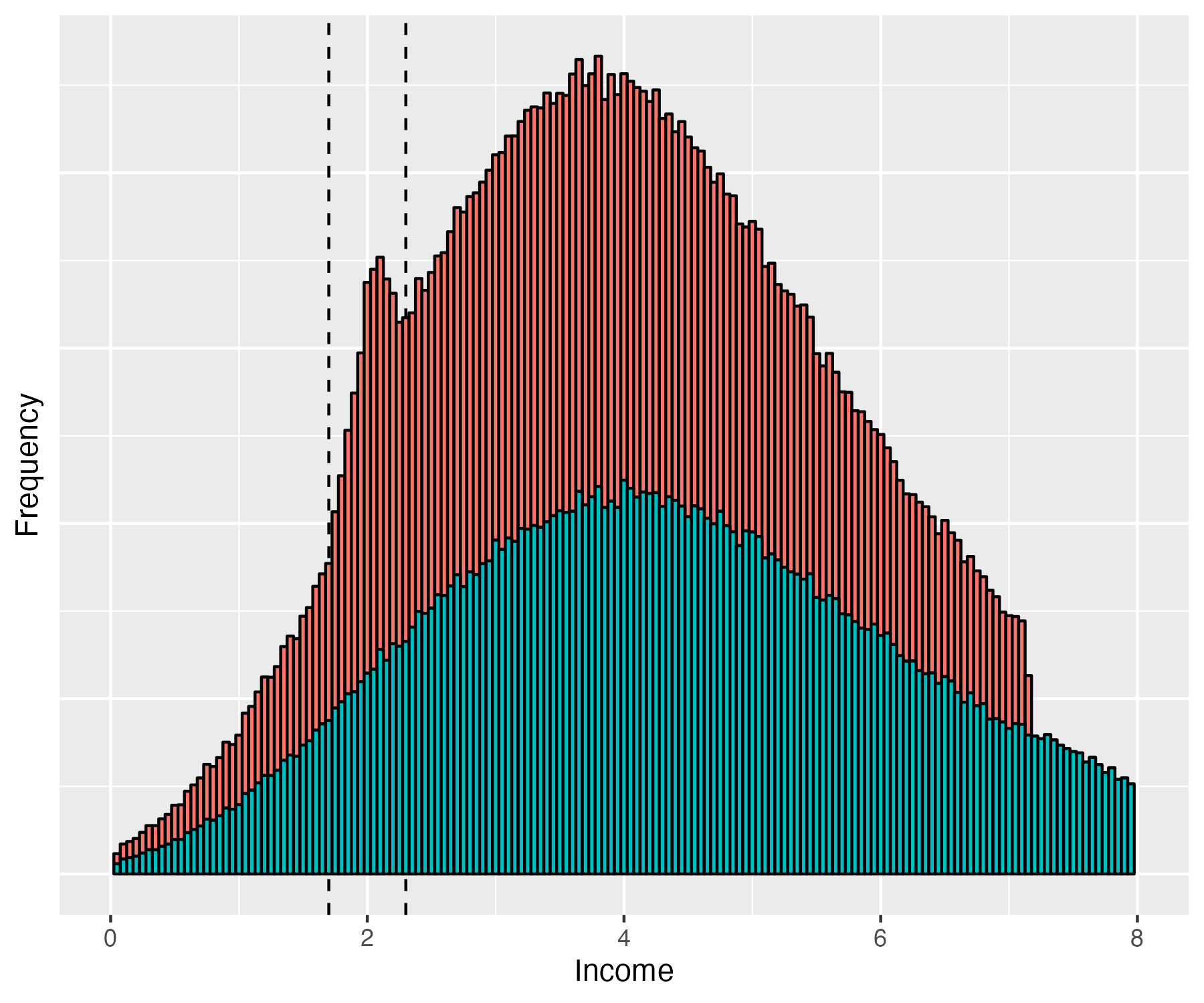}
    \end{subfigure}
    \hspace{2em}
    \begin{subfigure}[b]{0.45\linewidth}
        \caption{DGP $2$ (Gaussian mixture)}
        \includegraphics[scale=0.425]{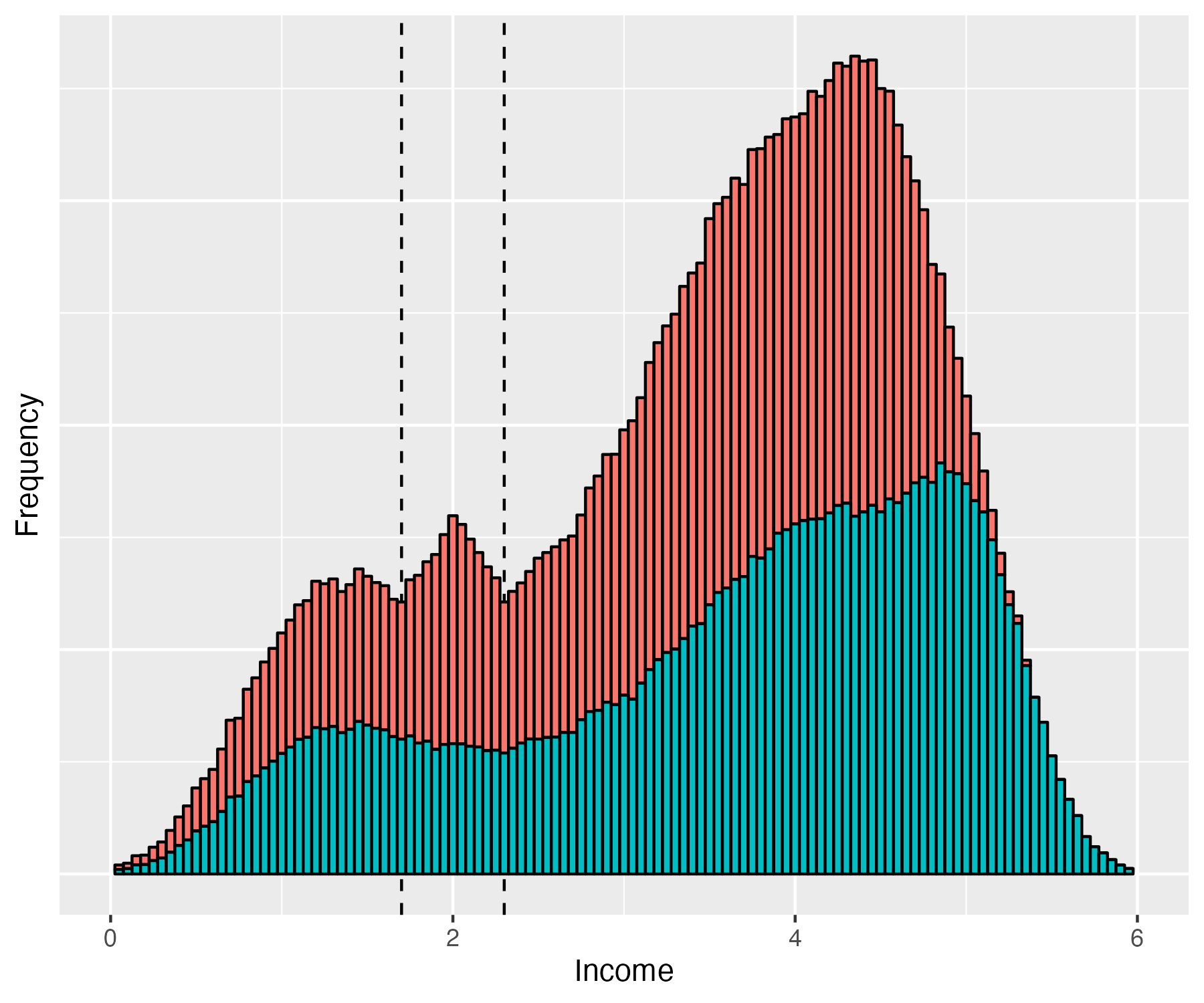}
    \end{subfigure}
	\end{center}

\begin{footnotesize}
Notes: Each panel displays the data-generating process used in the experiment.
  The histogram in red represents the empirical distribution of the actual income and the blue histogram shows the counterfactual income distribution prior to the introduction of the kink.
  For better distinction, the areas of the red and blue histograms are normalized to different values.
  The measurement unit is $10,000$ USD.
  The demarcated region indicates the optimization error window, which is set to $[1.7,2.3]$.
  \end{footnotesize}
\end{figure}

To impose the analyticity condition on $f_{\eta} = f_{\ys{0}}$, we consider two types of distributions: a polynomial density and a Gaussian mixture density, which are referred to as DGP~1 and DGP~2, respectively.\footnote{Details about the construction of these DGPs are relegated to \OA.}
Figure~\ref{fig:monte-carlo-dist-figures} exhibits each DGP through the histograms of $\y$ and $\ys{0}$, where $\theta_0 = 0.5$ and $\omega_0=0$.
The left panel for DGP~1 is based on a seventh-degree polynomial fitted to the U.S. taxable income data from 1960 to 1969, which was used in \citet{saezTaxpayersBunchKink2010}.
This corresponds to the counterfactual density estimated using the polynomial strategy with the iterative proportional adjustment.
DGP~2 is obtained by fitting a flexible location-scale Gaussian mixture distribution to a mixed skewed generalized error distribution.
DGP~2 exemplifies an analytic density not covered by well-known parametric families.

We compare the performance of our inferential method using the generalized polynomial strategy with that of the polynomial estimator, which are referred to as GPS and PE, respectively, hereafter.
The approximation order is set to $\tc = 5$ in the GPS method.
This choice is seemingly reasonable as it is capable of controlling approximation biases.
We adopted critical values disregarding potential approximation bias, which effectively performs inference on the pseudo-true value for $\tc=5$.
The selected polynomial degrees range from $7$ to $11$.
The size of the extrapolation norm $\chi_{\kap}^{-1}$ is monitored and bounded as $\chi_{\kap}^{-1} \approx 10 = n^{1/5}$ for all selected $\kap$.
We trimmed the lower 1\% and upper 5\% tails of the distribution for each DGP.
Such trimming may be necessitated in practice for various reasons: to prevent the counterfactual density from hitting the zero bound, to restrict the support of the data, or to exclude regions where singularities are suspected.

\begin{figure}[hb!]
	\caption{Power curves derived from GPS and PE methods}
	\label{fig:power-curves-compare}
    \vspace{-1em}
	\begin{center}
  \begin{subfigure}[b]{0.90\linewidth} 
        \caption{DGP~1}
        \includegraphics[scale = .65]{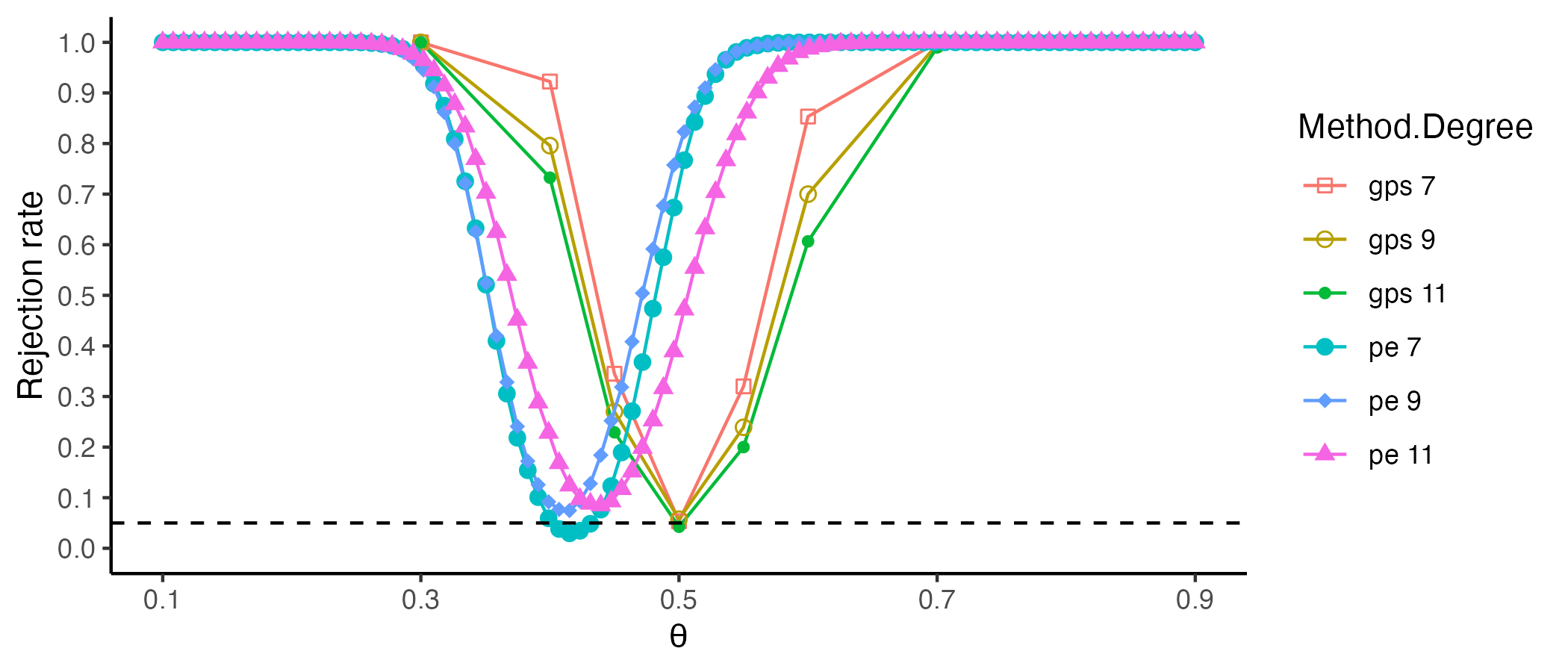}
    \end{subfigure}
    \\
    \begin{subfigure}[b]{0.90\linewidth}
        \caption{DGP~2 with 94\% of sample used}
        \includegraphics[scale = .65]{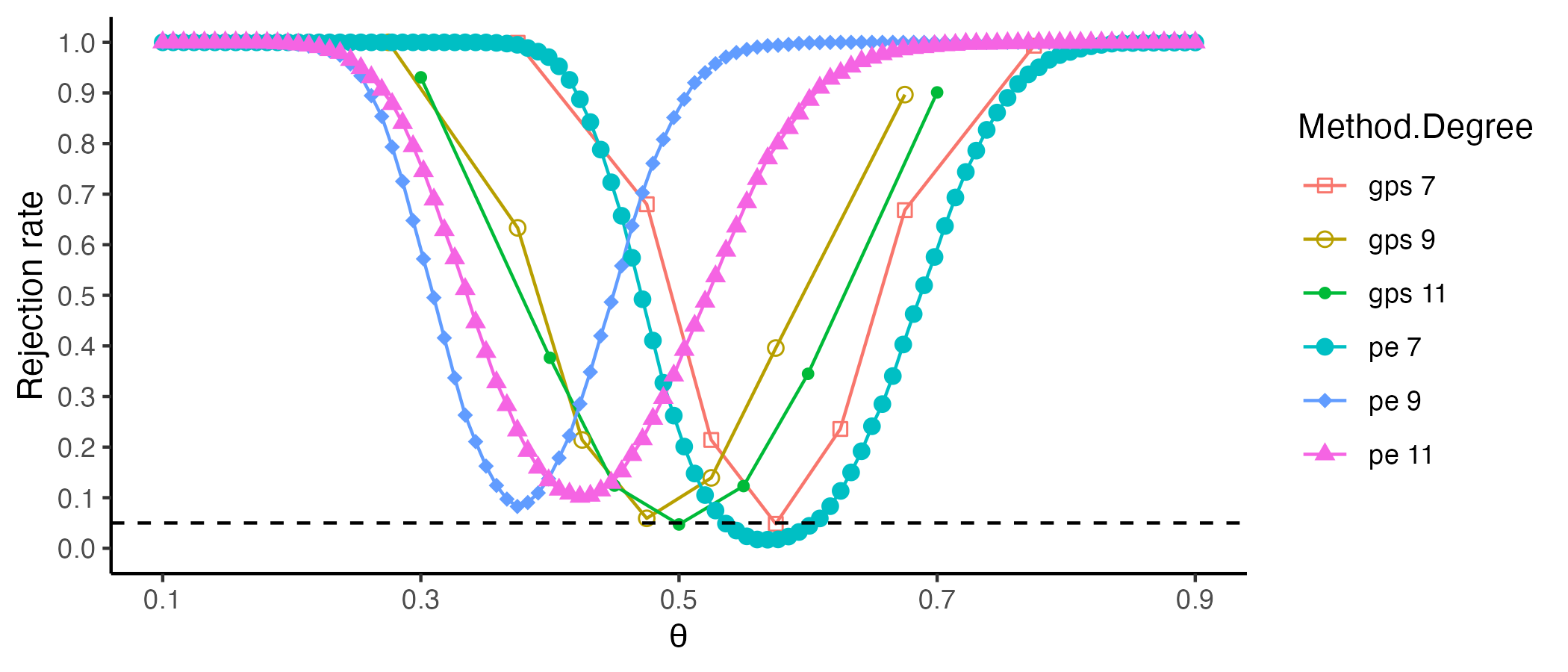}
    \end{subfigure}
     \begin{subfigure}[b]{0.90\linewidth}
        \caption{DGP~2 with 59\% of sample used}
        \includegraphics[scale = .65]{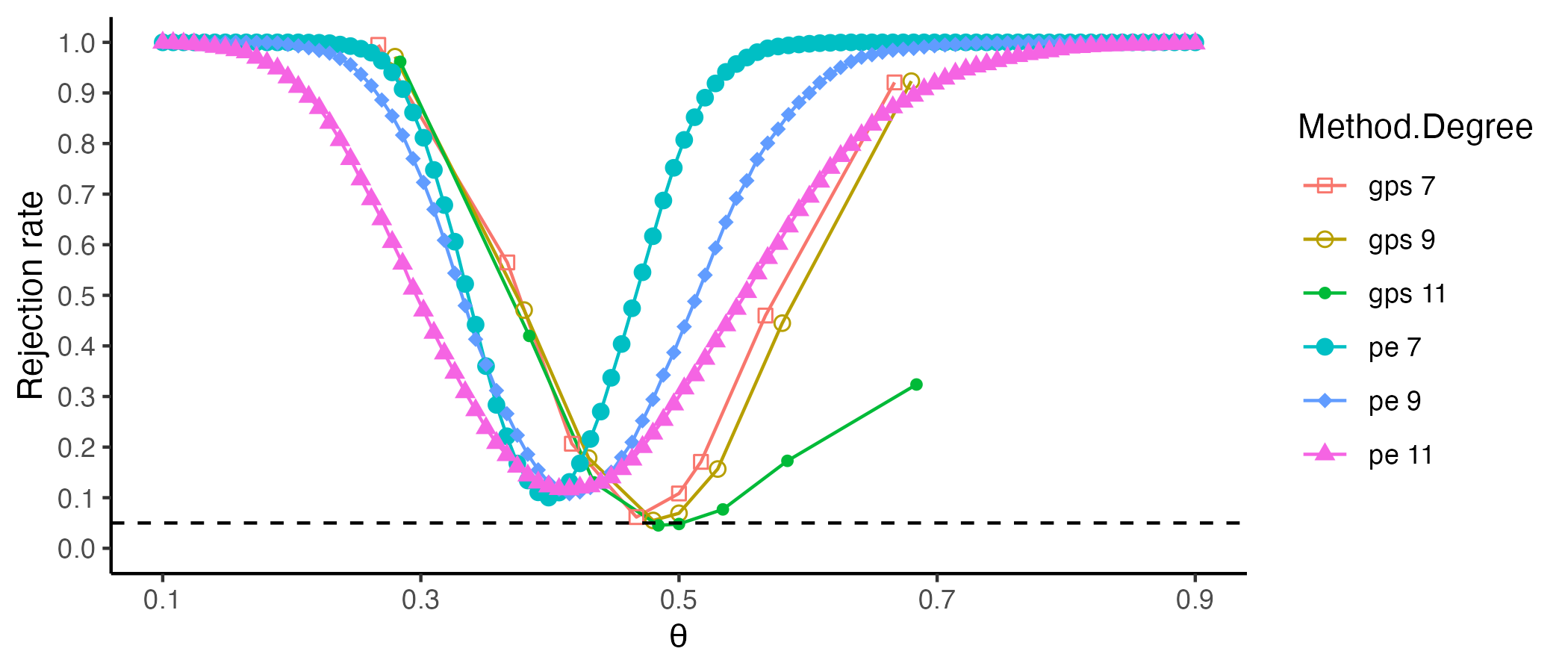}
    \end{subfigure}
	\end{center}

\begin{footnotesize}  
Notes:
  Each panel shows the power curves derived from GPS and PE methods with varying degrees of fitted polynomials.
  The dashed black line indicates the nominal size of 5\%.
  The bin width is set to 0.05 in the PE method.
  Each figure is based on 1200 replications for GPS and 5000 for PE, respectively.
\end{footnotesize}
\end{figure}

For the inference of $\theta_0 > 0$, where $\omega_0 = 0$ is treated as known, all tests are based on the unweighted bunching moment, $\P(\yii \in [\uk, \ok])$.
Figure~\ref{fig:power-curves-compare} presents the results of the size and power comparison.
To make this comparison, we used a version of the polynomial estimator that closely follows the procedure specified in \citet{chettyAdjustmentCostsFirm2011}.
In \OA, we show that the iteratively adjusted polynomial estimator can be formulated using the IV (instrumental variable) estimator in the following regression:
\begin{equation*}
    f_j = z_k(c_j)' \gamma + \sum_{l =\uj+1}^{\oj} \beta_{l} \left( \I{j = l} - \frac{f_j}{\sum_{j' > \oj} f_{j'}} \I{j > \oj}\right)+ e_j,\quad j = 1,\ldots, |\CJ|,
\end{equation*}
where $\CJ = \{ 1, \ldots, |\CJ| \}$ represents the set of equispaced bins partitioning the data support, sorted from left to right.
The dependent variable is the observed fraction of the data, $f_j = \frac{1}{n}\sum_{i=1}^n \I{\yii \in j}$, at bin $j$.
The first regressor is an order $k$ polynomial in the center of bin $j$, denoted by $c_j$.
This term represents the counterfactual fraction of bin $j$ before the introduction of the kink.

Let $\uj+1, \ldots, \oj$ be the set of bins located within the window.
The second set of regressors is coupled with the coefficients $\beta_l$, $\uj < l\le \oj$, which represent the amount of excess fraction relative to the counterfactual at bin $l$.
According to the given specification, the observed fraction exceeds the counterfactual fraction by $\beta_j$ if bin $j$ is located within the window.
If bin $j$ falls to the right of the window, $f_j$ falls short of the counterfactual fraction by $(\sum_{l=\uj+1}^{\oj} \beta_l)/(n^{-1}\sum_{i}\I{\yii > \ok}) f_j$, in proportion to the counterfactual fraction.
This assumption ensures that the excess and the missing fractions net to $0 = \sum_{\uj < l\le \oj} \beta_l - (\sum_{\uj < l\le \oj} \beta_l)/(n^{-1}\sum_i \I{\yii > \ok}) \sum_{j > \oj} f_j$, thereby the resulting estimate of the counterfactual density satisfies the integral constraint $\sum_{j \in \CJ} z_k(c_j)'\gamma = 1$ without adjustments.

The endogeneity introduced by the adjustment terms, $- \I{j > \oj}/(\sum_{j'> \oj}f_{j'})f_j$, which are correlated with the dependent variable, is addressed using an equal number of IVs
$$
\left[\I{j = \uj+1},\ldots, \I{j = \oj} \right]',
$$
which consist of the indicators for the bins within the window.
Let $(\what \beta',\what \gamma')'$ be the coefficient estimated by the IV estimator using these instruments.
We construct the counterfactual density and the fraction of excess bunching using the formulas
\begin{equation*}
    \what f := \frac{z_k(c_{j^*})' \what \gamma}{\op{mesh}(\CJ)},\qquad \what B := \sum_{j=\uj+1}^{\oj} \what \beta_{j},
\end{equation*}
where $j^*$ denotes the bin containing the cutoff and $\op{mesh}(\CJ)$ represents the width of each bin.\footnote{The original polynomial estimator by \citet{chettyAdjustmentCostsFirm2011} used $\what f = \frac{1}{\oj - \uj}\sum_{\uj + 1}^{\oj} z_k(c_j)'  \gamma / \op{mesh}(\CJ)$.
We modified it to more closely follow the identification strategy based on the small kink approximation.
Under both of our specifications, there were no significant differences between the original and modified estimators and the resulting CIs.}
This leads to the final estimate 
$
    \what \theta = \frac{\what B / \what f }{\k \log \left( \frac{1-\tau_0}{1-\tau_1}\right)}.    
$
The standard errors are computed based on the delta method\footnote{When $|\mathcal{J}|$ is sufficiently large, $\varepsilon_j$, $j=1,\ldots, |\mathcal{J}|$, can be treated as approximately independent errors.} and heteroscedasticity-robust standard errors, which allow us to construct an $\alpha$-level confidence intervals about $\theta_0$ as $[\what \theta \pm z_{1-\alpha/2} \op{se}(\what \theta)]$.

Figure~\ref{fig:power-curves-compare} presents our findings in the isoelastic model.
The top panel shows that our tests maintain correct size control across all selected degrees, whereas the power deteriorates as the order increases.
In terms of the average CI lengths, degree 11 is approximately 30\% wider than degree 7.
In contrast, the tests based on the polynomial estimator suffer from negative biases, leading to over-rejection of the correct null hypothesis.
A significant source of such biases is suspected to be the lack of an appropriate counterfactual adjustment.
Figure~\ref{fig:monte-carlo-dist-figures}(a) suggests that the PE method likely overestimates the counterfactual density, as the polynomial is fitted to the observed distribution rather than $f_{\ys{0}}$.
Since bunching fundamentally arises from the counterfactual distribution compressing near the cutoff, a proper adjustment should reconstruct it before this compression.

Figure~\ref{fig:power-curves-compare}(b) shows more variability compared to Figure~\ref{fig:power-curves-compare}(a), which is seemingly due to the fact that the counterfactual distribution in DGP~2 exhibits more hunches than in DGP~1.
This poses greater difficulty for precise counterfactual estimation.\footnote{
Some preliminary analysis shows that polynomials of degree 7 or 9 are quite poor at fitting the density in DGP~2 globally, especially near the hunch at the right peak.
The polynomial of degree 11 exhibits a noticeably improved fit, and the 14th-degree polynomial achieves an almost perfect fit.}
When we restrict the data to the 1st through the 60th percentiles, issues with the fitness of the estimated density are mostly resolved.
Based on 59\% of the sample, we obtain Figure~\ref{fig:power-curves-compare}(c) that exhibits a pattern similar to Figure~\ref{fig:power-curves-compare}(a).
Together with Figure~\ref{fig:power-curves-compare}(b), this illustrates that our method has a reduced sieve-approximation bias as the degree increases or the estimated region shrinks.
At degree $11$, our method achieves correct size control when 94\% of the sample is used.
In contrast, the PE methods are consistently oversized under all specifications.
Regarding the power properties, the average widths of the CIs, measured by the area between the power curve and the $y = 1$ line, increase with the selected polynomial degree.
This implies the bias-variance trade-off in selecting the polynomial degree.

In the augmented model, we test the joint hypothesis $H_0 : \theta_0=0.5, \omega_0 = \omega$ for various values of $\omega$.
The test requires at least two pieces of moments.
This is facilitated by an additional bunching moment, $\E[\exp(\x_i) \I{\yii \in [\uk, \ok]}]$, which assigns higher weights to those with higher $\theta_i$.
The Wald statistic is constructed based on the weighted and the unweighted bunching moments.
The critical value is set to $\chi^2_{2}(0.95)$ at the nominal level of 5\%, disregarding approximation bias.

\begin{figure}[ht!]
	\caption{Power curves across various values of $\omega$}
	\label{fig:power-curves-omega}
	\begin{center}

  \begin{subfigure}[b]{0.45\linewidth} 
        \caption{DGP~1}
        \includegraphics[scale = .85]{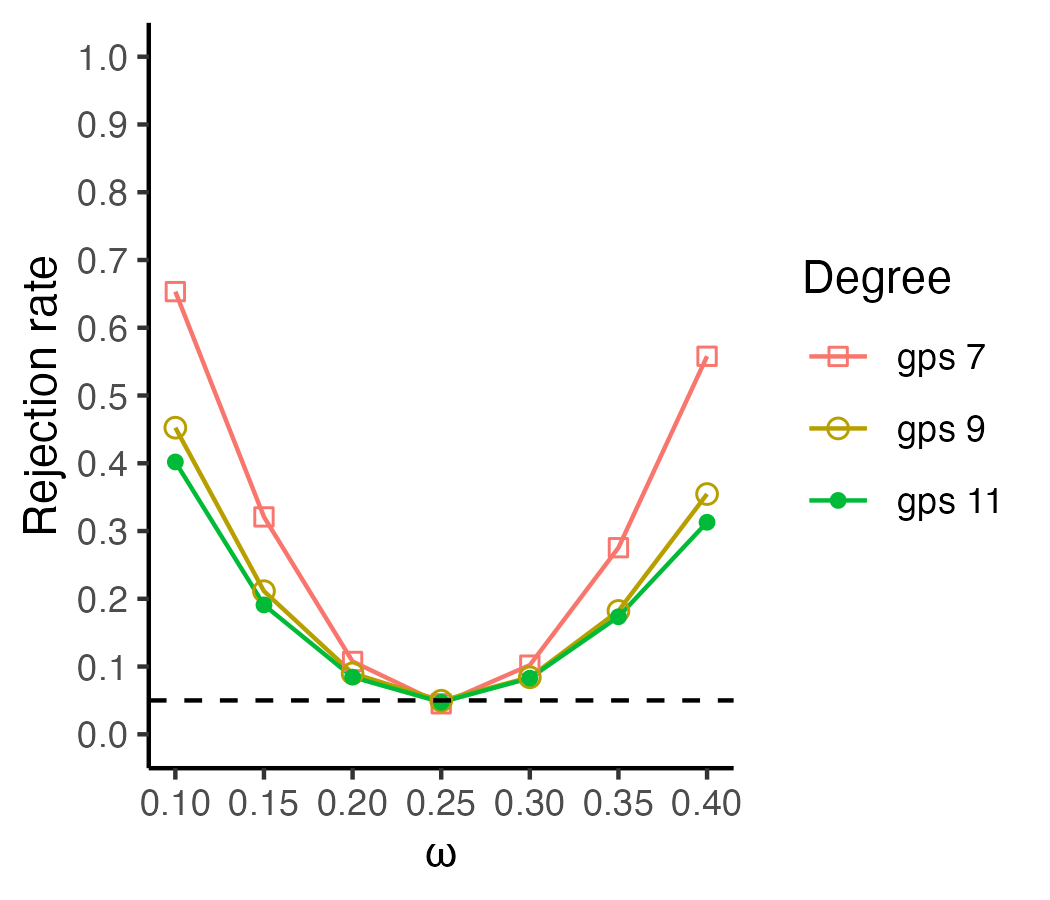}
    \end{subfigure}
    \begin{subfigure}[b]{0.45\linewidth}
        \caption{DGP~2}
        \includegraphics[scale = .85]{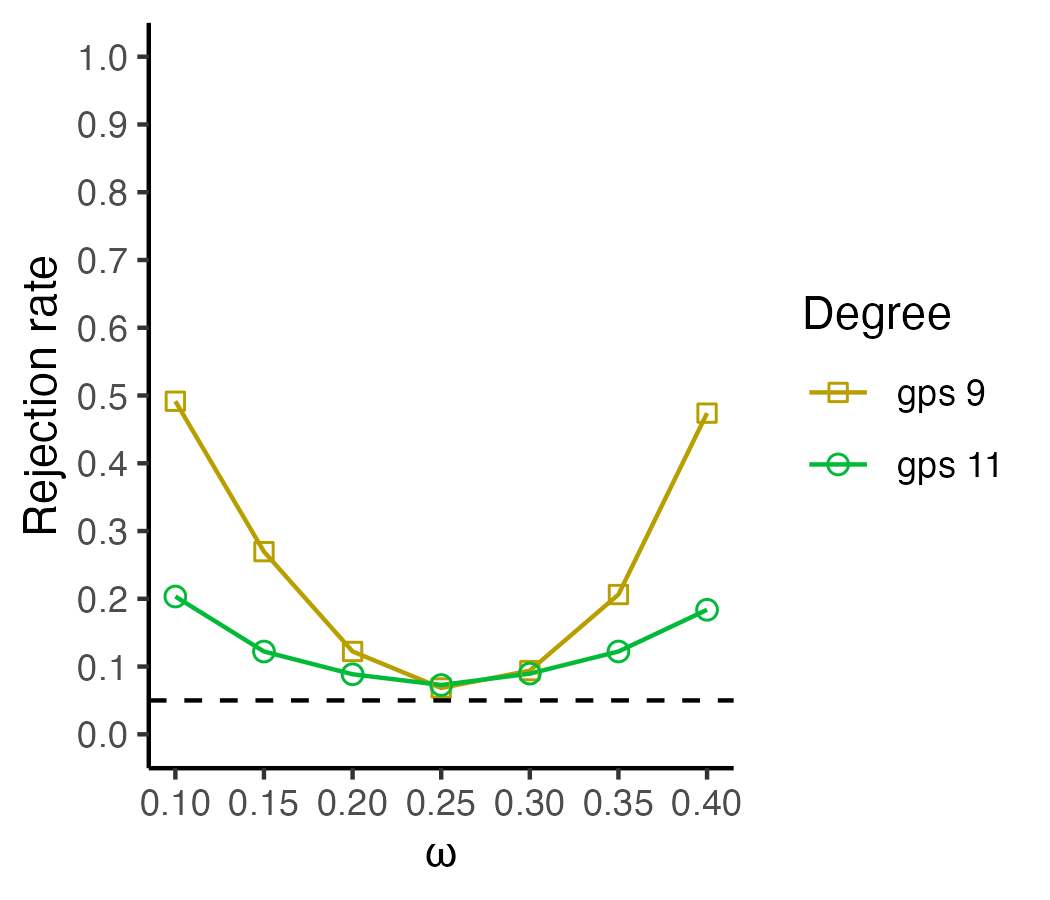}
    \end{subfigure}
	\end{center}

\begin{footnotesize}  
Note:
The true value is $\omega_0 = 0.25$ in both panels.
\end{footnotesize}
\end{figure}

Figure~\ref{fig:power-curves-omega} presents the power curves for the joint test under DGPs~1 and 2.
In both panels, U-shaped power curves are observed, indicating nearly correct size control and the nontrivial power of the test.
The actual size was approximately 7\% under DGP~2 when degree $11$ was selected, which appears to be a moderate degree of size distortion.
In Figure~\ref{fig:power-curves-omega}(b), we omitted the power curve for degree $7$, as the over-rejection at the true value is already demonstrated in Figure~\ref{fig:power-curves-compare}(b).
In light of near size control when degrees $9$ and $11$ are chosen, we reaffirm a trade-off between the size and the power of the test when selecting the polynomial degree.

\section{Empirical Application}
\label{sect:empirical}
As an empirical illustration, we revisit \citet{saezTaxpayersBunchKink2010}, who studied the elasticity of taxable income to marginal tax rate changes assuming the isoelastic model.
Specifically, we applied our method to the U.S. tax records data for joint married filers from 1960 to 1969, used for Figure~6A in \citet{saezTaxpayersBunchKink2010}.
Due to limited accessibility to the original data, we instead used the aggregate data attached to the replication files of \citet{blomquistBunchingIdentificationTaxable2021}.
For our purposes, these aggregate data are allegedly sufficient.
For details on the institutional setting, we refer the readers to Table~3 in \citet{saezTaxpayersBunchKink2010}.

By fitting a seventh-degree polynomial to the data outside the range $\k \pm \$4000$, where $\k = \$20000$, \citet{blomquistBunchingIdentificationTaxable2021} found a point estimate of 0.454 using the polynomial estimator by \citet{chettyAdjustmentCostsFirm2011}.
We conduct an empirical analysis of the same dataset based on our method and the polynomial estimator by \citet{chettyAdjustmentCostsFirm2011} along with the comprehensive robustness checks across choices of the polynomial degree, the excluded window, and the fitted range of the sample.
The approximation order remains $\tc = 5$, as in the previous section, which is seemingly reasonable given the valid size control in the data-based Monte Carlo studies.
We apply our testing method and the Wald test based on the polynomial estimator introduced in Section~\ref{sect:monte-carlo}.
The degrees are chosen from $\{ 7,9,11 \}$.
For the polynomial estimator, we used an equispaced grid with a width of $\$500$ following the analyses in \citet{blomquistBunchingIdentificationTaxable2021} and \citet{saezTaxpayersBunchKink2010}.
The excluded window is set either to $\k \pm \$4000$ or $\k \pm \$5000$, each serving as the main specification and the robustness check.
These windows are selected by examining the observed income distribution and identifying the points where the distribution begins to rise and returns to a normal level (see Figure~\ref{fig:prop-and-counterfactual-corrections}).
The sample is truncated from the 1st to the $u$th percentile, where $u$ ranges over $\{ 60,80,90,95 \}$.
The upper boundary $u$ represents the degree of locality in the counterfactual estimation.

The CIs from our analysis are presented in Table~\ref{tab:empirical-CIs}.
Several crucial observations can be made from this table.
First, as the polynomial degree increases, the GPS CIs become less sensitive to sample truncation.
This suggests potential issues with the model fitness of the seventh and ninth-degree polynomials over the entire data range.
In contrast, the eleventh-degree polynomial provides relatively stable CIs as more data are used, indicating its fitness for the empirical distribution.
The GPS CIs appear to converge on the same range as the estimation becomes more local to the window.
Consequently, $[0.250, 0.450]$ appears to be the most credible GPS CI robust to specifications.

On the other hand, the PE CIs show greater variability compared to the GPS CIs across the chosen degrees and the truncation levels.
Given a fixed level of truncation, the selected polynomial degree affects the final result quite significantly.
This makes it more involved to choose an appropriate polynomial degree in the PE method than in the GPS method.
The degrees 7 and 9 appear to be poor at fitting the entire range of distribution in light of the sensitivity to the truncated region.
Consequently, it seems reasonable to select one of the CIs with degree 11 among the PE CIs.
However, the resulting PE CI would significantly differ from the GPS CI.

In Panel~B, we repeated the same exercises as above except with the excluded window set to $\k \pm \$ 5000$.
We observe that the GPS CIs are mildly widened as the excluded window is expanded.
This may be attributed to increases in the standard errors.
Parallel to Panel~A, there is little ambiguity in the resulting GPS CIs when 60\% of the sample is used.
This substantiates our analysis using the main specification of the window, which is unlikely driven by an inappropriate guess on the error range.
However, this observation is unclear for the PE method.

Summing up, we found a CI of $[0.25, 0.45]$ for the taxable income elasticity, with $0.350$ being the most likely value.
This result differs from the previous estimates of 0.390 and 0.454, reported in \citet{blomquistBunchingIdentificationTaxable2021} using the trapezoidal approximation and polynomial estimator of degree 7, respectively.
These differences may well matter for the empirical implementation of the optimal tax rates.
Calculations based on \citet{saezUsingElasticitiesDerive2001} reveal that these differences in the compensated income elasticity translate to 2.7\%p and 6.4\%p increases in the revised optimal top tax rates, respectively.\footnote{\label{ftnote:saez} According to \citet{saezUsingElasticitiesDerive2001}, the optimal top rate in the absence of income effect is given by $\tau^* = (1-\bar{g})/(1-\bar{g} + a \theta)$, where $a>0$ denotes the Pareto parameter associated with the tail income distribution and $\bar{g} \ge 0$ represents the social welfare weight assigned to the top bracket tax-payers. In all calculations in this paper, these parameters are calibrated as $a = 2$ and $\bar{g} = 0$, respectively.}
If these elasticities apply only to a small neighborhood around the first kink, the 11\% and 30\% decreases in bunching estimates translate to approximately the same percentage increases in the odds of optimal local tax rates, assuming all else remains equal.

Our analysis reveals that the polynomial estimator may be sensitive to the choice of various user-chosen parameters, such as the polynomial degree, excluded window, or the fitted region.
It suggests that the counterfactual correction method reduces the variability of estimation results across polynomial degrees, enhancing their reliability.
Our findings demonstrate that the researcher can obtain credible confidence regions for the structural parameter by combining bunching evidence with principled methodological practices.

\begin{table}[ht!]
    \normalfont
   \begin{center}
   \caption{95\% Confidence invervals under various tuning parameter specifications}
   \label{tab:empirical-CIs}
   \medskip
   \begin{tabular}{C{0.15\linewidth}C{0.175\linewidth}C{0.175\linewidth}C{0.175\linewidth}C{0.175\linewidth}}\toprule
  &   \multicolumn{4}{c}{Percentage of sample used (from 1st to $(p+1)$-th percentile)}  \\
   \cmidrule(lr){2-5}
   & 94\%  & 89\% & 79\% & 59\% \\
    \midrule
    \multicolumn{5}{l}{Panel A. {\small $\k \pm \$ 4000$ }} \\[1ex]
  {\ GPS 7}  & $ [.390, .555] $ & $ [.340, .490] $ & $ [.335, .485] $ & $ [.275, .445] $ \\
  {\ GPS 9}  & $ [.315, .475] $ & $ [.310, .480] $  & $ [.215, .405] $  & $ [.250, .455] $  \\
  {\ GPS 11} & $ [.245, .440] $  & $ [.250, .450] $  & $ [.250, .465^*] $  & $ [.250, {.540}^{*}] $   \\
  {\ PE 7} & $  [.320, .484] $  & $ [.283, .446] $  & $ [.252, .422] $  & $ [.216, .369] $   \\
  {\ PE 9} & $  [.275, .421] $  & $ [.245, .409] $  & $ [.132, .312] $  & $ [.169, .351] $   \\
  {\ PE 11} & $ [.133, .329] $  & $ [.174, .350] $  & $ [.184, .392] $  & $ [.132, .507] $   \\
  [1.5ex]


\multicolumn{5}{l}{Panel B. {\small $\k \pm \$ 5000$ }} \\[1ex]

  {\ GPS 7}  & $ [.400,.660] $  & $ [.305, .525] $ & $ [.295, .500] $ & $ [.245, .475] $ \\
  {\ GPS 9}  & $ [.285, .500] $  & $ [.285, .505] $  & $ [.200, .450] $  & $ [.240, .490] $  \\
  {\ GPS 11}  & $ [.215, .475] $  & $ [.245, .535] $  & $ [.270, .735^{*}] $  & $ [.190^{**}, \op{NA}] $  \\
  {\  PE 7}  & $ [.322, .580] $  & $ [.266, .524] $  & $ [.216, .466] $  & $ [.151, .448] $  \\
  {\  PE 9}  & $ [.252, .475] $  & $ [.205, .477] $  & $ [.015, .338] $  & $ [.026, .426] $  \\
  {\  PE 11}  & $ [.000, .338] $  & $ [.034, .383] $  & $ [.060, .489] $  & $ [.000, 1.243] $  \\
    \bottomrule
   \end{tabular}
   \end{center}
   \smallskip
   \footnotesize{Notes: 
   The equispaced grid with a width 0.005 is used for the test inversion.
   In the GPS method, $\chi_\kap^{-1}$ exceeding $20$ is marked by an asterisk (*) at either edge of the confidence interval. 
   In addition, $\chi_\kap^{-1}$ greater than 45 is indicated by **.
    }
\end{table}


\section{Conclusion}
\label{sect:conclusion}

Bunching in the choice data offers a valuable opportunity to infer a structural parameter by leveraging it as a collective response to quasi-experimental variation in the incentive schedule.
Since its initial proposal by \citet{saezTaxpayersBunchKink2010} and subsequent development into a formal framework (\citeay{chettyAdjustmentCostsFirm2011}; \citeay{klevenUsingNotchesUncover2013}), it has found a broad range of applications in economic studies.
However, despite its growing popularity, the econometric foundations of its theoretical and practical aspects have been lacking until recently.

This paper develops an econometric framework and tools for identification and inference in bunching designs.
Our approach can incorporate observably heterogeneous responses into the model, a capability that was only partially permitted in previous methods.
Our identification scheme hinges on the analyticity condition of the counterfactual, which does not necessitate an ex-ante parametric family of distributions.
We develop a suite of tools for counterfactual estimation and inference, termed the generalized polynomial strategy.
The proposed method restores the merits of the traditional polynomial strategy while addressing weaknesses in the widespread practice.
In an alternative approach to partial identification, the assumption of analytic counterfactual is relaxed, which could enhance credibility of the procedure.

In this paper, our focus has primarily been on the setting where the counterfactual can be deduced from the observed distribution in the data.
In other important applications, such an assumption may not be realistic, for example, due to responses in participation margin, limited data availability, etc.
Exploring viable methodologies in these contexts could be a fruitful avenue for future research.

\bibliographystyle{plainnat}

\bibliography{My_Library}

\newpage

\appendix
\addtocontents{toc}{\protect\setcounter{tocdepth}{-1}}
\addtocontents{toc}{\protect\setcounter{tocdepth}{2}}
\addappheadtotoc
\section*{\Huge{Appendices}}

\section{Extended Framework for Notch Designs}

The notch design is another popular design that utilizes bunching, which applies to cases where the payoff schedule exhibits jumps at specific cutoffs, referred to as notches.
These notches are known to cause both clustering and gaps in the observed distribution of the running variable.
This section extends our econometric framework established in Section~\ref{sect:econometric-framework} to the notch design setup.

\subsection{Utility Model}
We continue to denote $\x_i \in \mathcal{X}\subseteq \mathbb{R}^{d_x}$ as observed characteristics and $\eta_i \in \mathcal{H} \subseteq \mathbb{R}$ as a scalar preference heterogeneity.
We consider the counterfactual payoff function $\yv \mapsto U_d(\yv,\xv,\eta)$ for each $d \in \{ 0,1 \}$, which are smooth and strictly concave functions of $\yv$.
Then, the counterfactual choice subject to policy $d \in \{ 0,1 \}$ is defined as
\begin{equation*}
    \yis{d} = \argmax{\yv \in \mathcal{Y}} U_d(\yv, \x_i, \eta_i).    
\end{equation*}
Without loss of generality, we assume that the payoff schedule displays a nonpositive (negative) notch across the cutoff $\k$, meaning that for all $(\xv, \eta) \in \mathcal{X}\times \mathcal{H}$ and $\yv \ge \k$,
\begin{equation}
\label{eq:notch-condition}
    U_0(\yv, \xv, \eta) - U_1(\yv, \xv, \eta) \ge 0.
\end{equation}
Typically, $\k$ is called a notch if the above difference is positive.
For theoretical purposes, however, we allow for a notch whose effect may be null.
This condition implies that policy $0$ becomes more beneficial than policy $1$ when $\yv$ exceeds $\k$.
Thus, introduction of policy $1$ incentivizes a reduction in the desired choice of $\y^*(0)$ on the right side of the notch.
Condition~\eqref{eq:notch-condition} in the notch design can be viewed as a relaxed, one-sided version of Condition~\eqref{asm:kinked-payoff}, which was necessary for inducing bunching in the kink design.
Such negative notches can be found commonly in many regulatory policies that impose a discrete liability or penalty upon exceeding a cutoff.

A negative notch in the compound payoff schedule is known to create an excess mass precisely at the cutoff and a missing mass (gap) immediately above the cutoff.
To formalize this notion, we define the counterfactual values associated with each policy $d \in\{ 0,1 \}$ as follows:
\begin{align*}
    \vis{0} &= \max_{\yv \le \k}U_0(\yv, \x_i, \eta_i) = U_0(\min(\yis{0},\k), \x_i, \eta_i), \\
    \vis{1} &= \max_{\yv \ge \k} U_1(\yv, \x_i, \eta_i) = U_1(\max(\yis{1},\k), \x_i, \eta_i).    
\end{align*}
Note that the counterfactual value of each policy is derived under the corresponding choice set constraint.
Let 
$$
\yiis = \argmax{\yv \in \mathcal{Y}} U(\yv,\x_i,\eta_i)
$$
denote the desired choice subject to the actual payoff schedule, which can be written as 
$$
U(\yv,\xv,\eta) = U_0(\yv, \xv, \eta)\I{\yv \le \k} + U_1(\yv, \xv, \eta)\I{\yv > \k}.
$$
Suppose that ties are broken by choosing the lowest value.
Then, we have the following link between $\yiis$ and the set of counterfactual quantities $(\yis{0}, \yis{1}, \vis{0}, \vis{1})$.

\begin{propA}
\label{propA:notch-design}
Assume that each counterfactual payoff $U_d$ is strictly concave in $\yv \in \mathcal{Y}$ for each $(\xv, \eta) \in \mathcal{X}\times \mathcal{H}$.
Then, it holds
\begin{align*}
    \yiis  = \begin{cases}
\yis{0}  & \text{ if }\ \yis{0} < \k\\
\yis{1}  & \text{ if }\ \vis{0} < \vis{1}   \\
\k       & \text{ if }\ \yis{0} \ge \k, \ \vis{0} \ge \vis{1}
\end{cases}.
\end{align*}
These three events correspond to $\{ \yiis < \k \}$, $\{ \yiis > \k \}$, and $\{ \yiis = \k \}$, respectively.
Consequently, $\{ \vis{0} < \vis{1} \} \subseteq \{ \yis{0} \ge \k \} \cap \{ \yis{1} > \k \}$.
\end{propA}
The proof of Proposition~\ref{propA:notch-design} is deferred to the end of this section.
Proposition~\ref{propA:notch-design} establishes a concise link between the actual choice and the counterfactual choices and values in the notch design.
It differs from Proposition~\ref{prop:bunching-characterization} since $\yis{1}$ is now subject to censoring by the comparison of the counterfactual values.
This censoring creates a noticeable hole in the distribution of $\yiis$, and makes it challenging to infer the counterfactual distribution of $\yis{1}$ from $\yiis$ observed on the right side of the cutoff.

Proposition~\ref{propA:notch-design} implies that no $\yiis$ (if directly observable) can be located immediately above $\k$.
For the reasoning, assume that $\yiis = \yis{1} = \k + \varepsilon$ as $\varepsilon \to 0+$.
By Proposition~\ref{propA:notch-design}, this implies $\vis{1} = U_1(\k + \varepsilon, \x_i, \eta_i) > \vis{0} \ge U_0(\k, \x_i, \eta_i)$ as $\varepsilon \to 0$, contradicting Condition~\eqref{eq:notch-condition}.
The range of $\yv \in [\k, \oy]$ such that $U_0(\k,\xv,\eta) > U_1(\yv, \xv, \eta)$ regardless of $\eta \in \mathcal{H}$, in particular, is referred to as the dominated region in the literature.

\subsection{Structural Equations}
Assume we have specified a correct utility model, from which the structural equations for the counterfactual choices and values can be derived as follows:
\begin{align*}
    \yis{d} &= m(d, \x_i, \eta_i, \theta_0),\\    
    \vis{d} &= v(d, \x_i, \eta_i, \theta_0).
\end{align*}
To proceed, we make an assumption analogous to Assumption~\ref{asm:invertible-structural-equations}(i): both $m$ and $v$ are continuous and strictly increasing with respect to $\eta$ for each $d \in \left\{ 0,1\right\}$ and $(\xv,\theta) \in \mathcal{X} \times \Theta$.
Note that $v(d,.)$ is strictly increasing with $\eta$ given that $\partial U_d/\partial \eta > 0$ for each $d \in \{ 0,1 \}$.
Moreover, for each $(\xv,\theta) \in \mathcal{X} \times \Theta$, we assume there exist $H(\xv,\theta) \in \mathcal{H}$ such that 
\begin{equation}
\label{eqA:single-crossing}
    v(1,\xv,\eta,\theta)-v(0,\xv,\eta,\theta)
    \begin{cases}
>0 & \ \ \text{if }\eta > H(\xv,\theta)\\
\le0 & \ \ \text{if } \eta \le H(\xv,\theta)
\end{cases}.
\end{equation}
The single crossing condition~\eqref{eqA:single-crossing} is satisfied under the assumption that $\partial [v(1,\xv,\eta,\theta)-v(0,\xv,\eta,\theta)]/\partial \eta > 0$ for all $(\xv, \eta,\theta)$, which, in turn, holds provided
\begin{equation}
\label{eqA:suff-condition-for-single-crossing}
    \frac{\partial}{\partial \eta}[U_1(\k, \xv, \eta, \theta)-U_0(\k, \xv, \eta, \theta)] > 0 \ \ \text{and}\ \ \frac{\partial^2}{\partial \eta \partial\yv}U_d(\yv, \xv, \eta, \theta) \ge 0
\end{equation}
for each $d$ and $\theta$.
To see this, note that by the envelope theorem,
\begin{align*}
    {\partial v(1,\xv,\eta,\theta)}/{\partial \eta} &= \frac{\partial}{\partial \eta}U_1(\yv, \xv, \eta, \theta)|_{\yv=\max(m(1,\xv,\eta,\theta), \k)}\\
    & \ge \frac{\partial}{\partial \eta}U_1(\k, \xv, \eta, \theta) \\
    & > \frac{\partial}{\partial \eta}U_0(\k, \xv, \eta, \theta) \\
    &\ge \frac{\partial }{\partial \eta}U_0(\yv, \xv, \eta, \theta)|_{\yv=\min(m(0,\xv,\eta,\theta), \k)} = {\partial v(0,\xv,\eta,\theta)}/{\partial \eta}.
\end{align*}
Condition~\eqref{eqA:suff-condition-for-single-crossing} is met, for instance, in the standard isoelastic model that features a negative notch at $\k$.

Condition~\eqref{eqA:single-crossing} implies that those with $\eta_i > H(\x_i,\theta_0)$ will strictly prefer policy $1$ over policy $0$.
The value $H(\xv,\theta)$ identifies marginal individuals who are indifferent between two policies among those who share the observed characteristics $\xv$.

Proposition~\ref{propA:notch-design} imposes two restrictions on the model in terms of $H(\xv,\theta)$ and the structural equations $m(d,\xv,\eta,\theta)$.
Due to the mutual exclusivity between the three events in Proposition~\ref{propA:notch-design}, it must hold that $m(0,\xv,H(\xv,\theta), \theta) \ge \k$ and $m(1,\xv,H(\xv,\theta), \theta) \ge \k$ across all values of $\theta \in \Theta$.

For each value of $\theta$, we define 
$$
\mathcal{K}(\xv,\theta) := m(1,\xv,H(\xv,\theta), \theta) \ge \k
$$
as the upper limit of $\yis{1}$ among those engaged in bunching who share the same observed characteristics $\xv$.
Thus, this also identifies the location of marginal individuals in the distribution of $\yiis$.
This allows us to express bunching at the notch as
\begin{align*}
    \{ \yiis = \k \} &= \{ \yis{0}\ge \k,\ \yis{1} \le \mathcal{K}(\x_i, \theta_0) \} \\
    &= \{ \k \le \yis{0} \le \conv(\mathcal{K}(\x_i, \theta_0), \x_i, \theta_0) \},
\end{align*}
where the reversion $\conv$ is defined as in \eqref{eq:def-reversion}.
The above equation shows that the role of $\conv(\k,\x_i,\theta)$ in the kink design is now played by
$
\conv(\mathcal{K}(\x_i,\theta),\x_i,\theta)
$
in the notch design with $\k$ replaced by $\mathcal{K}(\x_i,\theta_0)$.

The standard bunching identification in the literature relies on the first-order (small-notch) approximation
\begin{equation*}
    \P(\yiis = \k) \simeq (\conv(\mathcal{K}(\theta_0),\theta_0)-\k) f_{\ys{0}}(\k),
\end{equation*}
which serves as an identifying restriction for $\theta_0$ assuming a model without observable heterogeneity.
However, our identification method accounts for higher-order effects when $f_{\ys{0}}$ is not flat or when $\conv(\mathcal{K}(\xv,\theta), \xv,\theta)$ depends on covariates.


\subsection{Optimization Frictions}
\label{subsectA:optim-frictions}

In reality, the observed distribution of $\yii$ often exhibits a non-trivial amount of mass located immediately above the cutoff.
The presence of a positive mass in the dominated region is typically attributed to optimization frictions in the literature (\citeay{klevenUsingNotchesUncover2013}), modeled by the inattention parameter $a\in [0,1]$.
While the standard method could also be incorporated into our approach, we propose a more conservative treatment of optimization frictions: we assume that they can arbitrarily shuffle the distribution within a narrow window surrounding the hole in the distribution of $\yiis$, as this better aligns with the analysis presented in the paper.

Let $\uk \le \k$ be a value in the interior of $\CY_0$, which represents the lower edge of the excluded window.
Moreover, let $\oe \ge 0$ be a fixed constant representing the distance of the upper edge of the excluded window from the location of the marginal individuals in the distribution $\yiis$ given $\x_i$.
That is, $\ok(\x_i) = \mathcal{K}(\x_i,\theta) + \oe$ serves as the upper bound of the window subject to the impact of optimization errors.
Since $\mathcal{K}(\x_i,\theta) \equiv \k$ holds in the kink design, this can be considered an analogue of Assumption~\ref{asm:data-optimization-errors} allowing $\ok$ to depend on $\x_i$ and $\theta$.
Then, we make the following assumption on the data-generating process of the actual choices $(Y_i)_{i=1}^n$.
\begin{asmA}
\label{asmA:optimization-frictions}
For each unit $i \in \{ 1,\ldots, n \}$, the following hold.
\begin{enumerate}[leftmargin = 0.04\linewidth]

\item     
If $\yiis \in [\uk, \mathcal{K}(\x_i,\theta_0) + \oe]$, then $\yii \in [\uk, \mathcal{K}(\x_i,\theta_0)+\oe]$.
\item
If $\yiis \notin [\uk,\mathcal{K}(\x_i,\theta_0) + \oe]$, then $\yii = \yiis$.

\end{enumerate}

\end{asmA}
Note that since $\yiis \in (\k, \mathcal{K}(\x_i,\theta_0)]$ is empty, $\yiis \notin [\uk, \k]\cup [\mathcal{K}(\x_i,\theta_0),\mathcal{K}(\x_i,\theta_0) + \oe]$ is equivalent to $\yiis \notin [\uk, \mathcal{K}(\x_i,\theta_0) + \oe]$.
Assumption~\ref{asmA:optimization-frictions} requires that the optimization frictions can impact the distribution of $\yii$ by relocating $\yiis$ only within the window $[\uk, \mathcal{K}(\x_i,\theta_0) + \oe]$.
This accounts for two potential mistakes committed by those engaged in bunching.
They may either be located within a small window $[\uk,\k]$ below the notch point or end up within $[\k, \mathcal{K}(\x_i,\theta_0) + \oe]$, possibly due to mistaking policy $1$ for their preferred policy.
In contrast, those with $\yiis < \uk$ or $\yiis > \mathcal{K}(\x_i,\theta_0) + \oe$ are assumed to stick with their desired choices.
This allows for fuzzy bunching with a positive mass remaining in the dominated region.

\subsection{Point Identification}

Our extension is based on the following observation.
Under Assumption~\ref{asmA:optimization-frictions}, the fuzzy bunching moment can be expressed as
\begin{align*}
\Q(\uk \le \yii \le \mathcal{K}(\x_i, \theta_0)+\oe) &= \Q(\yiis \in [\uk, \mathcal{K}(\x_i,\theta_0) + \oe]) \\
&= \Q(\yis{0} \in [\uk, \conv(\mathcal{K}(\x_i, \theta_0) + \oe, \x_i,\theta_0)]).    
\end{align*}
Define 
$$
\widetilde \conv(\oe,\x_i,\theta) := \conv(\mathcal{K}(\x_i, \theta) + \oe, \x_i,\theta) \ge \k
$$
as the generalized reversion for each $\theta \in \Theta$.
We can derive a bunching moment equation for $\theta_0$ using a variant of Theorem~\ref{thm:point-identification}.
More specifically, assume the same conditions as in Theorem~\ref{thm:point-identification}, except that $\{ \quant^{(\Q)}_{\widetilde \conv(\oe,\x_i,\theta)|\ys{0}}(\qua,.) \}_{\qua \in (0,1)}$ now satisfies the analyticity condition~(ii) in place of $\{ \quant^{(\Q)}_{\conv(\ok, \x,\theta)|\ys{0}}(\qua,.) \}_{\qua \in (0,1)}$ for each $\theta \in \Theta$.
Then, $\theta_0$ arises as a solution to the following equation:
\begin{align*}
   & \E[\ti \I{\uk \le \yii \le \mathcal{K}(\x_i, \theta) + \oe}]\\
    = &\ \sum_{j=1}^\infty \frac{1}{j!}\dif_{\yv}^{j-1}[\E[\ti (\widetilde \conv(\oe, \x_i,\theta)-\uk)^j|\yis{0} = \yv]f_{\ys{0}}(\yv)]_{\yv=\uk}.
\end{align*}
This is similar to Theorem~\ref{thm:point-identification} except that the bunching moment on the left-hand side is now dependent on $\theta$ and $\conv(\ok,\x_i,\theta)$ is replaced with $\widetilde \conv(\oe,\x_i,\theta)$.
Our counterfactual estimation and inference procedure depicted in Section~\ref{sect:confidence-region} can be straightforwardly modified to test this relationship under $H_0 : \theta_0 = \theta$ for each hypothesized value of $\theta$.

Unlike the standard practice in the notch design, our method does not assume a common inattention parameter $a \in [0,1]$ governing the share of non-optimizing individuals.
Our method does not require separate adjustments for non-optimizing behavior, as the fuzzy bunching moment accounts for it by incorporating $\uk$ and $\oe$.
Our assumption allows for any reallocation of mass within the distribution hole, including patterns driven by individual-specific inattention parameters.

\subsection{Proof of Proposition~\ref{propA:notch-design}}
The proof proceeds in the following steps:
\begin{enumerate}[leftmargin = 0.04\linewidth]

    \item If $\yis{0} < \k$, then $\yiis = \yis{0} < \k$.
    Conversely, if $\yiis < \k$, then $\yis{0} < \k$ and is equal to $\yiis$.

    \item If $\vis{1} > \vis{0}$, then $\yiis = \yis{1} > \k$.
    If $\yiis > \k$, then $\vis{1} > \vis{0}$ and $\yiis = \yis{1}$.

    \item Conclude by noting that Steps~1 and 2 together imply that $\yiis = \k$ if and only if $\yis{0} \ge \k$ and $\vis{0} \ge \vis{1}$.
\end{enumerate}
\noindent \tbf{Step 1}:
Assume $\yis{0} < \k$.
We will first show that $\yiis = \argmax{\yv} U(\yv, \x_i, \eta_i)$ is less than $\k$.
Assume to the contrary that $\yiis \ge \k$ is a global maximum of $U$.
Then, by the negative notch condition, we must have $U(\yiis, \x_i,\eta_i) = U_1(\yiis, \x_i,\eta_i) < U_0(\yiis, \x_i, \eta_i) \le U_0(\yis{0}, \x_i, \eta_i) = \vis{0}$.
This leads to a contradiction since $\vis{0} \le U(\yiis, \x_i,\eta_i)$ must hold.
Thus $\yiis < \k$.
When $\yiis < \k$, it then must coincide with the constrained maximizer of $U$ subject to the constraint $\yv < \k$, which is $\yis{0}$.

To prove the converse, assume $\yiis < \k$.
If $\yis{0} \ge \k$, then by the strict concavity of $U_0$, $U_0$ must be strictly increasing on the left-hand side of $\k$.
This implies $U(\yii,\x_i,\eta_i) = U_0(\yiis,\x_i,\eta_i) < U_0(\yiis+\varepsilon,\x_i,\eta_i) = U(\yiis + \varepsilon, \x_i, \eta_i)$ for any $\varepsilon \in (0, \k - \yiis)$.
This contradicts that $\yiis = \argmax{\yv \in \mathcal{Y}} U(\yv, \x_i, \eta_i)$.
The equality $\yis{0} = \yiis$ must hold by the first part.
\medskip
\\
\noindent \tbf{Step 2}:
Suppose $\vis{1} > \vis{0}$.
This implies $\yis{1} > \k$ because otherwise $\vis{1} = U_1(\k, \x_i, \eta_i) < U_0(\k, \x_i,\eta_i) \le \vis{0}$, contradicting the previous assumption.
Since $\vis{1} > \vis{0}$, unit $i$ will choose $\yiis > \k$, or more specifically $\yii = \yis{1}$.
Now, suppose $\yiis > \k$.
This is only possible when $\vis{1} \ge \vis{0}$.
We can further rule out the possibility of equality since the ties are broken to be the lowest.
Thus $\vis{1} > \vis{0}$, which implies $\yis{1} < \k$ and $\yiis = \yis{1}$, as established previously.


\section{Partial Identification under Nonparametric Shape Restrictions}
This section illustrates our applications of Theorem~\ref{thm:partial-identification} to the isoelastic model under shape restrictions discussed in the literature.
We consider two nonparametric shape restrictions: monotonic densities in \citet{blomquistBunchingIdentificationTaxable2021} and Lipschitz bounded densities in \citet{bertanhaBetterBunchingNicer2023}.
We retrieve their identified bounds under the respective set of assumptions and extend those of \citet{bertanhaBetterBunchingNicer2023} to a setting that allows for optimization frictions considered in this paper.

\subsection{Replication of bounds in \citet{blomquistBunchingIdentificationTaxable2021}}
We assume that $\eta_i$ possesses a positive and continuous density $f_\eta$ in the interior of its support containing $[\ueta, \oeta]$.
\citet{blomquistBunchingIdentificationTaxable2021} imposed the following shape restriction on $f_\eta$: for all $\eta \in [\ueta, \oeta]$ and some $0 < \sigma \le 1 \le \bar{\sigma}$,
\begin{equation}
\label{eqA:blomquist-assumption}
    \sigma \min(f_\eta(\ueta), f_\eta(\oeta)) \le f_\eta(\eta) \le \bar{\sigma} \max(f_\eta(\ueta), f_\eta(\oeta))
\end{equation}
where $[\ueta, \oeta] = [\uk (1-\tau_1)^{-\theta_0}, \ok(1-\tau_0)^{-\theta_0}]$ denotes the bunching interval.
In particular, when $\sigma = \bar{\sigma} = 1$, this condition imposes that the maximum and minimum of $f_\eta$ occur at the boundary of $[\ueta, \oeta]$.
This assumption encompasses all monotonic densities.

This section demonstrates that, assuming Condition \eqref{eqA:blomquist-assumption} and the window of optimization errors, $[\uk, \ok]$, it holds (\citeay{blomquistBunchingIdentificationTaxable2021}, Theorem~2)
\begin{equation}
\label{eqA:blomquist-bounds}
    \sigma \min(D^-(\theta_0),D^+(\theta_0)) \le \P(\yii \in [\uk, \ok]) \le \bar{\sigma} \max(D^-(\theta_0),D^+(\theta_0)),
\end{equation}
where, for all $\theta > 0$,
\begin{align*}
    D^-(\theta) &= f_{\y}(\uk) \left(\left( \frac{\rho_0}{\rho_1}\right)^{\theta} \ok - \uk \right)   ,\\
    D^+(\theta) &= f_{\y}(\ok) \left(\ok -\left(\frac{\rho_0}{\rho_1}\right)^{-\theta}\uk \right).  
\end{align*}
From here on, we will denote the net-of-tax rates subject to the prekink and postkink policies as $\rho_d = 1-\tau_d$, $d \in\{ 0,1 \}$, respectively, for notational convenience.
To prove \eqref{eqA:blomquist-bounds}, we first observe the following:

\begin{propA}
\label{propA:observed-density-from-unobservable}
It holds
$
f_{\ys{d}}(\yv) = 
\rho_d^{-\theta_0} f_\eta(\rho_d^{-\theta_0} \yv)
$
for each $d \in \{ 0,1 \}$.
Morevoer, $\y$ permits the density
\begin{align*}
     f_{\y}(\yv) = 
     \begin{cases}
     \rho_0^{-\theta_0} f_\eta(\rho_0^{-\theta_0} \yv) & \yv < \uk\\
\rho_1^{-\theta_0} f_\eta(\rho_1^{-\theta_0} \yv) & \yv > \ok
\end{cases}
\end{align*}
outside $[\uk, \ok]$.
\end{propA}
\begin{proof}
This is an immediate result of \eqref{eq:solution-isoelastic} and the change of variables formula for PDFs.
\end{proof}
As a direct consequence of Proposition~\ref{propA:observed-density-from-unobservable}, it follows
\begin{equation*}
    f_{\ys{0}}(\yv) \in [\sigma \rho_0^{-\theta_0} \min(f_\eta(\ueta),f_\eta(\oeta)) , \bar{\sigma}\rho_0^{-\theta_0}\max(f_\eta(\ueta),f_\eta(\oeta))]
\end{equation*}
for all $\yv \in [\uk,\ook] = [\uk, (\rho_0/\rho_1)^{\theta_0} \ok]$ by Condition~\eqref{eqA:blomquist-assumption}, and that
\begin{align*}
    f_\eta(\ueta) = \rho_0^{\theta_0} f_{\y}(\uk),\quad 
    f_\eta(\oeta) = \rho_1^{\theta_0} f_{\y}(\ok).
\end{align*}
We apply Theorem~\ref{thm:partial-identification} with constant envelope functions, $\uf = \sigma\rho_0^{-\theta_0} \min(f_\eta(\ueta),f_\eta(\oeta))$ and $\of = \bar{\sigma} \rho_0^{-\theta_0}\max(f_\eta(\ueta),f_\eta(\oeta))$, and $\ti = 1$.
From the fact that $\conv(\ok,\xv,\theta) - \uk = (\rho_0/\rho_1)^\theta \ok - \uk \equiv c$, we have
\begin{align*}
    \underbar{B}(\theta_0) &= c \uf = \sigma \min(c f_\y(\uk), c (\rho_0/\rho_1)^{-\theta_0}f_\y(\ok)) = \sigma \min(D^-(\theta_0), D^+(\theta_0)),\\
    \bar{B}(\theta_0) &= c \of = \bar{\sigma} \max(c f_\y(\uk), c (\rho_0/\rho_1)^{-\theta_0}f_\y(\ok)) = \bar{\sigma} \max(D^-(\theta_0), D^+(\theta_0)).
\end{align*} 
Since $\P(\yii \in [\uk, \ok]) \in [\underbar{B}(\theta_0),   \bar{B}(\theta_0)]$, this completes the proof of \eqref{eqA:blomquist-bounds}.

\subsection{Extension of bounds in \citet{bertanhaBetterBunchingNicer2023}}
This section extends the partial identification bounds in \citep{bertanhaBetterBunchingNicer2023}{Theorem~2} in the presence of optimization frictions.
In the absence of optimization frictions, the same nonparametric bounds as in \citet{bertanhaBetterBunchingNicer2023} are recovered.

Following \citet{bertanhaBetterBunchingNicer2023}, we assume that the density of $\log (\eta_i)$ is Lipschitz continuous on $[\log \ueta, \log \oeta]$ with a Lipschitz constant given by $M > 0$.
Given this assumption, we will establish that
\begin{align}
\label{eqA:bertanha-bounds}
    \theta_0 \in \begin{cases}
\varnothing & \text{if}\quad B <   \left|   \frac{f_y (\ukk-)^2-f_y (\okk+)^2}{2M}    \right|\\
[\utheta,\otheta] & \text{if}\quad \left|   \frac{f_y (\ukk-)^2-f_y (\okk+)^2}{2M}    \right| \le B < \frac{f_y (\ukk-)^2+f_y (\okk+)^2}{2M} \\
[\utheta, \infty) & \text{if}\quad  \frac{f_y (\ukk-)^2+f_y (\okk+)^2}{2M}  \le B \\
\end{cases},
\end{align}
where $f_\yv$ denotes the observed density of logged income $\yv_i = \log (\yii)$, $B$ denotes the bunching fraction, and 
\begin{align*}
    \utheta &= \frac{\frac{-\frac{f_y(\ukk-)+f_y(\okk+)}{2} +\sqrt{\frac{f_y (\ukk-)^2+f_y (\okk+)^2}{2} +MB}}{M/2} - (\okk - \ukk)}{\log(\rho_0/\rho_1)},\\    
    \otheta &= \frac{\frac{\frac{f_y(\ukk-)+f_y(\okk+)}{2} - \sqrt{\frac{f_y (\ukk-)^2+f_y (\okk+)^2}{2} -MB}}{M/2} - (\okk - \ukk)}{\log(\rho_0/\rho_1)}.    
\end{align*}
Note that when $\ukk = \okk = k$, these bounds coincide with those in Theorem~2 of \citet{bertanhaBetterBunchingNicer2023}.

Since the Lipschitz boundedness is imposed on the density of $\log(\eta_i)$, denoted by $f_{\log \eta}$, we treat $y_i^*(d) = \log (\yis{d}) = \theta_0 \log \rho_d + \log (\eta_i)$ as the log counterfactual income subject to each policy $d\in\{ 0,1 \}$.
According to the log transform, we define the reversion as $\conv(\yv,\theta) = \yv + \theta \log(\rho_0/\rho_1)$.
Similarly, we denote $k_0 \le k_1$ as the edges of the optimization error window, and $\bar{k}_1 = \conv(\okk, \theta_0)= \okk + \theta_0 \log(\rho_0/\rho_1)$.

\begin{figure}[ht!]
	\caption{Upper and lower envelope functions for $f_{y^*(0)}$}
	\label{figA:envelope-functions}
	\begin{center}
		\includegraphics[scale=0.6]{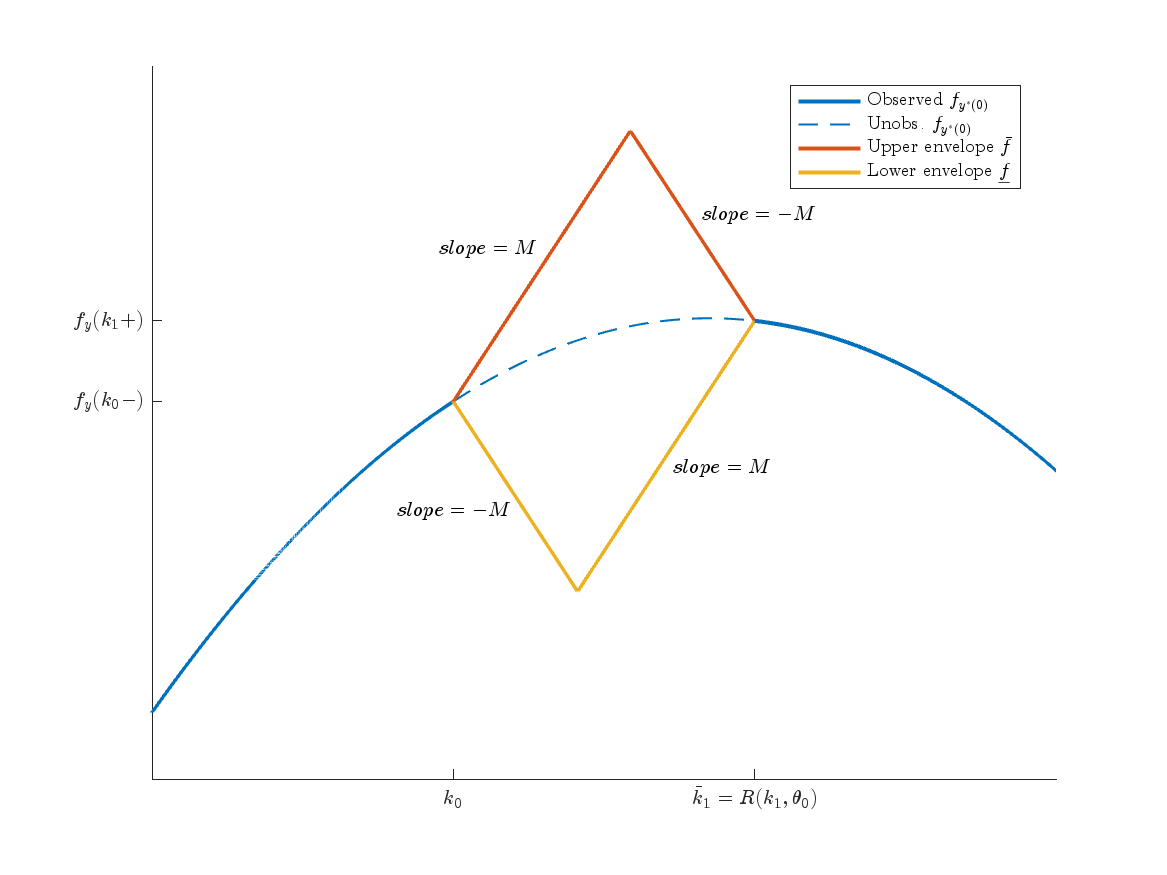}
	\end{center}
\end{figure}

From the fact that $f_{y^*(0)}(\yv) = f_{\log \eta}(\yv - \theta_0 \log \rho_0)$, $f_{y^*(0)}$ must also be a Lipschitz function on $[\ukk, \bar{k}_1]$ with the same Lipschitz bound.
Note that $f_{y}(k_0-) = f_{y^*(0)}(k_0)$ and $f_{y}(k_1+) = f_{y^*(1)}(k_1) = f_{y^*(0)}(\bar{k}_1) = f_{y^*(0)}(\conv(\okk,\theta_0))$.
Consequently, the upper and lower envelopes for $f_{y^*(0)}$ are given by the following functions, which together form an enclosed region in the shape of a parallelogram:
\begin{align*}
    \of(\yv) &=  M(\yv - \ukk) + f_{y}(\ukk-) -2M(\yv - k_{mid} - \Delta)\mathbbm{1}\{\yv > k_{mid} + \Delta\},\\   
    \uf(\yv) &=  -M(\yv - \ukk) + f_{y}(\ukk-) +2M(\yv - k_{mid} + \Delta)\mathbbm{1}\{\yv > k_{mid} - \Delta\},
\end{align*}
where $k_{mid} = (\ukk + \conv(\okk,\theta_0))/2$ and $\Delta = \frac{f_{y}(\okk-) - f_{y}(\ukk+)}{2M}$.
Figure~\ref{figA:envelope-functions} provides a graphical representation of these functions.
Henceforth, we will drop the left and right limits in the notation for simplicity of notation.
Since $\of$ and $\uf$ exhibit kinks at $k_{mid} \pm \Delta$, respectively, we need an extension of Theorem~\ref{thm:partial-identification} that allows for a finite number of break points in the envelope functions.
Such an extension is presented in the following theorem.


\begin{thmA}
\label{thmA:extension-partial-iden}
Assume the same set of assumptions as in Theorem~\ref{thm:partial-identification} except that $\of$ and $\uf$ are piecewise analytic functions such that
\begin{align*}
    \of(\yv) &= \sum_{s=1}^{S} \of_{s}(\yv)\I{\yv \in [\knot_{s-1},\knot_{s})} ,    \\
    \uf(\yv) &= \sum_{s=1}^{S} \uf_{s}(\yv)\I{\yv \in [\knot_{s-1},\knot_{s})} ,   
\end{align*}
where $\uk = \knot_0 < \knot_1 < \cdots < \knot_{S} = \ook$ are knots in $[\uk, \ook]$.
Moreover, for each $s =1,\ldots, S$, assume that $\of_{s}$ and $(\qb(\qua,\theta,.))_{\qua \in (0,1)}$ satisfy the analyticity condition for each $\theta \in \Theta$ with uniform smoothness constants $(\rad, \bar{\beta},\delta)$ such that $\delta \in [0,1)$, and that $\uf_{s}$ and $(\qb(\qua,\theta,.))_{\qua \in (0,1)}$ satisfy the same condition with smoothness constants $(\rad, \underbar{\beta},\delta)$.
Then, it holds that $Q(\yii \in [\uk, \ok]) \in [\ubf(\theta_0), \obf(\theta_0)]$ where
\begin{align*}
             \bar{B}(\theta) &= \sum_{s=0}^{S-1}\left( \sum_{j=1}^\infty \frac{1}{j!} \dif_{\yv}^{j-1}[\E_\Q[(\conv(\ok, \x_i,\theta) - \knot_{s})_+^j|\yis{0} = \yv] \Delta \of_s(\yv)]_{\yv = \knot_{s}}\right),\\
        \underbar{B}(\theta) &= \sum_{s=0}^{S-1} \left( \sum_{j=1}^\infty \frac{1}{j!} \dif_{\yv}^{j-1}[\E_\Q[(\conv(\ok, \x_i,\theta) - \knot_{s})_+^j|\yis{0} = \yv] \Delta \uf_s(\yv)]_{\yv = \knot_{s}}\right),
\end{align*}
where $\Delta \of_s = \of_{s+1} - \of_{s}$ and $\Delta \uf_s = \uf_{s+1} - \uf_{s}$.
Both $\of_{0}$ and $\uf_{0}$ are set to $0$.
\end{thmA}
\begin{proof}
We apply Lemma~\ref{lemA:prob-integral-to-moment-series} to the integral $\int_{\knot_{s-1}}^{\knot_{s}} \E_\Q[\I{\yis{0} \le \conv(\ok,\x_i,\theta_0)}|\yis{0} = \yv] \of_s(\yv) d\yv$ to get
\begin{align*}
&     \int_{\knot_{s-1}}^{\knot_{s}} \E_\Q[\I{\yis{0} \le \conv(\ok,\x_i,\theta_0)}|\yis{0} = \yv] \of_s(\yv) d\yv \\
     = &\ \sum_{j=1}^\infty \frac{1}{j!} \left( \dif_{\yv}^{j-1}[\E_\Q[(\conv(\ok, \x_i,\theta) - \knot_{s-1})_+^j|\yis{0} = \yv] \of_s(\yv)]_{\yv = \knot_{s-1}}\right.\\
     & \ \ \ \ \ \ \ \ \ \ \left. -\dif_{\yv}^{j-1}[\E_\Q[(\conv(\ok, \x_i,\theta) - \knot_{s})_+^j|\yis{0} = \yv] \of_s(\yv)]_{\yv = \knot_{s}}\right).    
\end{align*}
Then, add both sides over $s = 1,\ldots, S$.
This implies $Q(\yii \in [\uk, \ok]) \le \obf(\theta_0)$.
The rest of the proof is omitted for brevity.
\end{proof}

Now, we derive the expressions for $\otheta$ and $\utheta$ by applying Theorem~\ref{thmA:extension-partial-iden} based on the bunching moment weighted by $\ti = 1$.
Following the notation adopted in \citet{bertanhaBetterBunchingNicer2023}, we denote $B = \P(y_i \in [\ukk, \okk])$ as the bunching mass, where $y_i = \log (\yii)$.
\medskip
\\
\noindent \tbf{Evaluation of $B \le \bar{B}(\theta_0)$:}
Let $v = \conv(\okk, \theta_0) - \ukk$.
According to Theorem~\ref{thmA:extension-partial-iden}, we have
\begin{align*}
      \bar{B}(\theta_0) &= \sum_{j=1}^\infty \frac{1}{j!} \dif_{\yv}^{j-1}[(\conv(\okk, \theta_0) - \ukk)^j (M(\yv - \ukk) + f_y(\ukk))]_{\yv = \ukk} \\
      & \qquad +   \sum_{j=1}^\infty \frac{1}{j!} \dif_{\yv}^{j-1}[(\conv(\okk, \theta_0) - k_{mid} - \Delta)_+^j (-2M)(\yv - k_{mid} - \Delta)]_{\yv = k_{mid} + \Delta} \\
      &= f_y(\ukk)(\conv(\okk, \theta_0) - \ukk) + \frac{M}{2}(\conv(\okk, \theta_0) - \ukk)^2 \\
      & \qquad -M (\conv(\okk, \theta_0) - k_{mid} - \Delta)_+^2 \\
      &= f_y(\ukk) v + \frac{M}{2}v^2 - \frac{M}{4}(v - 2 \Delta)^2
\end{align*}
where we used $k_{mid} + \Delta \le \conv(\okk, \theta_0)$ to simplify $(\conv(\okk, \theta_0) - k_{mid} - \Delta)_+ = \conv(\okk, \theta_0) - k_{mid} - \Delta = v/2 - \Delta$.
Thus, $v$ must satisfy the quadratic inequality
\begin{equation*}
    Q_1(v) :=  \frac{M}{4}v^2 + (f_y (\ukk) + M \Delta) v - (B + M \Delta^2) \ge 0.
\end{equation*}
Since $Q_1$ has both positive and negative roots, we can solve $Q_1(v) \ge 0$ as
\begin{align}
\label{eqA:bertanha-intermediate-1}
  \conv(\okk, \theta_0) - \ukk = v &\ge  \frac{-\frac{f_y(\ukk)+f_y(\okk)}{2} +\sqrt{2M^2 \Delta^2 + 2Mf_y(\ukk)\Delta +f_y(\ukk)^2 +MB}}{M/2} \nonumber \\
    &=  \frac{-\frac{f_y(\ukk)+f_y(\okk)}{2} +\sqrt{    \frac{f_y (\ukk)^2+f_y (\okk)^2}{2}     +MB}}{M/2} =: \underbar{v}.
\end{align}
Rearranging this for $\theta_0$ gives $\theta_0 \ge \utheta$.
\medskip
\\
\noindent \tbf{Evaluation of $B \ge \underbar{B}(\theta_0)$:} $\underbar{B}(\theta_0)$ can be expressed as a quadratic function of $v$ in a similar manner to the previous part.
We skip the details.
Consequently, we have that
\begin{equation*}
    Q_2(v) := \frac{M}{4}v^2 - (f_y (\ukk) + M \Delta) v + (B - M \Delta^2) \ge 0. 
\end{equation*}
According to the sign of the discriminant, defined as
\begin{equation*}
       D =  (f_y (\ukk) + M \Delta)^2 - M(B - M \Delta^2) = \frac{f_y (\ukk)^2+f_y (\okk)^2}{2} - MB,
\end{equation*}
we consider two cases.
\medskip
\\
\noindent \tbf{Case 1}:
Assume $D \le 0$. Then, $Q_2(v) \ge 0$ holds for all $v \in \mathbb{R}$.
Together with \eqref{eqA:bertanha-intermediate-1}, this leads to 
$
    \theta_0 \in [\utheta, \theta),
$
which corresponds to the last case in \eqref{eqA:bertanha-bounds}.
\medskip
\\
\noindent \tbf{Case 2}:
If $D > 0$, then $Q_2$ has two real roots, say $v_l < v_u$. 
Accordingly, $Q_2 \ge 0$ can be solved as
$
    v \in (-\infty, v_l] \cup [v_u, \infty),
$
where
\begin{align*}
    v_l &= \frac{\frac{f_y(\ukk)+f_y(\okk)}{2} - \sqrt{\frac{f_y (\ukk)^2+f_y (\okk)^2}{2} -MB}}{M/2},\\    
    v_u &= \frac{\frac{f_y(\ukk)+f_y(\okk)}{2} + \sqrt{\frac{f_y (\ukk)^2+f_y (\okk)^2}{2} -MB}}{M/2}.
\end{align*}

Now, we argue that $\min_{\yv \in [\ukk, \bar{k}_1]} \uf(\yv) = \uf(k_{mid} - \Delta) \ge 0$ holds if $D > 0$.
Once we have shown that, from the fact that
$$
\uf(k_{mid} - \Delta) = - M v/2 + (M \Delta + f_y(\ukk)) = -M/2\left(v - \frac{f_y(\ukk)+f_y(\okk)}{M}\right)\ge 0,
$$
we obtain $v \le v_l$.

Assume to the contrary that $\uf(k_{mid} - \Delta) < 0$.
We denote $(\ukk + \frac{f(\ukk)}{M}, \conv(\okk,\theta_0) - \frac{f(\okk)}{M}) \subseteq [\ukk, \bar{k}_1]$ as the middle segment where $\uf$ becomes negative.
Since $f_{\ys{0}} \ge 0$, we have $\max(\uf(\yv), 0)$ as a lower envelope function for $f_{\ys{0}}$ that is tighter than $\uf$.
Since it holds that
\begin{equation*}
    B = \int_{\ukk}^{\bar{k}_1} f_{\ys{0}} d\yv \ge \int_{\ukk}^{\bar{k}_1} \max(\uf(\yv), 0)d\yv =\frac{M}{2} \left( \left( \frac{f(\ukk)}{M}\right)^2+ \left( \frac{f(\okk)}{M}\right)^2\right)= \frac{f_y (\ukk)^2+f_y (\okk)^2}{2M},
\end{equation*}
this leads to a contradiction to $D > 0$.
\\

Putting it all together, if $v_l \ge \underbar{v}$ and $D > 0$, we have 
\begin{equation*}
    \frac{-\frac{f_y(\ukk)+f_y(\okk)}{2} +\sqrt{\frac{f_y (\ukk)^2+f_y (\okk)^2}{2} +MB}}{M/2} \le v \le \frac{\frac{f_y(\ukk)+f_y(\okk)}{2} - \sqrt{\frac{f_y (\ukk)^2+f_y (\okk)^2}{2} -MB}}{M/2}.    
\end{equation*}
Note that $v_l \ge \underbar{v}$ holds iff $B \ge \left| \frac{f_y (\ukk)^2-f_y (\okk)^2}{2M} \right|$.
Solving the above inequality for $\theta_0$, we obtain $\theta_0 \in [\utheta,\otheta]$, which corresponds to the second case in \eqref{eqA:bertanha-bounds}.

Otherwise, if $D>0$ and $v_l < \underbar{v}$, there are no values of $\theta_0$ satisfying the restriction $B \in [\underbar{B}(\theta_0),\bar{B}(\theta_0)]$.
This corresponds to the first case in \eqref{eqA:bertanha-bounds}.

\section{Characterizing Iterative Proportional Adjustment Polynomial Estimator as IV Estimator}
We provide an alternative formulation of the iterated proportional adjustment polynomial estimator proposed by \citet{chettyAdjustmentCostsFirm2011}.
We establish that the iterative limit of the polynomial estimator can be computed at once based on this alternative formulation, which takes the form of an IV regression.

We adopt the following notation.
Let $\{ \yii : i = 1,\ldots,n \}$ be the income data supported on $[\underbar{Y},\bar{Y}]$, which contains the excluded window $[\uk, \ok]$ in its interior.
Let $j = 1,\ldots, |\CJ|$ denote equi-spaced bins partitioning $[\underbar{Y},\bar{Y}]$, sorted from left to right.
The width of each bin is denoted by $b>0$.
For each bin $j$, the proportion of $\yii$ falling within that bin and its center are denoted by $f_j = \frac{1}{n} \sum_{i=1}^n \I{\yii \in j}$ and $c_j$, respectively.
Let $\uj+1$ and $\oj$ be the first and last bins contained in the excluded window $[\uk, \ok]$.
For simplicity, we assume that the right edge of $j_0$ equals $\uk$, and that of $j_1$ equals $\ok$.
Denote by $\P_R = \frac{1}{n} \sum_{i=1}^n \I{\yii > \ok}$ the proportion of $\yii$ located to the right of the excluded window.

Initialize $\hat{B}^{(0)} = 0$.
Given $\hat{B}^{(s-1)}$, obtained from $(s-1)$-th iteration step, we implement the following OLS regression in $s$-th iteration step:
\begin{align}
\label{eqA:chetty-regression}
    f_j\left( 1 +  \frac{\hat{B}^{(s-1)}}{\P_R}\I{j > \oj} \right)= z_k(c_j)' \hat{\gamma}^{(s)} + \sum_{l = \uj + 1}^{\oj} \hat{\beta}^{(s)}_l \I{j = l} + e_j^{(s)}.
\end{align}
The dependent variable is the observed proportion $f_j$ at bin $j$, multiplied by $1+\hat{B}^{(s-1)}/\P_R$ if bin $j$ is located to the right of the window.
This multiplier represents the iterative proportional adjustment that ensures the estimated counterfactual density integrates to 1.
The first set of regressors, $z_k(c_j) = (1, c_j, \ldots, c_j^{k-1})' \in \mathbb{R}^k$, is an order $k$ polynomial in the center of bin $j$, which is associated with the coefficient $\hat{\gamma}^{(s)}$.
This term represents the fitted counterfactual density, which would have been observed before the introduction of the kink.
The second set of regressors includes the fixed effects for the bins falling within the window, where each $\hat{\beta}_l^{(s)}$ measures the fraction of the excess bunching at bin $l$.
Note that, according to the Frisch-Waugh-Lovell theorem, $\hat{\gamma}^{(s)}$ can be computed from the same regression excluding bins $j = \uj + 1,\ldots, \oj$ from the data.
Each fixed effect estimate $\hat{\beta}_l^{(s)}$ can subsequently be computed as the difference of the dependent variable at bin $l$ from the value of the fitted counterfactual density $z_k(c_l)'\hat{\gamma}^{(s)}$ at bin $l$.

At the end of $s$-th iteration, the estimate for the gross excess bunching fraction is updated as $\hat{B}^{(s)} = \sum_{l= \uj + 1}^{\oj} \hat{\beta}_l^{(s)}$.
Then, we substitute $\hat{B}^{(s)}$ into $\hat{B}^{(s-1)}$ on the left-hand side of \eqref{eqA:chetty-regression}.
We proceed as in the previous step, moving to the $(s+1)$-th iteration.

Iterating this procedure, we obtain a sequence of estimates $(\hat{B}^{(s)})_{s =0}^\infty$ for the excess bunching fraction.
The iteration continues until $\hat{B}^{(s)}$ converges to a limit $\hat{B} = \lim_{s \to \infty}\hat{B}^{(s)}$, which finalizes the estimate for the excess bunching fraction.
The estimated counterfactual density can be iterated alongside $B^{(s)}$ in an obvious manner, yielding
the final estimate defined as $\hat{f}(\yv) = \lim_{s\to\infty} z_k(\yv)'\hat{\gamma}^{(s)}$.
The construction of the algorithm ensures that the integral constraint $\sum_{j=1}^{|\mathcal{J}|} \hat{f}(c_j) = 1$ is satisfied when the limit exists.

Note that, provided the convergence of $\hat{B}^{(s)} \to \hat{B}$, the corresponding $\hat{\gamma}^{(s)}$, $\hat{\beta}_l^{(s)}$'s, and the residuals converge to their respective limits.\footnote{
The existence of a limit $\hat{B}^{(s)} \to \hat{B}$ requires that the linear updating rule $\hat{B}^{(s)} = \phi_0 + \phi_1 \hat{B}^{(s-1)}$ corresponds to a contraction mapping, i.e., $|\phi_1|<1$.
The linearity follows from the linearity of the overall procedure.
Our characterization can determine the limit $\phi_0/(1-\phi_1)$ in the more general cases where $\phi_1 \ne 0$.
}
Taking these limits, we have the following:
\begin{align}
\label{eqA:poly-est-equation-1}
     \sum_{j=1}^{|\mathcal{J}|} e_j z_k(c_j)  &=0, \\
\label{eqA:poly-est-equation-2}
    \sum_{j=1}^{|\mathcal{J}|} e_j \I{j=l}   &=0,  \ \ \forall l \in [\uj+1,\oj],
\end{align}    
where 
$$
e_j = f_j + \hat{B} \frac{f_j}{\P_R} \I{j > \oj} - z_k(c_j)' \hat{\gamma} - \sum_{l = \uj + 1}^{\oj} \hat{\beta_l} \I{j = l}
$$
and $\sum_{l = \uj + 1}^{\oj} \hat{\beta}_l = \hat{B}$.
Here, $e_j$ represents $\lim_{s\to\infty}e_j^{(s)}$, and similarly for $\hat{\gamma}$ and $\hat{\beta}_l$'s.
By substituting $\hat{B} = \sum_{l = \uj + 1}^{\oj} \hat{\beta}_l$ into the above equation, we get
$$
e_j = f_j - z_k(c_j)' \hat{\gamma} - \sum_{l = \uj + 1}^{\oj} \hat{\beta_l} \left( \I{j = l}-\frac{f_j}{\P_R} \I{j > \oj}\right),
$$
which implies, after rearranging,
\begin{equation*}
        f_j = z_k(c_j)' \hat{\gamma} + \sum_{l = \uj + 1}^{\oj} \hat{\beta_l} \left( \I{j = l}-\frac{f_j}{\P_R} \I{j > \oj}\right) + e_j.
\end{equation*}
By the orthogonality conditions from \eqref{eqA:poly-est-equation-1} and \eqref{eqA:poly-est-equation-2}, it follows that $(\what \beta',\what \gamma')'$ are characterized as the IV estimator with
\begin{equation*}
    inst_j = 
    \left[ \begin{matrix}
        z_k(c_j) \\
        \I{j= \uj + 1}\\ \vdots\\ \I{j = \oj}
\end{matrix} \right],\ \ j = 1,\ldots, |\mathcal{J}|,
\end{equation*}
used as the set of instruments.

\section{Details about DGPs Used in Monte Carlo Experiments}

This section provides details about the DGPs used in Section~\ref{sect:monte-carlo}: DGP~1 is a seventh-degree polynomial density fitted to the U.S. tax records and DGP~2 is a Gaussian mixture distribution approximating a certain skewed generalized error distribution.

In Sections~\ref{sect:monte-carlo} and \ref{sect:empirical}, we revisited the data used in \citet{saezUsingElasticitiesDerive2001}.
Due to accessibility issues, we used the collapsed income histogram data published alongside \citet{blomquistBunchingIdentificationTaxable2021}, instead of the original data.
These data consist of taxable income bins ranging from -$\$20000$ to $\$60000$ with a width of $\$100$.
Each bin is associated with the share of tax filers falling within that bin, adjusted by sampling weights.
For our empirical application, these collapsed data are believed to serve well.
In the application, we generated $\tilde{n} = 1,025,838$ i.i.d. random draws from the income histogram distribution.
The simulated sample size $\tilde{n}$ is six times larger than the original sample size, $n = 170,973$, to effectively control the simulation error.
The length of the confidence interval is adjusted to match the original sample size by multiplying the estimated standard error by $\sqrt{6}$.

We fitted a seventh-degree polynomial to the simulated income distribution as follows.
First, we shift the simulated income upward by $\$20000$ to make the values nonnegative.
All bins whose center is located within the excluded window, $\$20000 \pm \$4000$, are excluded from the polynomial fitting.
We then regress the proportion of each bin on a seventh-degree polynomial in the bin centers.
The fitted density, denoted $f_{\ys{0}}$, is given by
\begin{align*}
    f_{\ys{0}}(\yv) &= 10^{-3}\times (4.986,
   25.223,
   3.839,
   14.006,
  -5.542,
   0.612,
  -0.009,
  -0.001)  (1, \yv, \ldots, \yv^7)'.
\end{align*}

For DGP~2, we first introduce the class of skewed generalized error distributions (SGED).
The SGED is a flexible family of distributions, parametrized as
\begin{align}
\label{eqA:SGED-density}
    f_{\op{SGE}}(\yv|\mu,\sigma,k,\lambda) = \frac{C}{\sigma} \exp \left( - \left(\frac{|\yv - \mu + \delta \sigma|}{(1 - \lambda \sgn(\yv - \mu + \delta \sigma)) \theta \sigma }\right)^k \right)
\end{align}
where $C$, $\delta$ and $\theta$ are constants derived from the parameter $(\mu,\sigma,k, \lambda) \in \mathbb{R} \times \mathbb{R}_{++} \times \mathbb{R}_{++} \times (-1,1)$.
For details and the definition of these constants, we refer to \citet{theodossiouSkewedGeneralizedError2015}.
The mean and the standard deviation of this distribution are represented by $\mu$ and $\sigma$.
The other parameters $k > 0$ and $\lambda \in (-1,1)$ determine the kurtosis and the skewness of the distribution.
In particular, if $k = 2$ and $\lambda = 0$, the SGED distribution with the parameter $(\mu,\sigma,2,0)$ coincides with the normal distribution with mean $\mu$ and variance $\sigma^2$.

\begin{figure}[ht!]
	\caption{Densities of mixture SGED and fitted Gaussian mixture distribution}
	\label{figA:SGED}
	\begin{center}
		\includegraphics[scale=0.4]{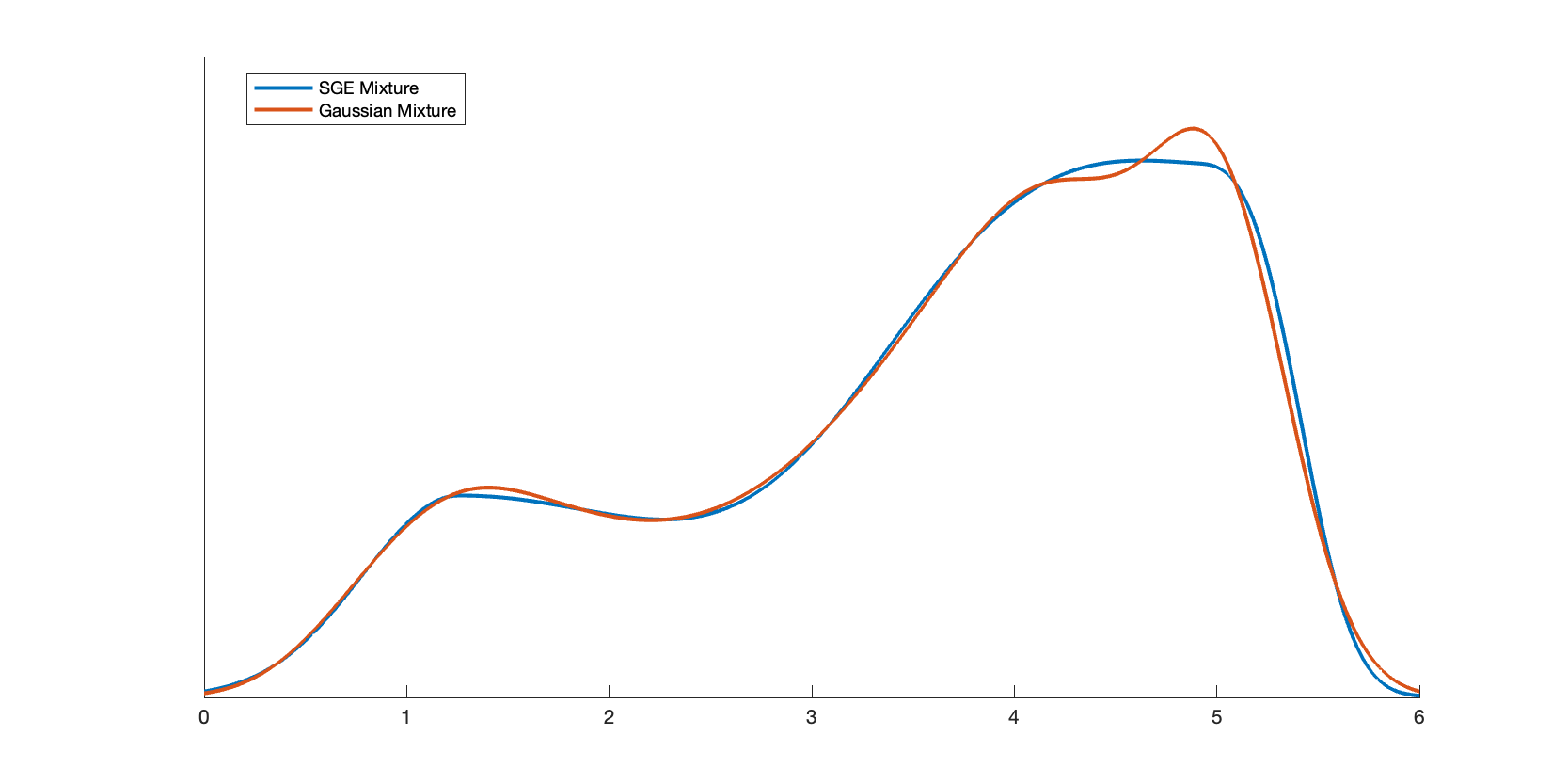}
	\end{center}
\end{figure}

The finite SGED mixture model was employed in \citet{bertanhaBetterBunchingNicer2023} for numerical experiments.
However, this family is not immediately suitable for our analysis, as the density in \eqref{eqA:SGED-density} exhibits a kink when $\lambda \ne 0$, violating the analyticity assumption.
To work around this issue, we have fitted a flexible location-scale Gaussian mixture model to the following two-component mixture of SGED:
\begin{equation*}
    0.3 f_{\op{SGE}}(\yv|0.5, 1, 2, -0.5) + 0.7 f_{\op{SGE}}(\yv|5.5, 0.75, 3, 0.5).
\end{equation*}
Figure~\ref{figA:SGED} exhibits the density for this mixture SGED and the fitted Gaussian mixture distribution.
The fitted Gaussian mixture distribution has 300 components, where each component $s = 1,\ldots,300$ is assigned a mean $\mu_s$, a variance $\sigma_s^2  \ge 0.1$, and a probability weight $\pi_s$.
The data for $(\mu_s,\sigma_s^2,\pi_s)_{s=1}^{300}$ is available upon request.

\section{Technical Lemmas and Proofs}
\label{sectA:tech-lemmas-proofs}
Section~\ref{sectA:tech-lemmas-proofs} is organized as follows.
First, we present technical and intermediary lemmas useful for proving the main results in the paper.
We borrow several concepts and tools from the complex analysis theory to develop an approximation theory of analytic functions tailored to our purposes.
When appropriate, we will occasionally use their Fourier-analytic counterparts, often expressed via the change of arguments $z = r e^{\i t} \in \mathbb{C}$.
The technical and intermediate lemmas are followed by their proofs.
The proof of the main results in the paper are presented at the end.

\subsection{Lemmas and Remarks}

We first characterize the class of analytic functions well-suited for our analysis.
Let $[a,b]$ be a nonempty compact interval and $R>0$.
\begin{define}
    Let $\mathcal{F}_R([\ubars,\bars])$ be a class of analytic functions $f:[\ubars,\bars] \to \mathbb{R}$ that can be analytically extended to the $R$-neighborhood of $[\ubars,\bars]$, equipped with a norm defined as 
$$
\|f\|_{\mathcal{F}_R([\ubars,\bars])} = \sup_{\vv \in [\ubars,\bars],0\le r \le R, x \in [0,1]} |f(\vv + r e^{2\pi \i x})|,
$$
where $f(\yv + r e^{2\pi \i x}) := f(\yv) + \sum_{j=1}^\infty \frac{1}{j!} \dif_{\yv}^{j}[f(\yv)] r^j e^{2\pi \i j x}$.
\end{define}
This norm quantifies the variability of the function $f$ alongside its derivatives within any compact subset of the $R$-neighborhood of $[\ubars,\bars]$.
Here, we define an $R$-neighborhood of a set $A \subseteq \mathbb{R}^d$ as $\{x  \in \mathbb{R}^d : \inf_{z \in A} \|x-z\| < R\}$.

In what follows, we denote $\P$ as a probability measure on the underlying measurable space and $\E$ as the expectation measured with respect to $\P$. Let $(\v, \s)$ denote a random vector with a compact support.
Lemmas~\ref{lemA:prob-integral-to-moment-series} and \ref{lemA:sufficient-analytic} play a central role in establishing Theorems~\ref{thm:point-identification} and \ref{thm:partial-identification} in the main text.
In proving these theorems, it is helpful to note that we will apply those lemmas to the random vector
$
(\v, \s) = (\yis{0}, \conv(\ok,\x_i,\theta)),
$
and the function, either $f_{\ys{0}}$, $\of$, or $\uf$, over the interval $[\ubars,\bars] =[\uk, \ook]$.


\begin{remark}
\label{rmk:analytic-condition}
The constant $\beta$ in Definition~\ref{def:analyticity-condition} represents the following quantity:
\begin{equation*}
\beta(\rad) = \sup_{\yv \in [\ubars,\bars]}  \int_0^{1} \left| \frac{\appftna{\alpha}(y + \rad e^{2\pi \i x})}{\appftna{\alpha}(y)} \right| dx \in [1,\infty),
\end{equation*}
which measures the relative variation of $\appftna{\alpha}$ in the $\rad$-neighborhood of $[\ubars,\bars]$.
By the mean-value property of holomorphic functions, $\beta(\rad)$ increases with $\rad\ge0$, and $\beta(0) = 1$.
Similarly, $\delta < 1$ regulates the local deviation of $\appftnb{\qua}(\yv)$ from $\yv$ by requiring that
\begin{equation*}
     \sup_{\yv \in [\ubars,\bars], |z| = \rad} \left| \frac{ \appftnb{\qua}(\yv + z)- \yv}{z} \right| \I{\yv \le \appftnb{\qua}(\yv)} \le \delta < 1 \ \ \text{for all }\qua \in (0,1).
\end{equation*}
A sufficient condition is given by 
\begin{equation*}
     \sup_{\yv \in [\ubars,\bars],|z| = \rho} \left| \frac{ \appftnb{\qua}(\yv + z)- \appftnb{\qua}(\yv)}{z} \right| \I{\yv \le \appftnb{\qua}(\yv)} + \frac{\appftnb{\qua}(\yv)-\yv}{\rad} \le \delta,
\end{equation*}
which requires $\appftnb{\qua}$ to be a contraction in $\yv$ around $\rad$-neighborhood of $[\ubars,\bars]$.

The magnitude of $\beta$ affects the finite-sample approximation bias in the counterfactual estimation.
On the other hand, $\delta < 1$ requires the upper edge of the bunching interval to be a contraction in $\yv$.
This condition is necessary to ensure that there are marginal individuals who are indifferent between bunching and nonbunching.
To see this point, note that the bunching indicator is given by $\{ \uk \le \yis{0} \le \conv(\ok, \x_i,\theta_0) \} = \{ \uk \le \yis{0} \le \qb(\qua_i, \theta_0, \yis{0}) \}$ where $\qua_i \in (0,1)$ denotes the $\Q$-quantile of $\conv(\ok, \x_i,\theta_0)$ for unit $i$ conditional on $\yis{0} = \yv$.
This equation utilizes the conditional version of the quantile transformation $X_i \overset d =  F_X^{-1}(\qua_i) = \quant_X (\qua_i)$ where $\qua_i \in (0,1)$ is uniformly distributed.
Since $\qb(\qua_i, \theta_0, \yv)$ is assumed to be a contraction in $\yv$, the bunching will only occur in the range $\yis{0} \in [\uk, U(\qua_i)]$, where $U(\qua_i) \le \oy$ denotes a unique solution to the equation $U(\qua_i) = \qb(\qua_i,\theta_0,U(\qua_i))$.
The existence and uniqueness of the solution is ensured by the intermediate value theorem and the condition $\delta < 1$. (See the proof of Theorem~\ref{thm:partial-identification} for more details.)
\end{remark}


\begin{lemA}
\label{lemA:prob-integral-to-moment-series}
Let $l \in \N$ and $f:[\ubars,\bars]\to\mathbb{R}$ be $(l-1)$-times continuously differentiable with $\dif^l f$ almost everywhere bounded in $[\ubars,\bars]$.
Assume that for each $1\le j \le l$, $\dif_\vv^{j-1}\E[(\s -\sv)_+^j |\v = \vv]$ is Lipschitz continuous in $(\sv,\vv) \in [\ubars,\bars] \times [\ubars,\bars]$.
Then, it holds that
\begin{align*}
&\int_{\ubars}^{\bars} \E\left[ \I{\s \ge \v}  | \v = \vv \right] f(\vv) d\vv \\
= &\ \sum_{j=1}^l \frac{1}{j!} \dif_\vv^{j-1} \left[ \E[(\s-\ubars)_+^j | \v =\vv] f(\vv) \right]_{\vv = \ubars} \\
& - \sum_{j=1}^l \frac{1}{j!} \dif_\vv^{j-1} \left[ \E[(\s-\bars)_+^j | \v =\vv] f(\vv) \right]_{\vv = \bars} + \frac{1}{l!}\int_{\ubars}^{\bars} \dif_{\vv}^{l} \left[ \E\left[(\s - \wv)_+^l | \v = \vv\right] f(\vv) \right]_{\wv = \vv} d \vv.
\end{align*}
\end{lemA}

\begin{lemA}
\label{lemA:sufficient-analytic}
Assume that $\s = \appftnb{}(\v,\tau)$ for some function $g$, where $\tau \in (0,1)$ is a random variable independent of $\v$ whose distribution is $\mu$.
Let $\appftna{}$ and $(\appftnb{}(., \tau))_{\tau \in (0,1)}$ satisfy the analyticity condition on $[\ubars,\bars]$ for some smoothness constants $(\rad, \delta, \beta)$. 
Then the following are true.

\begin{enumerate}[leftmargin = 0.05\linewidth]

\item [(i)] 
For each $j \in \N$, $\dif_\vv^{j-1}[ \E[(\s - \sv)_+^j |\v = \vv] f(\vv)]$ exists and is Lipschitz continuous in $(\sv,\vv)  \in [\ubars, \bars] \times [\ubars, \bars]$.

\item [(ii)] For each $m \in \Z_+$, $l \in \Z_+$, and $\sv \in [\ubars, \bars]$, it holds that, for all $R \in (0, \rad]$ and $\vv \in [\ubars, \bars]$,
\begin{align*}
&\frac{1}{l!} \left|\dif_{\vv}^l \left[ \E[(\s -\sv)_+^m | \v = \vv] f(\vv) \right] \right| \\
\le &\ R^{-l}\E_{\tau\sim \mu}\left[   \left| \int_0^{1} \re{\{ (\appftnb{}(\vv + R e^{2\pi \i x}, \tau) -\sv)^m f(\vv+ R e^{2\pi \i x}) e^{-2\pi \i l x} \}} dx \right|\I{\appftnb{}(\vv, \tau) - \sv > 0}  \right] \\
\le &\ R^{-l}\E_{\tau\sim \mu}\left[  \left( \sup_{x \in [0,1]}\left| \appftnb{}(\vv+R e^{2\pi \i x},\qua) - \sv\right| \right)^m \I{\appftnb{}(\vv,\qua) - \sv > 0} \right] \int_0^{1} |\appftna{}(\vv+ R e^{2\pi \i x})|dx.
\end{align*}
Here, the functions $f$ and $g$ are understood as the unique analytic continuations of the respective functions.

\item [(iii)] For each $l \in \N$, it holds
\begin{align*}
    & \frac{1}{l!} \int_{\ubars}^{\bars} \left|\dif_{\vv}^l \left [ \E[(\s-\wv)_+^l | \v = \vv] f(\vv) \right ]_{\wv = \vv}\right| d\vv \\
     \le &\ \rad^{-l}  \int_{\ubars}^{\bars} \E_{\tau\sim \mu}\left[   \left| \int_0^{1} \re{ \{ (\appftnb{}(\vv + \rad e^{2\pi \i x}, \tau) -\vv)^l f(\vv+ \rad e^{2\pi \i x}) e^{-2\pi \i l x} \}}dx \right|\I{\appftnb{}(\vv, \tau) - \vv > 0}  \right] d\vv  \\
     \le &\ \beta \delta^l \int_{\ubars}^{\bars} \E[\I{\s \ge \v} |\v = \vv] |f(\vv)| d\vv.
\end{align*}

\item [(iv)]
Let $f_j(\vv) := \E[(\s-\ubars)_+^j|\v = \vv]f(\vv)$ for each $j \in \N$.
Then, for any $0 < R < \rad$, $f_j$ can be analytically extended to the $R$-neighborhood of $[\ubars,\bars]$.
Moreover, it holds that 
$$
\sup_{\vv \in [\ubars,\bars],0\le r \le R, x \in [0,1]} |f_j(\vv + r e^{2\pi \i x})| \lesssim (\delta \rad)^j\ \ \text{for all }j \in \N.
$$



\end{enumerate} 
\end{lemA}







\begin{lemA}
\label{lem:poly-approx-of-analytic}

Let $\varepsilon > 0$ be a constant such that $\frac{1}{1 + \frac{2R}{b-a} + \frac{2}{b-a}\sqrt{R^2 + (b-a)R}} <\varepsilon< 1$.
Then, there exist some constants $c_1, c_2 > 0$ depending only on $\varepsilon$ such that for all $j \in \Z_+$, it holds
\begin{equation*}
\inf_{p \in \mathcal{P}_k} \sup_{\vv \in [a,b]} \frac{1}{j!}| \dif_{\vv}^j [f(\vv) - p(\vv)]| \le c_1  c_2^{j} \varepsilon^{k} \|f\|_{\mathcal{F}_R([\ubars,\bars])}, \quad \forall f \in \mathcal{F}_R([\ubars,\bars]).
\end{equation*}
Here, $\CP_k$ denotes the linear span of all polynomials of degree less than $k \in \N$.
\end{lemA}

The next lemma provides the variance order in the polynomial sieve estimation of the $j$th derivative, when the point of estimation lies in the interior or on the boundary of the support.
\begin{lemA}
\label{lem:Hilbert-matrix}
Let $\uy < a <  b \le \oy$.
Let $z_k(y) = (1,y -a,\ldots, (y-a)^{k - 1})'$ and $S = [\uy,a]\cup[b,\oy]$. Define $Q_k := \int_S z_k(y)z_k(y)'dy$ and $H_k = \int_{\uy}^{\oy} z_k(y)z_k(y)'dy$.
Then the following are true.
\begin{enumerate}[leftmargin = 0.05\linewidth]

\item [(i)]
Let $\varepsilon \in (0,1)$.
Then, there exist positive constants $c_1$, $c_2$, and $c_3$ depending only on $S$ and $\varepsilon$ such that \begin{align*}
j^{-1}(c_1 k/j)^{2j-1} \lesssim &[H_k^{-1}]_{j,j} \lesssim j^{-1}(c_2 k/j)^{2j-1},\\
& [Q_k^{-1}]_{j,j} \lesssim j^{-1}(c_3 k/j)^{4 j-2},       
\end{align*}
uniformly for all $k \in \N$ and $j\in\N$ such that $j/k\le \varepsilon$.
There exists a uniform constant $C>0$ such that $\|H_k\| \le C^k$ for all $k \in \mathbb{N}$.


\item [(ii)] Let $\chi_k = \lambda_{\min}(H_k^{-1/2}Q_kH_k^{-1/2})$. Then, $\chi_k \gtrsim c^{k}$ holds for some constant $c \in (0,1)$ that depends only on $S$.
Here, $\lambda_{\min}(A)$ denote the smallest eigenvalue of a square matrix $A$.


\end{enumerate}
\end{lemA}
\begin{lemA}
\label{lem:variance-bound}
Let $a > 0$ and $b \ge 0$.
Then, it holds
\begin{equation*}
    \sum_{j=1}^{l} j^{-aj+b}k^{aj-b-1} \le l^{-al+b-1}k^{al-b} \frac{1}{1-\left( \frac{el}{k}\right)^a}
\end{equation*}
for any $l \in \mathbb{N}$ such that $el < k$.
\end{lemA}
\begin{lemA}[Non-asymptotic bound on sum of i.i.d. random matrices]
\label{lem:Matrix-moment}
Let $p \in [1, \infty)$.
Let $(B_i)_{i=1}^n$ be a sequence of i.i.d. $d_1$ by $d_2$ random matrices such that $\E[B_i] = 0$ and $\E[\|B_i\|^p] < \infty$. Then, for some absolute constant $C_p>0$, it holds that
\begin{equation*}
\E\left[\left\|\frac 1 {\sqrt n}\sum_{i=1}^n B_i\right\|^p\right]^{1/p} \le C_p \E[\sigma^ p]^{1/p} \sqrt{\log(d_1 + d_2)},
\end{equation*}
where $\sigma = \max\{\|\frac 1 n \sum_{i=1}^n B_i B_i'\|^{1/2}, \|\frac 1 n \sum_{i=1}^n B_i'B_i\|^{1/2}\}$.
\end{lemA}


\subsection{Intermediary Lemmas}

\begin{lemA}
\label{lem:estimation-consistency}
Under the same set of assumptions as in Theorem~\ref{thm:asymptotic-normality}, as $n \to \infty$, it holds
\begin{align*}
\what\mu_n = \oP(1)\ \ \text{under}\ \ H_0.
\end{align*}
\end{lemA}

\begin{lemA}
\label{lem:average-consistency-fkj}
Under the same set of assumptions as in Theorem~\ref{thm:asymptotic-normality}, there exists an absolute constant $C > 0$ such that it holds
\begin{equation*}
\frac{1}{n}\sum_{i \in \nn} \sum_{j=1}^\tc |\what{f}_{\kap j i} - f_{\kap j i}|^2 \leP \frac{C^\tc \kap}{n},
\end{equation*}
where $f_{\kap j i}$ and $\what{f}_{\kap j i}$ are shorthands for $f_{\kap j}(\yi{0})$ and $\what{f}_{\kap j}(\yi{0})$, respectively, for units $i\in\nn$.
\end{lemA}

\subsection{Proofs of lemmas}

\subsubsection{Proof of Lemma~\ref{lemA:prob-integral-to-moment-series}}
Let $l \in \N$ be fixed.
Define an auxiliary function
\begin{equation*}
m(\vv) = - \sum_{j=1}^l \frac{1}{j!} \dif_{\wv}^{j-1} [ \E\left[  (\s-\vv)_+^j | \v = \wv\right] f(\wv) ]_{\wv = \vv}.
\end{equation*}
By the assumption, $m$ is Lipschitz continuous, and thus is absolutely continuous. By the fundamental theorem of calculus, we have that
\begin{align*}
    \int_{\ubars}^{\bars} m'(\vv)d\vv = m(\bars) - m(\ubars) &= \sum_{j=1}^l \frac{1}{j!} \dif_{\vv}^{j-1} [ \E\left[(\s-\ubars)_+^j | \v = \vv\right] f(\vv) ]_{\vv = \ubars} \\
&\ \ \ -\sum_{j=1}^l \frac{1}{j!} \dif_{\vv}^{j-1} [ \E\left[(\s-\bars)_+^j | \v = \vv\right] f(\vv) ]_{\vv = \bars}.
\end{align*}
To evaluate the left-hand side of the equation above, we note that by the chain rule of differentiation,
\begin{align*}
m'(\vv) &= -\sum_{j=1}^l \left[ -\frac{1}{(j-1)!} \dif_{\vv}^{j-1} [ \E\left[ (\s-\wv)_+^{j-1} | \v = \vv\right] f(\vv) ]_{\wv = \vv} \right.\\
& \quad \qquad \qquad \left. + \frac{1}{j!}\dif_{\vv}^{j} [ \E\left[  (\s- \wv)_+^{j} | \v = \vv\right] f(\vv) ]_{\wv = \vv}\right] \\
&= \E\left[  \I{\s \ge \v} | \v = \vv\right] f(\vv) - \frac{1}{l!}\dif_{\vv}^{l} [ \E\left[ (\s-\wv)_+^{l} | \v = \vv\right] f(\vv) ]_{\wv = \vv}.
\end{align*}
This gives
\begin{equation*}
 \int_{\ubars}^{\bars} m'(\vv)d\vv = \int_{\ubars}^{\bars}  \E\left[  \I{\s \ge \v} | \v = \vv\right] f(\vv) d\vv - \frac{1}{l!} \int_{\ubars}^{\bars} \dif_{\wv}^{l} [ \E\left[ (\s-\vv)_+^{l} | \v = \wv\right] f(\wv) ]_{\wv = \vv}d\vv    ,  
\end{equation*}
which completes the proof of Lemma~\ref{lemA:prob-integral-to-moment-series}.


\subsubsection{Proof of Lemma~\ref{lemA:sufficient-analytic}}
In this proof, the partial differential operators, $\dif_\sv$ and $\dif_\vv$, are understood in the weak sense.
Then, mixed partials are independent of the order of differentiation up to a set of measure zero.
Differentiation under the integral sign is also justified almost everywhere under some mild conditions that are met in our case.

To prove Lemma~\ref{lemA:sufficient-analytic}(i), let $q(\sv,\vv) := \dif_{\vv}^{j-1} \E[( \s - \sv)_+^j | \v = \vv]$.
Note that, by differentiating under the integral sign,
$$
q(\sv,\vv) =  \int \dif_\vv^{j-1}[ (\appftnb{}(\vv,\tau) - \sv)^j ] \I{\appftnb{}(\vv,\tau) - \sv > 0} d\mu(\tau)
$$
for \textit{all} $(\sv,\vv) \in [\ubars, \bars] \times [\ubars, \bars]$.
Thus, for almost every $(\sv,\vv) \in [\ubars, \bars] \times [\ubars,b]$, we have weak derivatives
\begin{align*}
&\dif_{\vv} q(\sv,\vv) =  \int \dif_\vv^{j}[ (\appftnb{}(\vv,\tau) - \sv)^j ] \I{\appftnb{}(\vv,\tau) - \sv > 0} d\mu(\tau), \\
&\dif_{\sv} q(\sv,\vv) =  -j \int \dif_\vv^{j-1}[ (\appftnb{}(\vv,\tau) - \sv)^{j-1} ] \I{\appftnb{}(\vv,\tau) - \sv > 0} d\mu(\tau).
\end{align*}
On the right-hand side, the integrals are a.e. bounded in $[\ubars, \bars] \times [\ubars, \bars]$, which implies that $q$ has a Lipschitz continuous version.
Since $q$ is continuous itself, it must agree with that version.
Thus, $q$ is Lipschitz continous, completing the proof of Lemma~\ref{lemA:sufficient-analytic}(i).

To prove Lemma~\ref{lemA:sufficient-analytic}(ii),
we first define $\tilde{\appftnb{}}(\vv, \tau) = \appftnb{}(\vv, \tau) - \sv$.
By differentiating under the integral sign, we observe that 
\begin{align*} 
    \left|\dif_\vv^l \{  \E[ (\s-\sv)_+^m | \v = \vv]f(\vv) \}\right| &= | \E[ \dif_\vv^l  \{\tilde{\appftnb{}}(\vv,\tau)^m  \I{\tilde{\appftnb{}}(\vv, \tau) > 0}f(\vv) \}] | \\
& \le  \E[|\dif_{\vv}^l \{ \tilde{\appftnb{}}(\vv,\tau)^m f(\vv) \}| \I{\tilde{\appftnb{}}(\vv, \tau) > 0}].
\end{align*}
Let us first focus on the term $|\dif_{\vv}^l \{ \tilde{\appftnb{}}(\vv,\tau)^m f(\vv) \}|$ inside the expectation.
For each $\tau \in (0,1)$, the function $z \mapsto \tilde{\appftnb{}}(z,\tau)^m f(z)$ can be extended to a complex-analytic function on the $\rad$-neighborhood of $[\ubars,\bars]$ by the analyticity condition.\footnote{The $\rad$-neighborhood of $[\ubars,\bars]$ is defined as $\{ z\in \C : \op{dist}(z, [\ubars,\bars]) \le \rad\}$ where $\op{dist}(z, [\ubars,\bars]) = \inf_{t \in [\ubars,\bars]} |z-t|$ for $z \in \mathbb{C}$.}
With slight abuse of notation, let us denote its analytic continuation again by $\tilde{\appftnb{}}(z,\tau)^m f(z)$.
It follows by Cauchy's integral theorem (\citeay{steinComplexAnalysis2010}) that
\begin{align} 
\frac{1}{l!} \dif_{\vv}^l [\tilde{\appftnb{}}(\vv,\tau)^m f(\vv)] &= \frac{
1}{2\pi i} \int_{|z - \vv| = R} \frac{\tilde{\appftnb{}}(z,\tau)^m f(z)}{(z - \vv)^{l+1}}  dz \nonumber \\
&= R^{-l}\int_0^{1} \tilde{\appftnb{}}(v + R e^{2\pi \i x},\tau)^m f(v + R e^{2\pi \i x}) e^{-2\pi \i l x} dx\\
\label{eqA:remainder-bound}&= R^{-l}\int_0^{1} \re{\left\{ \tilde{\appftnb{}}(v + R e^{2\pi \i x},\tau)^m f(v + R e^{2\pi \i x}) e^{-2\pi \i l x}\right\}} dx.
\end{align}
for any $0 < R \le \rad$, where the substitution $z = v + R e^{2\pi \i x}$, $x \in [0,1]$, was made.
This proves the first inequality in Lemma~\ref{lemA:sufficient-analytic}(ii) by the triangle inequality.

By the triangle inequality, \eqref{eqA:remainder-bound} implies
\begin{align*}
\frac{1}{l!} |\dif_{\vv}^l [\tilde{\appftnb{}}(\vv,\tau)^m f(\vv)]| 
\le R^{-l} \sup_{x \in [0,1]} |\appftnb{}(\vv + Re^{2\pi \i x},\tau) - \sv|^m \int_0^{1} |f(v + R e^{2\pi \i x})| dx.        
\end{align*}
Multiplying by $\I{\appftnb{}(\vv,\tau)-\sv> 0}$ and integrating with respect to $\tau \sim \mu$, we obtain the last inequality.

The second line in Lemma~\ref{lemA:sufficient-analytic}(iii) is a direct consequence of Lemma~\ref{lemA:sufficient-analytic}(ii) by substituting $m = l$, $R = \rad$, and $\sv = \yv$ and then integrating with respect to $\yv$.
The last line follows from the assumed analyticity condition.

To prove Lemma~\ref{lemA:sufficient-analytic}(iv), without loss of generality, assume $[\ubars,\bars] =[-1,1]$.
It suffices to upper bound the power series $\sum_l \frac{1}{l!}|\dif_\vv^l f_j(\vv)|R^l$ uniformly in $\vv \in [-1,1]$ for any $R < \rad$.
Applying Lemma~\ref{lemA:sufficient-analytic}(ii) with $m = j$, $s = \ubars$, and $R = \rad$, we obtain
\begin{align*}
   \frac{1}{l!}|\dif_\vv^l  f_j(\vv)| &\le \rad^{-l} \E\left[ \left(\sup_{x \in [0,1]} \left| {\appftnb{}}(\vv, \tau) -a +\sum_{n = 1}^\infty \frac{\dif_\vv^n {\appftnb{}}(\vv, \tau)}{n!} \rad^n e^{2\pi \i n x} \right|\right)^j \I{\appftnb{}(\vv,\tau) - a > 0}\right]\\
   &\qquad\qquad \times \int_0^{1} \left|f(\vv + \rad e^{2\pi \i x})\right|dx,
\end{align*}
where the expectation is computed with respect to $\tau \sim \mu$.
It follows that for all $\vv \in [a,b]$,
\begin{align*}
\sum_{l=0}^\infty \frac{1}{l!}|\dif_\vv^l  f_j(\vv)| R^l 
& \le \frac{1}{1-R/\rad}\E\left[ \left(\sup_{x \in [0,1]} \left| {\appftnb{}}(\vv, \tau) -a +\sum_{n = 1}^\infty \frac{\dif_\vv^n {\appftnb{}}(\vv, \tau)}{n!} \rad^n e^{2\pi \i n x} \right|\right)^j\right]\\
   &\qquad\qquad \times \int_0^{1} \left|\frac{f(\vv + \rad e^{2\pi \i x})}{f(\vv)}\right|dx\cdot f(\vv)\\
&\le \frac{1}{1-R/\rad} \beta C^j f(\vv)\\
&\lesssim C^j,
\end{align*}
where $C$ is a constant such that $\sup_{x \in [0,1]} \left| {\appftnb{}}(\vv, \tau) -a +\sum_{n = 1}^\infty \frac{\dif_\vv^n {\appftnb{}}(\vv, \tau)}{n!} \rad^n e^{2\pi \i n x} \right| \le C$ for all $\tau \in (0,1)$ and $\vv \in [\ubars,\bars]$.


\subsubsection{Proof of Lemma~\ref{lem:poly-approx-of-analytic}}
Without loss of generality, we may assume $\|f\|_{\mathcal{F}_R([\ubars,\bars])} < \infty$.
We first observe the following fact.
If $f \in \mathcal{F}_R([\ubars,\bars])$, then $f \in \CA_r([a,b])$ where $r := \frac{b-a}{2}+R+\sqrt{R^2 + (b-a)R}$, as $C_{r,\ubars,\bars}$ is contained in the $R$-neighborhood of $[\ubars,\bars]$.
Using the results in \citep{timanTheoryApproximationFunctions1994}{pp.~130-132,~280-281}, it can be demonstrated that
\begin{equation}
\label{eq:pf-lemma-A3}
\inf_{p \in \mathcal{P}_k} \| f-p\|_{\mathcal{F}_{R'}([\ubars,\bars])} \le c_1 \varepsilon^k \|f\|_{\CA_r([\ubars,\bars])}
\end{equation}
where $R'>0$ is any constant such that $\varepsilon = \frac{b-a}{2r}(1+2R'/(b-a)) < 1$ and $c_1$ is a constant depending only on $\varepsilon$.
Since $\inf_{p \in \mathcal{P}_k} \sup_{\vv \in [\ubars,\bars]} \| f(\vv) - p(\vv)\| \le \inf_{p \in \mathcal{P}_k} \| f - p\|_{\mathcal{F}_{R'}([\ubars,\bars])}$ and $\|f\|_{\CA_r([\ubars,\bars])} \le \|f\|_{\mathcal{F}_R([\ubars,\bars])}$, the given assertion is true for $j = 0$.

To address the case where $j\ge 1$, we note that Cauchy's integral theorem implies $\frac{1}{j!}\|\dif^j f\|_{\CF_{R_2}([a,b])} \le (R_1-R_2)^{-j} \|f\|_{\CF_{R_1}([a,b])}$ for all $j \ge 1$ and $0 \le R_2 < R_1$.
Thus, by the inequality \eqref{eq:pf-lemma-A3}, letting $R_1 = R'$ and $R_2 = 0$, we have
\begin{align*}
\frac{1}{j!}\inf_{p \in \mathcal{P}_k} \sup_{y \in [a,b]} |\dif_y^j[f(y) - p(y)]| &= \inf_{p \in \mathcal{P}_k} \frac{1}{j!}\|\dif_\yv^j [f-p]\|_{\CF_{0}([a,b])}\\
&\le (R')^{-j} \inf_{p \in \mathcal{P}_k}\|f-p\|_{\mathcal{F}_{R'}([\ubars,\bars])}\\
& \le c_1 (R')^{-j} \varepsilon^k \|f\|_{\CA_r([\ubars,\bars])}\\
& \le c_1 (R')^{-j} \varepsilon^k \|f\|_{\mathcal{F}_R([\ubars,\bars])}.
\end{align*}
This completes the proof of Lemma~\ref{lem:poly-approx-of-analytic}.


\subsubsection{Proof of Lemma~\ref{lem:Hilbert-matrix}}
We begin with some preliminary results.
\paragraph{Preliminary 1.} 
For $i,j=1,\ldots, k$, define $[\CH_k]_{ij} = \frac{1}{2}\int_{-1}^1 t^{i+j-2} dt = \frac{1+(-1)^{i+j}}{2(i+j-1)}$, to which we relate $H_k$.
Then, $\mathcal{H}_k^{-1}$ corresponds to the asymptotic variance of the OLS estimator arising from the following linear regression:
\begin{equation*}
    \y = \beta_0 + \beta_1 \x + \cdots + \beta_{k-1} \x^{k-1} + \epsilon,    
\end{equation*}
where the covariate $\x$ is uniformly distributed over $[-1,1]$ and $\epsilon$ is independent of $\x$ with unit variance.

Let $P_m(x) = \sum_{t = 0}^\infty p_{mt} x^t$ ($m \in \Z_+$) denote the $m$th normalized Legendre polynomial on $[-1,1]$ such that $\frac{1}{2}\int_{-1}^1 P_m(t)P_{m'}(t)dt = \I{m = m'}$.
The closed-form expression for $P_{m}$ can be derived from the well-known Rodrigues' formula, which is given by
\begin{equation*}
    P_m(x) = \frac{1}{2^m} \sqrt{\frac{2m+1}{2}} \sum_{0\le s \le m/2}  \frac{(-1)^s (2m -2s)!}{s!(m-s)!(m-2s)!}x^{m-2s}.
\end{equation*}
It follows that $p_{mt} = \frac{1}{2^{t+2r}} ({2r+t+\frac{1}{2}})^{1/2} \frac{(-1)^{r} (2t+2r)!}{t!r!(t+r)!}$ for $m = t + 2r$.
Since $\mathcal{H}_k^{-1} = P'P$, where $P = (p_{i-1,j-1})_{i,j=1}^k$, this implies that, for all $k > j \ge 0$,
\begin{equation*}
   [\CH_k^{-1}]_{j+1,j+1} = \sum_{0\le m \le k-1} p_{mj}^2 = \frac{1}{4^j (j!)^2}\sum_{0\le r \le (k-j-1)/2} \left( 2r+j+\frac{1}{2}\right)\frac{1}{16^r}\left( \frac{(2j+2r)!}{r!  (j+r)!}\right)^2.
\end{equation*}
Here, the substitution $r = (m-j)/2 \ge 0$ has been made.

Let $\varepsilon \in (0,1)$ be a constant.
We need to establish that the above expression is on the same order as $(j+1)^{-1}(\frac{k}{j+1})^{2j+1}$ uniformly in $k \ge 1$ and $0 \le j/k \le \varepsilon$.
We first note that
\begin{align*}
    \frac{(2j+2r)!}{r! (j+r)!} &= \frac{(2r)!}{(r!)^2} \frac{\prod_{m=1}^j (2r + 2m-1) \prod_{m=1}^j (2r + 2m)}{\prod_{m=1}^j (r+m)}\\
    &=\frac{(2r)!}{(r!)^2} 4^j \prod_{m=1}^j (r + m-1/2).
\end{align*}
Further, using $(2r)!/(r!)^2 = (-4)^r \binom{-1/2}{r}$ for all $r \ge 0$ and $\binom{-1/2}{r}^2 \asymp 1/(2r+ 1)$, we have
\begin{align*}
    [\CH_k^{-1}]_{j+1,j+1} & \asymp \frac{4^j}{(j!)^2} \sum_{0\le r \le (k-j-1)/2} \left( 2r + j + \frac{1}{2}\right)\left( r+\frac{1}{2}\right)\left( \prod_{m=1}^{j-1} (r+ m + 1/2)\right)^2.
\end{align*}
for all $k \ge 1$ and $j\ge 1$ such that $j/k \le \varepsilon$.
We will show that the last expression is comparable with
\begin{equation*}
    \frac{4^j}{j!(j+1)!}\left( \frac{k-j-1}{2}\right) \prod_{r=1}^j \left( \frac{k-j-1}{2} + r\right)^2.
\end{equation*}
Define $S_{n,j} = \prod_{m=1}^j (n+m+1/2)^2$ so that $S_{n,j}-S_{n-1,j} = \prod_{m=1}^{j-1} (n+m+1/2)^2 (2n+j+1) j$ for all $n \ge -1$ and $j \ge 1$.
Let $N = \lfloor k/2 - j/2 -1/2\rfloor$.
Then, for all $k\ge 1$ and $1\le j < k$,
\begin{align*}
    &   \sum_{0\le r \le (k-j-1)/2} \left( 2r + j + \frac{1}{2}\right)\left( r+\frac{1}{2}\right)\left( \prod_{m=1}^{j-1} (r+ m + 1/2)\right)^2  \\
    \le & \ j^{-1} \sum_{r=0}^N (S_{r,j} - S_{r-1, j}) (r + 1/2) \\
    =&\  j^{-1} \left( (N+1/2) S_{N,j} - \sum_{n=0}^{N-1} S_{n,j} - S_{-1,j}/2\right)\\
    \le &\ j^{-1} (N+1/2) S_{N,j} \\
    \le & \ j^{-1}\left( \frac{k-j}{2}\right) \prod_{m=1}^j \left( \frac{k-j}{2} + m\right)^2.
\end{align*}
The case where $j = 0$ can be separately addressed without difficulties.
This establishes the upper bound.

To establish the lower bound, note that $S_{n-1,j+1}-S_{n-2,j+1} = (n+1/2)^2\prod_{m=1}^{j-1} (n+m+1/2)^2 (2n+j) (j+1)$.
It follows that
\begin{align*}
    &   \sum_{0\le r \le (k-j-1)/2} \left( 2r + j + \frac{1}{2}\right)\left( r+\frac{1}{2}\right)\left( \prod_{m=1}^{j-1} (r+ m + 1/2)\right)^2  \\
    \ge & \ (j+1)^{-1} \sum_{n=0}^N (S_{n-1,j+1} - S_{n-2, j+1}) \frac{1}{n + 1/2} \\
    \ge &\ (j+1)^{-1} \frac{\left( \sum_{n=0}^N (S_{n-1,j+1} - S_{n-2, j+1})\right)^2}{\sum_{n=0}^N (S_{n-1,j+1} - S_{n-2, j+1}) (n + 1/2)} \\
    \ge &\ (j+1)^{-1} \frac{ (S_{N-1,j+1} - S_{-2, j+1})^2}{(N+1/2) S_{N-1,j+1}} \\
    \gtrsim &\ (j+1)^{-1}\frac{S_{N-1,j+1}}{(N+1/2)}\\
    = &\ (j+1)^{-1}\left( \frac{k-j}{2}\right) \prod_{r=1}^j \left( \frac{k-j}{2} + r\right)^2.
\end{align*}
Putting these inequalities together and using the Stirling's formula, we obtain
\begin{align*}
    [\CH_k^{-1}]_{j+1,j+1} &\asymp \frac{4^j}{j!(j+1)!}\left( \frac{k-j}{2}\right) \prod_{r=1}^j \left( \frac{k-j}{2} + r\right)^2 \\
    & \asymp \frac{(k+j)^{k+j}}{(k-j)^{k-j-1} (j+1)^{2j+1}} \\
    &    \asymp \frac{1}{j+1}\left(\frac{k}{j+1}\right)^{2j+1}
\end{align*}
uniformly in $k \ge 1$ and $j\ge 1$ such that $j/k \le \varepsilon$.

\paragraph{Preliminary 2.}
Let $\mathcal{A}_k$ denote the $k\times k$ unscaled Hilbert matrix, defined as $[\mathcal{A}_k]_{ij} = \int_0^1 t^{i+j-2}dt = \frac{1}{(i+j-1)}$, $i,j=1,\ldots, k$.
It is well known that
\begin{equation*}
    [\mathcal{A}_k^{-1}]_{j+1,j+1} = (2j+1){k+j \choose k-j-1}^2 {2j \choose j}^2.
\end{equation*}
(see \citeay{schechter1959inversion}, for instance.)
Using the Stirling's formula, we have
\begin{align*}
   [\mathcal{A}_k^{-1}]_{j+1,j+1} \asymp \frac{1}{j+1} \left( \frac{k}{j+1}\right)^{4j+2}.      
\end{align*}

\paragraph{Proof of (i):} 
Let $c_1 = \min(a-\uy, \oy-a)>0$ and $c_2 = \max(a-\uy, \oy-a)>0$.
Define diagonal matrices $D_1 = \diag(1,c_1,\ldots, c_1^{k-1})$ and $D_2 = \diag(1,c_2,\ldots, c_2^{k-1}).$
Then, by change of variables formula, we have $D_1 \mathcal{H}_k D_1 = \int_{a-c_1}^{a+c_1} z_k(\yv)z_k(\yv)'d\yv \le H_k \le D_2 \mathcal{H}_k D_2=\int_{a-c_2}^{a+c_2} z_k(\yv)z_k(\yv)'d\yv$, whence it follows
\begin{equation*}
    c_2^{-2j}  [\mathcal{H}_k^{-1}]_{j+1,j+1} \le  [H_k^{-1}]_{j+1,j+1} \le c_1^{-2j} [\mathcal{H}_k^{-1}]_{j+1,j+1}.
\end{equation*}
The fact that $\|H_k\| = O(\max(1,c_2)^{2k})$ follows from $\sup_{k \in \mathbb{N}}\|\mathcal{A}_k\| < \infty$ (see Part~(ii)), as it holds
\begin{equation*}
    2 \mathcal{H}_k = \mathcal{A}_k + \tilde{D} \mathcal{A}_k\tilde{D},
\end{equation*}
where $\tilde{D} = \diag(1,(-1),\ldots, (-1)^{k-1})$.

Lastly, we have $D_3 \mathcal{A}_k D_3 = \int_{\uy}^a z_k(\yv)z_k(\yv)'d\yv \le Q_k$ where $D_3 = \diag(1,(\uy-a),\ldots, (\uy-a)^{k-1})$.
Therefore,
\begin{equation*}
        [Q_k^{-1}]_{j+1,j+1} \le (a-\uy)^{-2j} [\mathcal{A}_k^{-1}]_{j+1,j+1}.
\end{equation*}
This completes the proof of (i).
\\

\paragraph{Proof of (ii):}
It is well-known that the largest and smallest eigenvalues of $\mathcal{A}_k$ satisfy (\citeay{schechter1959inversion})
\begin{equation*}
    \lambda_{\max}(\mathcal{A}_k) = O(1), \quad \lambda_{\min}(\mathcal{A}_k) \gtrsim c^k
\end{equation*}
for some $c \in (0,1)$.
Let $D_4 = \diag(1,\oy-a,\ldots, (\oy-a)^{k-1})$.
Since $Q_k \ge D_3\mathcal{A}_k D_3 \ge C_1^k I_k$ and $H_k = D_3 \mathcal{A}_k D_3 + D_4 \mathcal{A}_k D_4 \lesssim C_2^k I_k$ for some constants $C_1>0$ and $C_2>0$, it follows that $\chi_k \gtrsim (C_1/C_2)^k$. 
\subsubsection{Proof of Lemma~\ref{lem:variance-bound}}
Note that
$$
\sum_{j=1}^{l} j^{-aj+b}k^{aj-b-1} \le l^{b}k^{-b} \sum_{j=1}^{l} j^{-aj}k^{aj-1}.
$$
By substituting $j = l - m$,
\begin{equation*}
        \sum_{j=1}^{l} j^{-aj}k^{aj-1} = \sum_{m=0}^{l-1} (l-m)^{-a(l-m)} k^{a(l-m)-1} = l^{-al}k^{al-1}\sum_{m=0}^{l-1} l^{al}(l-m)^{-a(l-m)} k^{-am}.
\end{equation*}
Using the fact that $l^{al}(l-m)^{-a(l-m)} = l^{am} \left( \frac{l}{l-m}\right)^{a(l-m)}$ and $l/(l-m) = 1 + m/(l-m)\le e^{m/(l-m)}$, we have
\begin{equation*}
        \sum_{m=0}^{l-1} l^{al}(l-m)^{-a(l-m)} k^{-am} \le \sum_{m=0}^\infty l^{am} e^{am}k^{-am} = \frac{1}{1- \left( \frac{el}{k}\right)^a}.
\end{equation*}
This completes the proof.

\subsubsection{Proof of Lemma~\ref{lem:Matrix-moment}}
A proof can be found in \citet{troppUserFriendlyTailBounds2012}.




\subsubsection{Proof of Lemma~\ref{lem:estimation-consistency}}
\textit{Notational remarks.} Throughout this proof, $C$ denotes a generic absolute constant that may differ in each occurrence.
When we use the symbols $\oP(.)$ and $\OP(.)$, they are understood to be uniform in $\P \in \mathfrak{P}_0$.
All derived bounds are non-asymptotic with hidden factors depending on $\P$ only through $\con_{k}$'s appearing in Assumptions.
Since $\yi{0}$ is not available outside the estimation sample, $\sum_{i=1}^n$ will often be used interchangeably with $\sum_{i\in\nn}$ when the summand includes a function of $\yi{0}$, such as $z_{\kap i} = z_{\kap}(\yi{0})$.
The same applies to $\E[.]$ and $\E[.\I{i \in \nn}]$ when expectations are taken with respect to functions involving $\yi{0}$, where $\I{i \in \nn} = \I{\yi{0} \in \S}=\I{\yis{0} \in \S}$.
We also identify $\nn$ and $\S$ with their probability limits, as this does not undermine the validity of the asymptotic analysis.
\\

\noindent \textit{Proof.} Without loss of generality, assume $\mathcal{Y}_0 = [\uy,\oy] = [0,1]$ and denote the sup-norm as $\|f\|_\infty = \sup_{y \in \mathcal{Y}_0}|f(y)|$.
Let $j \in [1, \tc]$ and $(\rho,\beta,\delta)$ be the smoothness constants in Assumption~\ref{asm:regularity-for-estimation}.
Let $f_{\kap j}(\yv) = z_{\kap}(\yv)' \gamma_{\kap j}$ denote the order $\kap$ approximation of $f_j$ defined as the solution to the population problem \eqref{eq:population-sieve-problem}.

Pick $\frac{1}{1 + \rad + \sqrt{\rad^2 + 2\rad}} < \zeta <(1-\epsilon_0)(\liminf_{k\to\infty} \chi_k^{1/k})^{1/\con_{5}}$ for a sufficiently small $\epsilon_0 \in (0,1)$, which is possible by Assumption~\ref{asm:regularity-for-estimation}.
By Lemma~\ref{lemA:sufficient-analytic}(iv), we know that $\|f_j\|_{\CF_{\rad'}([0,1])} \lesssim (\delta \rad)^j$ where $\rad' \in (0, \rad)$.
Since we may take such $\rad'$ that $\frac{1}{1 + \rad + \sqrt{\rad^2 + 2\rad}} < \frac{1}{1+\rad' + \sqrt{(\rad')^2+2\rad'}} < \zeta$, by Lemma~\ref{lem:poly-approx-of-analytic}, there exists $\tilde{\gamma}_{\kap j} \in \R^{\kap}$ such that
\begin{equation*}
\|\dif_\yv^{j} [f_j(\yv) -  z_{\kap}(\yv)'\tilde{\gamma}_{\kap j} ]\|_\infty \lesssim j!C^{j}\zeta^{\kap}
\end{equation*}
uniformly in $j \in [1,\tc]$.\footnote{Here and throught the proof, the uniformity means that the implicit proportionality constant under $\lesssim$ is uniform and does not depend on $j$.}
Expand the Taylor series of $f_j(\yv)-  z_{\kap}(\yv)'\tilde{\gamma}_{\kap j}$ around $\yv = \uk$ up to degree $j-1$, which is denoted by $z_{\kap}(\yv)' \tau_{\kap j}$.
Let $\gamma^*_{\kap j} = \tilde{\gamma}_{\kap j} + \tau_{\kap j}$.
Since $\dif_{\yv}^m [f_j(\yv) - z_{\kap}(\yv)' \gamma^*_{\kap j}]_{\yv = \uk} = 0$ for all $m =0,\ldots,j-1$, it follows from the Taylor's theorem that
\begin{align*}
    \|f_j(\yv) - z_{\kap}(\yv)' \gamma^*_{\kap j}\|_\infty &\lesssim \frac{C^j}{j!} \|\dif_\yv^{j}[ f_j(\yv) - z_{\kap}(\yv)'\gamma^{*}_{\kap j} ]\|_\infty \\
    &=\frac{C^j}{j!}\|\dif_\yv^{j} [f_j(\yv) -  z_{\kap}(\yv)'\tilde{\gamma}_{\kap j} ]\|_\infty   \lesssim C^{j} \zeta^{\kap},
\end{align*}
Let $r^*_{\kap j} := f_j - z_{\kap}'\gamma^*_{\kap j}$.
This implies that $\|r^*_{\kap j}\|_\infty\lesssim C^{j} \zeta^{\kap}$ uniformly in $j \le \tc$.
Using the same method, we also have 
$$
\frac{1}{j!}|\dif_\yv^{j-1} [f_j(\yv)]_{\yv = \uk} - (j-1)!\gamma^*_{\kap j,j}| \le \frac{1}{j!}\|\dif_\yv^{j-1} r^*_{\kap j}\|_\infty \lesssim \frac{1}{j!}\|\dif_\yv[\dif_\yv^{j-1} r^*_{\kap j}]\|_\infty\lesssim C^{j} \zeta^{\kap},
$$
uniformly in $1\le j\le\tc$, making $\frac{1}{j}\gamma_{kj,j}^*$ a good approximation of $\frac{1}{j!}\dif_\yv^{j-1} [f_j(\yv)]_{\yv = \uk}$.

Next, we turn to bounding the difference $|\gamma^*_{\kap j,j} - \gamma_{\kap j,j}|$, where $\gamma_{\kap j,j}$ denotes the pseudo-true value associated with the order $\kap$ polynomial sieve space.
Let us use the shorthands $f_{ji} = f_j(\yi{0})$ and $f_{\kap j i}=f_{\kap j}(\yi{0})$.
By definition of $f_j(\yv)=\E[\ti w_i^j|\yis{0}=\yv] f_{\ys{0}}(\yv)$ and the first-order condition for $f_{\kap j}$, it is straightforward to see that
\begin{equation*}
    \E\left[\ti w_i^j \frac{z_{\kap i}}{f_{j i}}\right] = \E\left[ \ti w_i^j \frac{z_{\kap i}}{f_{\kap j i}}\right] = \int_{\S} z_{\kap}(\yv)d\yv.
\end{equation*}
Rearranging this equation, we have $\E\left[\ti w_i^j \frac{z_{\kap i}(f_{j i}-f_{\kap j i})}{f_{j i} f_{\kap j i}}\right] =0$, or equivalently,
\begin{align*}
    \gamma_{\kap j} - \gamma^*_{\kap j} &= \left(\E\left[ \ti w_i^j \frac{z_{\kap i}z_{\kap i}'}{f_{\kap j i}f_{j i}}\right] \right)^{-1}\E\left[ \ti w_i^j \frac{z_{\kap i}r^*_{\kap j i}}{f_{\kap j i}f_{j i}}\right]\\
    &= \left(\int_{\S} \frac{z_{\kap}(\yv)z_{\kap}(\yv)'}{f_{\kap j}(\yv)}d\yv \right)^{-1}\int_{\S} \frac{z_{\kap}(\yv)r^*_{\kap j}(\yv)}{f_{\kap j}(\yv)}d\yv.
\end{align*}
By Cauchy-Schwarz inequality,
\begin{align*}
    (\gamma_{\kap j} - \gamma^*_{\kap j})(\gamma_{\kap j} - \gamma^*_{\kap j})' &\lesssim \left(\int_{\S} \frac{z_{\kap}(\yv)z_{\kap}(\yv)'}{f_{\kap j}(\yv)}d\yv \right)^{-1}\int_{\S} \frac{z_{\kap}(\yv)z_{\kap}(\yv)'r^*_{\kap j}(\yv)^2}{f_{\kap j}(\yv)^2}d\yv\left(\int_{\S} \frac{z_{\kap}(\yv)z_{\kap}(\yv)'}{f_{\kap j}(\yv)}d\yv \right)^{-1} \\
    &\lesssim C^{2j} \zeta^{2\kap} \left(\int_{\S} z_{\kap}(\yv)z_{\kap}(\yv)'d\yv \right)^{-1}\\
    &=C^{2j} \zeta^{2\kap} Q_{\kap}^{-1},
\end{align*}
where $Q_{\kap} = \int_{\S} {z_{\kap}(\yv)z_{\kap}(\yv)'}d\yv$.
Now, we know that
\begin{align*}
\sum_{j=1}^\tc \left|\gamma_{\kap j,j}-\gamma^*_{\kap j,j}\right|^2 &\lesssim \zeta^{2\kap} \sum_{j=1}^\tc C^{2j} [Q_\kap^{-1}]_{jj} \lesssim \zeta^{2\kap} C^\tc \sum_{j=1}^\tc j^{-4j+1}\kap^{4j-2}\\
&\lesssim \zeta^{2\kap} C^\tc \tc^{-4\tc+1}\kap^{4\tc-2} 
\end{align*}
by Lemmas~\ref{lem:Hilbert-matrix} and \ref{lem:variance-bound}.
Combining the above results, we have
\begin{equation*}
\left|\sum_{j=1}^\tc \frac{1}{j!}\dif_\yv^{j-1} [ f_j(\yv) ]_{\yv=\uk} - \sum_{j=1}^\tc \frac{1}{j}\gamma_{\kap j,j}\right| \lesssim C^\tc \zeta^{\kap} \tc^{-2\tc+1/2}\kap^{2\tc-1} = o(n^{-1/2}).
\end{equation*}
This holds due to the choice of $\tc=O(\log n/\log\log n)$ and $\kap \lesssim \log n$, which ensures that $\tc^{-1/2}(\kap/\tc)^{2\tc-1} = o(n^{\epsilon})$ for any $\epsilon > 0$, and the fact that $\zeta^{\kap} = o((1-\epsilon_0)^\kap \chi_\kap^{1/\con_{5}}) = o(n^{-1/2-\epsilon_0'})$, which holds for some $\epsilon_0' > 0$ by the choice of $\zeta$.
This addresses the bias term from the sieve approximation to be $o(n^{-1/2})$.

Let $B_n = \sum_{j= 1}^\infty \frac{1}{j!}\dif_\yv^{j-1}[f_j(\yv)]_{\yv = \uk} - \sum_{j=1}^{\tc} \frac{1}{j}\gamma_{\kap j,j}
$
denote the overall bias.
Then, we have that
\begin{align}
    \label{eq:bias-term}
|B_n| & \lesssim \left| \sum_{j= \tc + 1}^\infty \frac{1}{j!}\dif_\yv^{j-1}[f_j(\yv)]_{\yv = \uk} \right| + \left|\sum_{j=1}^\tc \frac{1}{j!}\dif_\yv^{j-1} [ f_j(\yv) ]_{\yv=\uk} - \sum_{j=1}^\tc \frac{1}{j}\gamma_{\kap j,j}\right|\\
& \lesssim \delta^{\tc} + o(n^{-1/2}) \to 0. \nonumber
\end{align}

Next, we turn to bounding the stochastic term $\sum_{j=1}^{\tc} \frac{1}{j}(\what \gamma_{\kap j,j}-\gamma_{\kap j,j})$.
To this end, we define the orthogonal polynomial basis as $o_{\kap}(\yv) = H_{\kap}^{-1/2} z_{\kap}(\yv)$ where $H_{\kap} = \int_{0}^{1} z_{\kap}(\yv)z_{\kap}(\yv)'d\yv$. 
Let $\co_{\kap j}$ denote the coefficient associated with $o_{\kap}(\yv)$ such that $o_{\kap}(\yv)'\co_{\kap j} = z_{\kap}'(\yv)\gamma_{\kap j}$.
Namely, we have $\co_{\kap j} = H_{\kap}^{1/2} \gamma_{\kap j}$.
We can write $\gamma_{\kap j,j} = e_{j}'H_{\kap}^{-1/2} \co_{\kap j}$, where $e_{j}$ is a $\kap\times 1$ vector of 0's and only 1 at the $j$-th position for $j=1,\ldots,\kap$.
We denote $c_{\kap j} = H_{\kap}^{-1/2} e_{j}$. 

Using the first-order condition for $\what\co_{\kap j} = H_{\kap}^{1/2} \what \gamma_{\kap j}$ and the fact that $ \E\left[ \ti w_i^j\frac{o_{\kap i}}{f_{\kap j i}}\right]=\int_{\S} o_\kap(\yv)d \yv$, we know that
\begin{equation*}
    \frac{1}{n} \sum_{i=1}^n \ti  w_i^j \left( \frac{o_{\kap i}}{o_{\kap i}'\what\co_{\kap j}}  -\frac{o_{\kap i}}{o_{\kap i}'\co_{\kap j}}  \right) = - \frac{1}{n} \sum_{i=1}^n \left\{ \ti  w_i^j\frac{o_{\kap i}}{f_{\kap j i}} - \E\left[ \ti w_i^j\frac{o_{\kap i}}{f_{\kap j i}}\right]\right\},
\end{equation*}
thereby after solving for $\what \co_{\kap j}- \co_{\kap j}$, we have
\begin{align}
\label{eq:noname2}
    \sqrt{n}( \what\co_{\kap j}-\co_{\kap j}) &= \Bigg( \underbrace{\frac{1}{n} \sum_{i=1}^n \ti  w_i^j \frac{o_{\kap i}o_{\kap i}'}{\what f_{\kap j i}f_{\kap j i}}}_{\widetilde{I}_{1jn}}\Bigg)^{-1}\ub{\frac{1}{\sqrt n} \sum_{i=1}^n \left\{ \ti  w_i^j\frac{o_{\kap i}}{f_{\kap j i}} - \E\left[ \ti w_i^j\frac{o_{\kap i}}{f_{\kap j i}}\right]\right\}}_{\widetilde{I}_{2jn}} \nonumber \\
    &=: \widetilde{I}_{1jn}^{-1} \cdot \widetilde{I}_{2jn}.    
\end{align}
To bound $\sqrt n \sum_{j=1}^{\tc} \frac{1}{j}c_{\kap j}' (\what \co_{\kap j}-\co_{\kap j}) = \sqrt{n} \sum_{j=1}^{\tc} \frac{1}{j}(\what \gamma_{\kap j,j}-\gamma_{\kap j,j})$, we focus on bounding the following:
\begin{align}
\label{eq:noname3}
\left( \sqrt n \sum_{j=1}^{\tc} \frac{1}{j} \left|c_{\kap j}' (\what \co_{\kap j}-\co_{\kap j}) \right|\right)^2 &\lesssim n \left( \sum_{j=1}^{\tc} \frac{1}{j^2}\right)\sum_{j=1}^{\tc} \left| c_{\kap j}' (\what \co_{\kap j}-\co_{\kap j}) \right|^2 \lesssim n \sum_{j=1}^{\tc} \left| c_{\kap j}' (\what \co_{\kap j}-\co_{\kap j}) \right|^2 \nonumber \\
&\lesssim \sum_{j=1}^{\tc}   c_{\kap j}' \widetilde{I}_{1jn}^{-1} \widetilde{I}_{2jn}\widetilde{I}_{2jn}' \widetilde{I}_{1jn}^{-1} c_{\kap j}.
\end{align}
Since $\max_{i\in\nn}(f_{\kap j i}\what  f_{\kap j i}) \lesssim C^j$, we have
\begin{equation*}
\widetilde{I}_{1jn}^{-1} \lesssim C^j \left( \frac{1}{n} \sum_{i=1}^n \ti  w_i^j o_{\kap i}o_{\kap i}'\right)^{-1}    
\end{equation*}
uniformly for $1 \le j \le \tc$.
Using the fact that $\max_{i\in\nn}\|o_{\kap i}\| \lesssim \kap$ (\citeauthor{neweyConvergenceRatesAsymptotic1997}, \citeyear{neweyConvergenceRatesAsymptotic1997}), it follows from Lemma~\ref{lem:Matrix-moment} that 
$$
\E\left[ \left\|\frac{1}{n} \sum_{i=1}^n \ti  w_i^j o_{\kap i}o_{\kap i}'-\E\left[ \ti  w_i^j o_{\kap i}o_{\kap i}'\right]\right\|\right] \lesssim n^{-1/2} C^j \kap \sqrt{\log \kap},
$$
uniformly in $1\le j\le \tc$, where $\|.\|$ is the spectral norm of matrices.
Let $M_{\kap} := \int_{\S} o_{\kap}(\yv)o_{\kap}(\yv)' d\yv = H_\kap^{-1/2}Q_\kap H_\kap^{-1/2}$.
The smallest eigenvalue of the matrix
$$
\E\left[ \ti  w_i^j o_{\kap i}o_{\kap i}'\right] \ge \inf_{\yv \in \S} f_j(\yv) \cdot \int_{\S} o_{\kap}(\yv)o_{\kap}(\yv)' d\yv \gtrsim c^{j} M_{\kap}
$$
is bounded away from $0$ by a multiple of $c^{j} \chi_{\kap}$ uniformly in $1\le j\le \tc$, which follows from Jensen's inequality:
$$
f_j(\yv) = \E[\ti w_i^j | \yis{0} = \yv] f_{\ys{0}}(\yv) \ge \left(\frac{ \E[\ti w_i | \yis{0} = \yv]}{\E[\ti|\yis{0} = \yv]} \right)^j \E[\ti|\yis{0} = \yv]f_{\ys{0}}(\yv) \gtrsim c^j,
$$
where $c>0$ is a constant depending on $\con_{k}$'s.
Since it holds by Assumption~\ref{asm:selection-tuning} that
$$
\sum_{j=1}^\tc c^{-j} \E\left\|\frac{1}{n} \sum_{i=1}^n \ti  w_i^j o_{\kap i}o_{\kap i}'-\E\left[ \ti  w_i^j o_{\kap i}o_{\kap i}'\right]\right\| \lesssim n^{-1/2} C^\tc \kap \sqrt{\log \kap} =o(\chi_{\kap}),
$$
it follows that for a sufficiently small $\epsilon > 0$,
$$
\left\|\frac{1}{n} \sum_{i=1}^n \ti  w_i^j o_{\kap i}o_{\kap i}' - \E\left[ \ti  w_i^j o_{\kap i}o_{\kap i}'\right]\right\| \le \epsilon c^j \chi_\kap \le \epsilon \lambda_{\min}(\E\left[ \ti  w_i^j o_{\kap i}o_{\kap i}'\right])
$$
uniformly in $1 \le j \le \tc$ with probability approaching $1$.
Therefore,
\begin{equation*}
    \left( \frac{1}{n} \sum_{i=1}^n \ti  w_i^j o_{\kap i}o_{\kap i}'\right)^{-1} \lesssim \E\left[ \ti  w_i^j o_{\kap i}o_{\kap i}'\right]^{-1}    
\end{equation*}
uniformly in $1\le j\le \tc$ with probability approaching $1$.
This implies
\begin{equation*}
\widetilde{I}_{1jn}^{-1} \lesssim C^j \left( \frac{1}{n} \sum_{i=1}^n \ti  w_i^j o_{\kap i}o_{\kap i}'\right)^{-1} \lesssim C^j \E\left[ \ti  w_i^j o_{\kap i}o_{\kap i}'\right]^{-1} \lesssim C^j \left( \int_{\S} o_{\kap}(\yv)o_{\kap}(\yv)' d\yv\right)^{-1}
\end{equation*}
uniformly for $1\le j\le \tc$.
This leads to
\begin{equation*}
    \sum_{j=1}^{\tc}   c_{\kap j}' \widetilde{I}_{1jn}^{-1} \widetilde{I}_{2jn}\widetilde{I}_{2jn}' \widetilde{I}_{1jn}^{-1} c_{\kap j} 
    \leP C^{\tc}\sum_{j=1}^{\tc} c_{\kap j}'M_{\kap}^{-1} \widetilde{I}_{2jn}\widetilde{I}_{2jn}'M_{\kap}^{-1} c_{\kap j},
\end{equation*}
which allows us to focus on bounding the right-hand side term.
Since $\E[\widetilde{I}_{2jn}\widetilde{I}_{2jn}']$ corresponds to the variance of $\ti w_i^j o_{\kap i}/f_{\kap j i}$, we have uniformly for all $1\le j\le \tc$,
\begin{align*}
\E[\widetilde{I}_{2jn}\widetilde{I}_{2jn}'] \le \E\left[ \ti ^2 w_i^{2j} \frac{o_{\kap i}o_{\kap i}'}{f_{\kap j i}^2}\right] \lesssim C^j M_{\kap}.
\end{align*}
This implies that
\begin{align}
\label{eq:noname4}
\E\left[ C^{\tc}\sum_{j=1}^{\tc}   c_{\kap j}'M_{\kap}^{-1} \widetilde{I}_{2jn}\widetilde{I}_{2jn}'M_{\kap}^{-1} c_{\kap j}\right] &\lesssim C^{\tc} \sum_{j=1}^{\tc} c_{\kap j}'M_{\kap}^{-1}  c_{\kap j} \nonumber \\
&= C^{\tc} \sum_{j=1}^{\tc} e_{j}'Q_{\kap}^{-1}  e_{j} \nonumber \\
&\lesssim C^{\tc}  \tc^{-4\tc+1} \kap^{4\tc-2}
\end{align}
by Lemmas~\ref{lem:Hilbert-matrix} and \ref{lem:variance-bound}.
We conclude that
\begin{equation*}
\sum_{j=1}^{\tc} \frac{1}{j}  \left| c_{\kap j}' (\what \co_{\kap j}-\co_{\kap j}) \right| \leP n^{-1/2}C^{\tc} \tc^{-2\tc+1/2} \kap^{2\tc-1} = \oP(1).
\end{equation*}



\subsubsection{Proof of Lemma~\ref{lem:average-consistency-fkj}}
We adopt the same notation as in the proof of Lemma~\ref{lem:estimation-consistency}.
We start with the observation that
\begin{align*}
\sum_{i \in \nn} \sum_{j=1}^\tc |\what{f}_{\kap j i} - f_{\kap j i}|^2 &= \frac{1}{n} \sum_{i \in \nn} \sum_{j=1}^{\tc}   o_{\kap i}' \widetilde{I}_{1jn}^{-1} \widetilde{I}_{2jn}\widetilde{I}_{2jn}' \widetilde{I}_{1jn}^{-1} o_{\kap i} \\
&= \opw{tr} \left(  \sum_{j=1}^{\tc} \widetilde{I}_{1jn}^{-1} \widetilde{I}_{2jn}\widetilde{I}_{2jn}' \widetilde{I}_{1jn}^{-1} \left( \frac{1}{n} \sum_{i \in \nn} o_{\kap i}o_{\kap i}'\right)\right).
\end{align*}
From the proof of Lemma~\ref{lem:estimation-consistency}, it follows that $\widetilde{I}_{1jn}^{-1} \lesssim C^\tc M_{\kap}^{-1}$ uniformly in $1\le j\le \tc$ as $n \to \infty$ with probability approaching $1$.
Using the same method as in that proof, we can also show that $\frac{1}{n} \sum_{i\in \nn} o_{\kap i}o_{\kap i}' \lesssim M_{\kap}$.
Therefore, we obtain
\begin{align*}
\sum_{i \in \nn} \sum_{j=1}^\tc |\what{f}_{\kap j i} - f_{\kap j i}|^2 &\leP C^\tc \opw{tr} \left(  \sum_{j=1}^{\tc} M_{\kap}^{-1} \widetilde{I}_{2jn}\widetilde{I}_{2jn}' \right)\\
    &\leP C^\tc \op{tr} (I_{\kap}) \lesssim C^\tc \kap.
\end{align*}


\subsection{Proofs of the main results}
\label{proof:prop-bunching-char}
\subsubsection{Proof of Proposition~\ref{prop:bunching-characterization}}
By Assumption~\ref{asm:kinked-payoff}, we have $U(\yv,\xv,\eta) = U_0(\yv,\xv,\eta) \I{\yv \le \k} + U_1(\yv,\xv,\eta)\I{\yv > \k} = \min(U_0(\yv,\xv,\eta),U_1(\yv,\xv,\eta))$.
The compound payoff function is again a strictly concave function of $\yv$.
Let $\dif_\yv f(\yv)$ denote the superdifferential of a concave function $f$ at $\yv$.

\paragraph{Case 1}: If $\yiis < \k$ occurs in the interior of $\CY$, then, by the FOC, it holds
$$
0 \in \dif_{\yv} U(\yv, \x_i,\eta_i)|_{\yv = \yiis}= \dif_{\yv} U_0(\yv, \x_i,\eta_i)|_{\yv = \yiis}.
$$
This implies $\yiis$ is the solution to $\max_{\yv \in \mathcal{Y}} U_0(\yv,\x_i,\eta_i)$ as well.

If $\yiis$ occurs at the lower limit of $\CY$, by the Kuhn-Tucher condition, it holds 
$$
\dif_{\yv} U(\yv, \x_i,\eta_i)|_{\yv = \yiis}= \dif_{\yv} U_0(\yv, \x_i,\eta_i)|_{\yv = \yiis} \subseteq (-\infty, 0],
$$
which implies $\yiis$ also solves $\max_{\yv \in \mathcal{Y}} U_0(\yv,\x_i,\eta_i)$.
Therefore, $\yiis = \yis{0} < \k$.

\paragraph{Case 2}: If $\yiis > \k$, then we must have $\yiis = \yis{1} > \k$ by the same arguments as in Case~1.

\paragraph{Case 3}: If $\yiis = \k$, it follows that
$$
0 \in \dif_{\yv} U(\yv, \x_i,\eta_i)|_{\yv = \k} = [\dif_{\yv} U_1(\yv, \x_i,\eta_i)|_{\yv = \k}, \dif_{\yv} U_0(\yv, \x_i,\eta_i)|_{\yv = \k}].
$$
This implies that $\yis{1} \le \k \le \yis{0}$ by monotonicity of superdifferentials.
Combining Cases~1 through 3, this completes the proof of Proposition~\ref{prop:bunching-characterization}.

\subsubsection{Proof of analyticity of $F_{\eta}(h) = \E_X\left[\Phi( \frac{h - X'\beta_0}{\sigma_0})\right]$ and $f_{\eta}(h) = \frac{1}{\sigma_0}\E_X\left[\phi( \frac{h - X'\beta_0}{\sigma_0})\right]$.}
Without loss of generality, assume that $\sigma_0 = 1$ and define $W = X'\beta_0$.
Define $G$ as the CDF of $W$.
Then, it holds
\begin{equation*}
    F_\eta(z) = \int_{\mathbb{R}} \Phi(z - w) d G(w),\ \ z \in \mathbb{R}.
\end{equation*}
We will show that $F_\eta$ can be extended to the complex plane $z \in \C$.
We first note that $z \mapsto \Phi(z) = \int_{-\infty}^z \frac{1}{\sqrt{2\pi}} e^{-\frac{t^2}{2}} dt$ can be analytically extended to an entire function on $\C$ (\citeay{steinComplexAnalysis2010}).
Now, it holds that for all $x, y \in \mathbb{R}$,
\begin{equation*}
    |\Phi(x+\i y)| = \left| \int_{-\infty}^x \frac{1}{\sqrt{2\pi}} e^{-\frac{(t+\i y)^2}{2}} dt  \right| \le e^{\frac{y^2}{2}} \Phi(x) \le C e^{\frac{y^2}{2}}\frac{e^{-\frac{x^2}{2}}}{|x| + 1},   
\end{equation*}
where we used the Mill's ratio.
This implies
\begin{align*}
&    \int_\mathbb{R} |\Phi(x-w + \i y)| dG(w)  \\
\le &\ C e^{\frac{y^2}{2}} \int_\mathbb{R} \frac{e^{-\frac{(x-w)^2}{2}}}{|x-w| + 1}dG(w) \\
\le &\ 2C (|x| + 1) e^{\frac{x^2 + y^2}{2}}\int_\mathbb{R} \frac{e^{-\frac{w^2}{4}}}{|w| + 1} dG(w) < \infty
\end{align*}
where we used $(|x-w|+1)^{-1} \le 2 \frac{|x|+1}{|w|+1}$ and $|xw| \le x^2 + \frac{w^2}{4}$.
Thus, $F_\eta(x + \i y) \in \C$ is finite for all $x,y \in \mathbb{R}$, implying that $F_\eta : \C \to \C$ is well-defined.

Now, we turn to proving that $F_\eta$ is an entire function.\footnote{A function $f:\mathbb{C} \to \mathbb{C}$ is said to be entire if $f$ is complex-analytic on $\mathbb{C}$.}
To this end, by Morera's theorem (\citeay{steinComplexAnalysis2010}), it suffices to show that $F_\eta$ is continuous and that
\begin{equation*}
    \oint_{\triangle} F_\eta(z) dz = 0    
\end{equation*}
for any triangle $\triangle \subseteq \C$.
The continuity follows from Lebesgue's dominated convergence theorem, coupled with the derived bound above.
Also, by Fubini's theorem, we have
\begin{align*}
     \oint_{\triangle} F_\eta(z) dz &= \oint_{\triangle} \left[ \int_\mathbb{R} \Phi(z-w) dG(w)\right] dz \\
     &= \int_\mathbb{R} \left[ \oint_{\triangle} \Phi(z-w) dz\right] dG(w) \\
     &= \int_\mathbb{R} 0 dG(w)\\
     &= 0.
\end{align*}
We conclude that $F_\eta$ is an entire function, and hence is analytic on $\mathbb{R}$.
By Cauchy's integral theorem and Fubini's theorem again,
\begin{equation*}
    F_\eta'(z) = \frac{1}{2\pi i}\int_{|\zeta-z| = c} \frac{F_\eta(\zeta)}{\zeta - z}d\zeta = \int_\mathbb{R} \frac{1}{2\pi i}\int_{|\zeta-z| = c} \frac{\Phi(\zeta-w)}{\zeta - z}d\zeta dG(w) =   \int_\mathbb{R} \phi(z-w) dG(w)
\end{equation*}
where $\phi(z) =\Phi'(z)= \frac{1}{\sqrt{2\pi}}e^{-\frac{z^2}{2}}$.
The analyticity of $f_{\eta} = F_\eta'$ follows immediately.


\subsubsection{Proof of Theorem~\ref{thm:point-identification}}

\paragraph{Proof of existence of a solution:}
Without loss of generality, we may assume $\E[\ti] = 1$ so that $\Q$, defined as $d\Q/d\P = \ti$, is a probability measure.
Using the quantile transform, we may write
\begin{align*}
    &\conv(\ok,\x_i,\theta) = \qb(\tau_i,\theta, \yis{0}),
\end{align*}
where $\tau_i$ is a uniform random variable on $(0,1)$ independent of $\yis{0}$ with respect to the probability measure $\Q$.

In light of Lemma~\ref{lemA:sufficient-analytic}(i), we can apply Lemma~\ref{lemA:prob-integral-to-moment-series} to $(\s, \v) = (\conv(\ok, \x_i, \theta), \yis{0})$ with $f = f^{(\Q)}_{\ys{0}}$ and $[\ubars, \bars] = [\uk, \ook(\theta)]$, where the reference measure is $\Q$.
This leads to
\begin{align*}
    & \E_\Q[\I{\uk \le \yis{0} \le \conv(\ok, \x_i, \theta)}] \\
    \equiv &\ \int_{\uk}^{\ook(\theta)} \E_\Q[ \I{\conv(\ok, \x_i, \theta) \ge \yis{0} }|\yis{0} = \yv]f^{(\Q)}_{\ys{0}}(\yv) d\yv \\
    = &\ \sum_{j=1}^l \frac{1}{j!} \dif_{\yv}^{j-1}[\E_\Q[ (\conv(\ok, \x_i, \theta) - \uk)^j |\yis{0} = \yv] f^{(\Q)}_{\ys{0}}(\yv)]_{\yv = \uk} \\ 
    \quad &\ + \frac{1}{l!}\int_{\uk}^{\ook(\theta)} \dif_{\yv}^{l}[\E_\Q[(\conv(\ok, \x_i, \theta) - \wv)_+^l |\yis{0} = \yv] f^{(\Q)}_{\ys{0}}(\yv)]_{\wv = \vv}d\vv.
\end{align*}
Here, we have used the fact that $\conv(\ok,\x_i,\theta) \ge \conv(\k, \x_i, \theta) \ge \k \ge \uk$ to simplify $(\conv(\ok,\x_i,\theta) - \uk)_+ = \conv(\ok,\x_i,\theta) - \uk$ in the summation.
Moreover, by Lemma~\ref{lemA:sufficient-analytic}(ii) and Assumption~\ref{asm:identification-regularity}(i), we have that
\begin{align*}
    & \frac{1}{l!} \left| \dif_{\yv}^{l}[\E_\Q[(\conv(\ok, \x_i, \theta) - \wv)_+^l |\yis{0} = \yv] f^{(\Q)}_{\ys{0}}(\yv)]_{\wv = \vv} \right|    \\
    \le &\ \beta \delta^l \E_\Q[\I{\conv(\ok, \x_i,\theta) > \yis{0}} |\yis{0} = \vv]f^{(\Q)}_{\ys{0}}(\vv)
\end{align*}
for all $\vv \in [\ubars, \bars]$, which implies
\begin{align*}
     &\left| \frac{1}{l!}\int_{\uk}^{\ook(\theta)} \dif_{\yv}^{l}[\E_\Q[(\conv(\ok, \x_i, \theta) - \vv)_+^l |\yis{0} = \yv] f^{(\Q)}_{\ys{0}}(\yv)]_{\yv = \vv}d\vv    \right| \\
     \le &\ \beta \delta^l \int_{\uk}^{\ook(\theta)}\E_\Q[\I{\conv(\ok, \x_i,\theta) > \yis{0}} |\yis{0} = \yv]f^{(\Q)}_{\ys{0}}(\yv) d\yv \\
     = &\ \beta \delta^l \E_\Q[\I{\uk \le \yis{0} \le \conv(\ok, \x_i, \theta)}].
\end{align*}
As $l \to \infty$, the remainder term above tends to 0, and hence we obtain the following equation
\begin{align*}
    \bf(\theta) &:= \lim_{l \to \infty} \sum_{j=1}^l \frac{1}{j!} \dif_{\yv}^{j-1}[\E_\P[\ti (\conv(\ok, \x_i, \theta) - \uk)^j |\yis{0} = \yv] f_{\ys{0}}^{(\P)}(\yv)]_{\yv = \uk}\\
    &= \lim_{l \to \infty} \sum_{j=1}^l \frac{1}{j!} \dif_{\yv}^{j-1}[\E_\Q[(\conv(\ok, \x_i, \theta) - \uk)^j |\yis{0} = \yv] f^{(\Q)}_{\ys{0}}(\yv)]_{\yv = \uk} \\
    & =   \E_\Q[\I{\uk \le \yis{0} \le \conv(\ok, \x_i, \theta)}]
\end{align*}
for all $\theta \in \Theta$.
Finally, since $\{ \yis{0} \le \conv(\ok, \x_i, \theta_0) \}$ is equivalent to $\{ \yis{1} \le \ok \}$, we have
\begin{align*}
\bf(\theta_0) &= \E[\ti \I{\uk \le \yis{0} \le \conv(\ok, \x_i, \theta_0)}]\\
 &= \E_\Q[\I{\uk \le \yis{0} \le \conv(\ok, \x_i, \theta_0)}]\\
& = \E_\Q[\I{\uk \le \yis{0}, \yis{1} \le \ok}] \\
& = \E_\Q[\I{\yiis \in [\uk, \ok]}] \\
& = \E_\Q[\I{\yii \in [\uk, \ok]}]
\end{align*}
by Assumption~\ref{asm:data-optimization-errors}.

\paragraph{Proof of uniqueness of a solution:} In the previous part, we have already established that $\bf(\theta) = \E[\ti \I{\uk \le \yis{0} \le \conv(\ok, \x_i, \theta)}]$ for all $\theta \in \Theta$.
If the function on the right-hand side is one-to-one in $\theta$, $\theta_0$ must be a unique solution to the equation $\bf(\theta) = \E[\ti \I{\yii \in [\uk, \ok]}]$.

\subsubsection{Proof of Theorem~\ref{thm:partial-identification}}
Let
\begin{align*}
\obf(\theta) & :=  \sum_{j=1}^\infty \frac{1}{j!} \dif_{\yv}^{j-1}[\E_{\Q}[(\conv(\ok, \x_i, \theta) - \uk)^j |\yis{0} = \yv] \of(\yv)]_{\yv = \uk}.
\end{align*}
Using the same method as in the proof of Theorem~\ref{thm:point-identification}, we have
\begin{equation*}
    \obf(\theta) =    \int_{\uk}^{\ook(\theta)}\E_{\Q}[\I{\conv(\ok, \x_i,\theta) > \yis{0}} |\yis{0} = \yv]\of(\yv) d\yv.
\end{equation*}
Moreover, the following bound on the approximation error is also true given $\of \ge 0$:
\begin{equation*}
      \left| \sum_{j=l+1}^\infty \frac{1}{j!} \dif_{\yv}^{j-1}[\E_\Q[\ti (\conv(\ok, \x_i, \theta) - \uk)^j |\yis{0} = \yv] \of (\yv)]_{\yv = \uk}   \right| \le \bar{\beta} \delta^l \obf(\theta).
\end{equation*}

It remains to to prove that $\Q(\yii \in [\uk, \ok]) \le \obf(\theta_0)$.
To this end, it suffices to show that
\begin{align}
\label{eq:goal-theorem-2}
\Q(\yii \in [\uk, \ok])& \equiv \int_{\uk}^{\ook(\theta_0)}\E_{\Q}[\I{\conv(\ok, \x_i,\theta_0) > \yis{0}} |\yis{0} = \yv]f^{(\Q)}_{\ys{0}}(\yv) d\yv \nonumber \\
&\le \int_{\uk}^{\ook(\theta_0)}\E_{\Q}[\I{\conv(\ok, \x_i,\theta_0) > \yis{0}} |\yis{0} = \yv]\of(\yv) d\yv.
\end{align}
Using the quantile transform, we may write $\conv(\ok, \x_i,\theta) = \qc(\qua_i, \theta_0, \yis{0})$, where $\qua_i$ is independent of $\yis{0}$ and uniformly distributed over $(0,1)$ under $\Q$.
Let us denote $\qc(\qua, \theta_0, \yv)$ as $\qc(\qua, \yv)$ for notational simplicity.
By Fubini-Tonelli's theorem, this yields
\begin{align*}
&    \int_{\uk}^{\ook(\theta_0)}\E_{\Q}[\I{\conv(\ok, \x_i,\theta_0) > \yis{0}} |\yis{0} = \yv]f^{(\Q)}_{\ys{0}}(\yv) d\yv    \\
=&\ \int_0^1 \left[ \int_{\uk}^{\ook(\theta_0)} \I{\qc(\qua,\yv) > \yv} f^{(\Q)}_{\ys{0}}(\yv) d\yv\right] d\qua.
\end{align*}
We will first show that $\I{\qc(\qua,\yv) - \yv > 0}$ takes the form of $\I{\yv \in [\uk, U(\qua))}$ for some value $U(\qua) \in [\uk, \oy]$ for each $\qua \in (0,1)$.
Specifically, this function $U$ is defined as the unique solution to the equation $\qc(\qua,\yv) = \yv$ s.t. $\yv \in [\uk,\oy]$, whose existence is ensured by the fact that $\qc(\qua,\uk) \ge \k \ge \uk$ and $\qc(\qua,\oy) \le \oy$ for all $\qua \in (0,1)$ and the intermediate value theorem.

The uniqueness is demonstrated as follows.
Let $\yv_0 \in [\uk, \oy]$ be any solution to $\qc(\qua,\yv_0) - \yv_0 = 0$.
By Assumption~\ref{asm:partial-identification-regularity}(ii), we have for all $z \in \C$ such that $|z| = \rad$,
\begin{equation*}
   \left| \sum_{j=1}^\infty \frac{\dif_{\yv}^j \qc(\qua,\yv_0)}{j!}z^{j-1} \right|  = \frac{\left|\qc(\qua,\yv_0 + z) - \qc(\qua,\yv_0) \right|}{|z|} = \frac{\left|\qc(\qua,\yv_0 + z) - \yv_0 \right|}{\rad} \le \delta.
\end{equation*}
By the maximum modulus principle (\citeay{steinComplexAnalysis2010}), it follows that, for all $0 \le |z|\le\rad$,
$$
\left| \sum_{j=1}^\infty \frac{\dif_{\yv}^j \qc(\qua,\yv_0)}{j!}z^{j-1} \right| \le \delta.
$$
This implies that, for all $0 < h \le \rad$,
\begin{align*}
        \qc(\qua,\yv_0+h) - (\yv_0+h) &= \sum_{j=1}^\infty \frac{\dif_{\yv}^j \qc(\qua,\yv_0)}{j!}h^j - h\\
        &\le \delta h - h < 0,
\end{align*}
which implies $[\yv_0, \yv_0+\rad] \cap \{\yv \in [\uk,\oy]: \qc(\qua,\yv) - \yv > 0 \} = \varnothing$.
Since $\{ \yv \in [\uk, \oy]:\qc(\qua,\yv) - \yv > 0 \}$ is open in $[\uk, \oy]$, this implies that $\{ \yv \in [\uk, \oy]:\qc(\qua,\yv) - \yv > 0 \}$ has only one connected component that takes the form of $[\uk, U(\qua))$, where $U(\qua) \in [\uk, \oy]$.

Now, by Assumption~\ref{asm:envelope-condition}, we have that
\begin{align*}
\int_0^1 \left[ \int_{\uk}^{\ook(\theta_0)} \I{\qc(\qua,\yv) > \yv} f^{(\Q)}_{\ys{0}}(\yv) d\yv\right] d\qua &= \int_0^1 \left[ \int_{\uk}^{\min\{U(\qua), \ook(\theta_0)\}} f^{(\Q)}_{\ys{0}}(\yv) d\yv\right] d\qua \\
&\le \int_0^1 \left[ \int_{\uk}^{\min\{U(\qua), \ook(\theta_0)\}} \of(\yv) d\yv\right] d\qua \\
&= \int_0^1 \left[ \int_{\uk}^{\ook(\theta_0)} \I{\qc(\qua,\yv) > \yv} \of(\yv) d\yv\right] d\qua.
\end{align*}
Since the last expression coincides with $\int_{\uk}^{\ook(\theta_0)}\E_{\Q}[\I{\conv(\ok, \x_i,\theta_0) > \yis{0}} |\yis{0} = \yv]\of(\yv) d\yv$, we conclude that \eqref{eq:goal-theorem-2} holds.
The lower bound $\ubf(\theta_0) \le \Q(\yii \in [\uk, \ok])$ can be established in a similar manner, which will be omitted for brevity.

\subsubsection{Proof of Theorem~\ref{thm:asymptotic-normality}}
Throughout this proof, we adopt the same notational convention as in the proof of Lemma~\ref{lem:estimation-consistency}.

Let $\tilde{H}_{\tc}$ be an $\tc\times\tc$ rescaled Hilbert matrix, defined as $[\tilde{H}_{\tc}]_{ij} = \bar{w}^{-1}\int_0^{\bar w} x^{i+j-2}dx$ for $i,j=1,\ldots, \tc$, where $\bar{w} < \infty$ denotes an upper bound of $w_i\ge0$.
For each $i\in\nn$, let $\ib_{\tc i} = \tilde{H}_{\tc}^{-1/2} (w_i,w_i^2/2,\ldots, w_i^{\tc}/\tc)' \in \R^{\tc}$. 
Since $(d/dw_i) \ib_{\tc i} = \tilde{H}_{\tc}^{-1/2} (1,w_i,\ldots, w_i^{\tc-1})'$, this vector represents an integrated orthonormal polynomial basis with respect to the uniform measure on $[0,\bar{w}]$, which in turn satisfies $\|\ib_{\tc i}\| \lesssim \|(d/dw_i) \ib_{\tc i}\| \lesssim \tc$.
Define $\tc \times \kap$ matrices as
\begin{equation*}
\widetilde U_{\kap \tc} = \tilde{H}_{\tc}^{1/2} \left( \begin{matrix}
c_{\kap 1}' \widetilde{I}_{11n}^{-1} \\
c_{\kap 2}' \widetilde{I}_{12n}^{-1} \\
\vdots \\
c_{\kap, \tc}' \widetilde{I}_{1\tc n}^{-1}
\end{matrix} \right),\quad 
U_{\kap \tc} = \tilde{H}_{\tc}^{1/2} \left( \begin{matrix}
c_{\kap 1}' I_{11}^{-1} \\
c_{\kap 2}' I_{12}^{-1} \\
\vdots \\
c_{\kap, \tc}' I_{1\tc}^{-1}
\end{matrix} \right),
\end{equation*}
where $I_{1j} = \E[\ti w_i^j \frac{o_{\kap i}o_{\kap i}'}{f_{\kap j i}^2}]$ represents the population counterpart of $\widetilde{I}_{1jn} = \frac{1}{n} \sum_{i=1}^n \ti  w_i^j \frac{o_{\kap i}o_{\kap i}'}{\what f_{\kap j i}f_{\kap j i}}$.
Also let $\what{I}_{1jn} = \frac{1}{n} \sum_{i=1}^n \ti  w_i^j \frac{o_{\kap i}o_{\kap i}'}{\what f_{\kap j i}^2}$ and $\bar{I}_{1jn} = \frac{1}{n} \sum_{i=1}^n \ti  w_i^j \frac{o_{\kap i}o_{\kap i}'}{f_{\kap j i}^2}$.


We define the empirical process as $G_n[f_i] = \frac{1}{\sqrt n} \sum_{i=1}^n (f_i - \E[f_i])$ where $f_i=f(\yis{0}, \x_i)$ represents the value of a function $f$ for unit $i$.
Denote $D_{\kap \tc i} = \diag(1/f_{\kap 1 i},\ldots, 1/f_{\kap \tc i})$.
We will first establish that
\begin{align}
\label{eq:empirical-process-approximation}
\sqrt{n} \what \mu_n 
&= \sqrt n\left( \frac{1}{n} \sum_{i=1}^n \ti  \I{\yii \in [\uk, \ok]}  - \sum_{j=1}^{\tc} \frac{1}{j}\what\gamma_{\kap j,j} \right) \nonumber\\
&= G_n\left[\ti (\I{\yii \in [\uk, \ok]} -   \ib_{\tc i}' U_{\kap \tc} D_{\kap \tc i}o_{\kap i} \I{i \in \nn})\right] + \sqrt n b_{\tc} + \oP(1)
\end{align}
where $b_{\tc} = \sum_{j = \tc + 1}^\infty \frac{1}{j!} \dif_{\yv}^{j-1} [f_j(\yv)]_{\yv = \k}$ denotes the approximation bias.

Recall from \eqref{eq:noname2} that
\begin{align*}
& \sqrt n \sum_{j=1}^{\tc} \frac{1}{j}(\what\gamma_{\kap j,j} -\gamma_{\kap j, j}) 
= \sqrt n \sum_{j=1}^{\tc} \frac{1}{j }c_{\kap j}' (\what \co_{\kap j} - \co_{\kap j}) 
=  \sum_{j=1}^{\tc} \frac{1}{j} c_{\kap j}' \widetilde{I}_{1jn}^{-1} G_n\left[\ti w_i^j \frac{o_{\kap i}}{f_{\kap j i}}\right]  \\
= &\ \sum_{j=1}^{\tc}\frac{1}{j} c_{\kap j}' I_{1j}^{-1}G_n\left[\ti w_i^j \frac{o_{\kap i}}{f_{\kap j i}}\right] + \underbrace{\sum_{j=1}^{\tc}\frac{1}{j} c_{\kap j}' (\widetilde{I}_{1jn}^{-1} - I_{1j}^{-1})G_n\left[\ti w_i^j \frac{o_{\kap i}}{f_{\kap j i}}\right]}_{=: \rem_n}\\
= &\ \frac{1}{\sqrt n} \sum_{i=1}^n \left\{\ti  \ib_{\tc i}' U_{\kap \tc} D_{\kap \tc i} o_{\kap i} - \E\left[\ti  \ib_{\tc i}' U_{\kap \tc} D_{\kap \tc i} o_{\kap i}\right]\right\} + \rem_n. 
\end{align*}
The first term in the last line corresponds to the influence function $G_n[\ti  \ib_{\tc i}' U_{\kap \tc} D_{\kap \tc i} o_{\kap i} \I{i \in \nn}]$ and the last term represents the remainder, which we denote by $\rem_n$.
Combining Theorem~\ref{thm:point-identification} and \eqref{eq:bias-term}, we know that 
\begin{align*}
\E\left[ \ti  \I{\yii \in [\uk, \ok]} \right] &= \sum_{j \in \N} \frac{1}{j!} \dif_{\yv}^{j-1} [f_j(\yv)]_{\yv = \uk} \\
&= \sum_{j = 1}^{\tc} \frac{1}{j} \gamma_{\kap j,j} + b_{\tc}  + o(n^{-1/2}) .      
\end{align*}
It follows
\begin{align*}
 \sqrt n \what \mu_n 
= &\ \sqrt n\left( \frac{1}{n} \sum_{i=1}^n \ti  \I{\yii \in [\uk, \ok]} - \sum_{j=1}^{\tc} \frac{1}{j}\what\gamma_{\kap j,j}  \right)\\
=&\ \sqrt n\left(  \frac{1}{n} \sum_{i=1}^n \bigg[  \ti  \I{\yii \in [\uk, \ok]}-\E\left[ \ti  \I{\yii \in [\uk, \ok]} \right] \bigg] - \sum_{j=1}^{\tc} \frac{1}{j}(\what\gamma_{\kap j,j}-  \gamma_{\kap j, j}) \right) \\
& \quad \quad + \sqrt n b_{\tc} + \oP(1)\\
= &\ G_n\left[ \ti ( \I{\yii \in [\uk, \ok]} -\ib_{\tc i}' U_{\kap \tc} D_{\kap \tc i}o_{\kap i} \I{i \in \nn} )\right] + \sqrt n b_{\tc} + \rem_n + \oP(1).
\end{align*}
It suffices to establish that $\rem_n = \oP(1)$ to show \eqref{eq:empirical-process-approximation}.
The remaining proof is divided into three steps.
In the first step, we demonstrate $\rem_n = \oP(1)$.
In the following steps, we derive normal approximation of the t-statistic and the consistency of the variance estimator.
\medskip
\\
\noindent\tbf{Step 1.}
Recall that $\bar{I}_{1jn} = \frac{1}{n} \sum_{i=1}^n \ti  w_i^j \frac{o_{\kap i} o_{\kap i}'}{f_{\kap j i}^2}$.
We first decompose $\frac{1}{j} c_{\kap j}' (\widetilde{I}_{1jn}^{-1} - I_{1j}^{-1})$ included in $\rem_n$ into two parts:
\begin{equation*}
    \frac{1}{j} c_{\kap j}' (\widetilde{I}_{1jn}^{-1} - I_{1j}^{-1}) = \frac{1}{j} c_{\kap j}' (\widetilde{I}_{1jn}^{-1} - \bar{I}_{1j}^{-1}) + \frac{1}{j} c_{\kap j}' (\bar{I}_{1jn}^{-1} - I_{1j}^{-1}).    
\end{equation*}
For the first part, since $\|o_{\kap i}\| \lesssim \kap$ uniformly in $i \in \nn$, we observe that
\begin{align*}
\|c_{\kap j}'(\widetilde{I}_{1jn} - \bar{I}_{1jn}) \| &= \left\|\frac{1}{n} \sum_{i \in \nn}  \ti  w_i^j \frac{\|o_{\kap i}\|^2 c_{\kap j}'(\what{\co}_{\kap j} - \co_{\kap j}) o_{\kap i}'}{f_{\kap j i}^2 \hat f_{\kap j i}} \right\| \\
&\lesssim C^j \kap^3 |c_{\kap j}'(\what{\co}_{\kap j} - \co_{\kap j})|.
\end{align*}
Using \eqref{eq:noname3} and \eqref{eq:noname4} from the proof of Lemma~\ref{lem:estimation-consistency}, it follows that
\begin{align}
\label{eq:noname1}
\sum_{j=1}^{\tc}\|c_{\kap j}'(\widetilde{I}_{1jn} - \bar{I}_{1jn}) \|^2 \leP C^\tc \kap^6 \sum_{j=1}^{\tc} |c_{\kap j}'(\what{\co}_{\kap j} - \co_{\kap j})|^2 \leP \frac{C^\tc \tc^{-4\tc+1}\kap^{4\tc +4}}{n}.
\end{align}
In the proof of Lemma~\ref{lem:estimation-consistency}, we established that $\widetilde{I}_{1jn}^{-1} \lesssim C^j M_{\kap}^{-1}\lesssim C^j \chi_\kap^{-1} I_{\kap}$ uniformly in $1\le j\le\tc$ with probability approaching $1$ as $n\to\infty$.
Using the same line of arguments, it can be shown that the same bounds hold for $\bar{I}_{1jn}^{-1}$ as well.
That is, $\bar{I}_{1jn}^{-1} \lesssim C^j \chi_\kap^{-1} I_{\kap}$ uniformly in $1\le j\le\tc$.
Combined with \eqref{eq:noname1}, this implies that
\begin{align*}
\sum_{j=1}^{\tc}\|c_{\kap j}'(\widetilde{I}_{1jn}^{-1} - \bar{I}_{1jn}^{-1}) \|^2 = \sum_{j=1}^{\tc}\|c_{\kap j}'\widetilde{I}_{1jn}^{-1}(\bar{I}_{1jn}-\widetilde{I}_{1jn})\bar{I}_{1jn}^{-1} \|^2 & \leP C^\tc \frac{\tc^{-4\tc+1}\kap^{4\tc +4}}{n}\chi_{\kap}^{-4}.
\end{align*}

Let us bound the remaining part $\sum_{j=1}^{\tc}\|c_{\kap j}'(\bar{I}_{1jn}^{-1} - I_{1j}^{-1}) \|^2$ where $I_{1j} = \E[\bar{I}_{1jn}]$ denotes the probability limit of $\bar{I}_{1jn}$.
Applying Lemma~\ref{lem:Matrix-moment} with $p = 2$, we have
\begin{align*}
\E\|\bar{I}_{1jn} - I_{1j}\|^2 &\lesssim \frac{C^j}{n} \kap^2 \log \kap
\end{align*}
uniformly in $j\le\tc$.
Using the fact that $I_{1j}^{-1} \lesssim C^j \chi_{\kap}^{-1} I_{\kap}$ uniformly in $j\le\tc$, this in turn implies that
\begin{align*}
\E\sum_{j=1}^{\tc}\|c_{\kap j}'(\bar{I}_{1jn}^{-1} - I_{1j}^{-1}) \|^2 \lesssim C^\tc\frac{ \chi_\kap^{-4}\kap^2 \log \kap}{n}.
\end{align*}
Putting it all together, we get
\begin{equation}
\label{eq:noname5}
\|\widetilde {U}_{\kap \tc} - U_{\kap \tc}\|_F^2 = \sum_{j=1}^{\tc}\|c_{\kap j}'(\widetilde{I}_{1jn}^{-1} - I_{1j}^{-1}) \|^2 \leP C^\tc \frac{\chi_k^{-4} \tc^{-4\tc+1}\kap^{4\tc + 4}}{n}.
\end{equation}
Next, we observe that
\begin{align*}
  \E \sum_{j=1}^{\tc}\frac{1}{j^2}\left\|G_n\left[\ti w_i^j \frac{o_{\kap i}}{f_{\kap j i}}\right] \right\| ^2 &\le    \sum_{j=1}^{\tc}\frac{1}{j^2}C^j\tr{\var\left(\ti w_i^j \frac{o_{\kap i}}{f_{\kap j i}}\right) }\\
  &\lesssim \sum_{j=1}^{\tc}\frac{1}{j^2}C^j \tr{M_{\kap}}\lesssim C^{\tc} \kap.
\end{align*}
By Cauchy-Schwarz inequality, it follows that
\begin{align*}
\left| \sum_{j=1}^{\tc}\frac{1}{j} c_{\kap j}' (\widetilde{I}_{1jn}^{-1} - I_{1j}^{-1})G_n\left[\ti w_i^j \frac{o_{\kap i}}{f_{\kap j i}}\right] \right| &\leP \|\widetilde {U}_{\kap \tc} - U_{\kap \tc}\|_F \left( \sum_{j=1}^{\tc} \frac{1}{j^2} \left\| G_n\left[\ti w_i^j \frac{o_{\kap i}}{f_{\kap j i}}\right] \right\|^2\right)^{1/2}\\
&\leP C^\tc \frac{\chi_k^{-2}\tc^{-2\tc+1/2}\kap^{2\tc + 5/2}}{\sqrt n} = \oP(1),
\end{align*}
since $\tc^{-2\tc+1/2}\kap^{2\tc + 5/2} = o(n^\varepsilon)$ for any $\varepsilon>0$ and $\chi_\kap^{-2} = O(n^{\con_{6}})$ with $\con_{6} < 2/5$ by Assumption~\ref{asm:selection-tuning}.
This proves $\rem_n = \oP(1)$, and hence \eqref{eq:empirical-process-approximation}.
We now turn to the normal approximation of the stochastic component.
\medskip
\\
\noindent\tbf{Step 2.} (Normal approximation)
Let us denote the influence function associated with $\what \mu_n$ by
\begin{align*}
    s_{ni} &:= \ti  (\I{\yii \in [\uk, \ok]} - \ib_{\tc i}' U_{\kap \tc} D_{\kap \tc i}o_{\kap i} \I{i \in \nn}) \\
    &\quad - \E\left[ \ti  (\I{\yii \in [\uk, \ok]} - \ib_{\tc i}' U_{\kap \tc} D_{\kap \tc i}o_{\kap i} \I{i \in \nn})\right]
\end{align*}
and $\sigma_n^2 = \var(s_{ni})$.
By Yurinskii's coupling (\citeay{pollardUserGuideMeasure2001}), for all $\delta > 0$, there exists $\xi \sim N(0,1)$ such that
\begin{equation*}
\P\left( \left| \frac{G_n[s_{ni}]}{\sigma_n} - \xi \right| > \delta\right) \lesssim B (1+|\log(1/B)|)
\end{equation*}
where $B = \delta^{-3}n^{-3/2}\sum_{i=1}^n \mathbb{E}[|s_{ni}|^3]/\sigma_n^3$.
We first show that
\begin{equation*}
|s_{ni}| \lesssim 1 + \|\ib_{\tc i}\|\|U_{\kap \tc}\|_F \|D_{\kap \tc i}\| \|o_{\kap i}\| \lesssim C^\tc \chi_{\kap}^{-1/2}\tc^{-2\tc+3/2}\kap^{2\tc}
\end{equation*}
where we used $\|\ib_{\tc i}\| \lesssim \tc$, $\|D_{\kap \tc i}\| \lesssim C^\tc$, and $\|o_{\kap i}\| \lesssim \kap$, and
\begin{align}
\label{eq:noname6}
\|U_{\kap \tc}\|_F^2 = \tr{U_{\kap \tc}U_{\kap \tc}'} \lesssim \|\tilde{H}_\tc\|^2 \sum_{j=1}^\tc c_{\kap j}' {I}_{1j}^{-2}c_{\kap j}  & \lesssim \chi_{\kap}^{-1} C^{\tc} \sum_{j=1}^\tc c_{\kap j}' M_{\kap}^{-1} c_{\kap j} \nonumber\\
&\lesssim C^\tc \chi_{\kap}^{-1} \sum_{j=1}^\tc e_j' Q_{\kap}^{-1} e_{j}  \nonumber \\     
&\lesssim C^\tc \chi_{\kap}^{-1} \sum_{j=1}^\tc j^{-1}(\kap/j)^{4j-2}  \nonumber \\     
&\lesssim C^\tc \chi_{\kap}^{-1} \tc^{-4\tc+1}\kap^{4\tc-2}
\end{align}
which follows from Lemmas~\ref{lem:Hilbert-matrix}(i) and \ref{lem:variance-bound}.
Since $\sigma_n \gtrsim 1$ by Assumption~\ref{asm:variance-lower-bound}, it follows
\begin{equation*}
B \lesssim \delta^{-3}n^{-1/2}C^\tc  \chi_{\kap}^{-3/2} \tc^{-6\tc+9/2}\kap^{6\tc} =O( \delta^{-3} n^{-1/5+\epsilon})
\end{equation*}
for any small $\epsilon > 0$.
Letting $\delta \asymp n^{-1/15+\epsilon}$, this implies that $G_n[s_{ni}]/\sigma_n - \xi = \oP(n^{-1/15+2\epsilon})$.
\medskip
\\
\noindent\tbf{Step 3.} (Variance estimation)
By Assumption~\ref{asm:variance-lower-bound}, it follows that
\begin{align*}
\sigma_n^2 &= \E [\ti ^2 (\I{\yii \in [\uk, \ok]} - \ib_{\tc i}' U_{\kap \tc} D_{\kap \tc i}o_{\kap i} \I{i \in \nn})^2] \\ 
    &\quad - \E[\ti  (\I{\yii \in [\uk, \ok]} - \ib_{\tc i}' U_{\kap \tc} D_{\kap \tc i}o_{\kap i} \I{i \in \nn})]^2 \\
    & \gtrsim 1.
\end{align*}
The second term is already negligible since
\begin{equation*}
\E[\ti  (\I{\yii \in [\uk, \ok]} - \ib_{\tc i}' U_{\kap \tc} D_{\kap \tc i}o_{\kap i} \I{i \in \nn})] = b_\tc + o(1) \to 0.
\end{equation*}
We estimate $\sigma_n^2$ by $\what \sigma^2_n := \frac{1}{n} \sum_{i=1}^n \what s_{n i}^2$, where
\begin{align*}
\what s_{n i} = \ti \left(  \I{\yii \in [\uk, \ok]} - \ib_{\tc i}' \what{U}_{\kap \tc} \what{D}_{\kap \tc i}o_{\kap i} \I{i \in \nn}\right).
\end{align*}
Here, $\what{U}_{\kap \tc}$ is defined as the sample analogue of $\widetilde{U}_{\kap \tc}$, obtained by replacing all $\widetilde{I}_{1jn}$ in $\widetilde{U}_{\kap \tc}$ with their feasible estimates
$
    \what{I}_{1jn} = \frac{1}{n} \sum_{i=1}^n \ti  w_i^j \frac{o_{\kap i}o_{\kap i}'}{\what f_{\kap j i}^2}.
$
We can establish $\|\what{U}_{\kap \tc} - U_{\kap \tc}\|_F^2 \leP C^\tc \frac{\chi_k^{-4} \tc^{-4\tc+1}\kap^{4\tc + 4}}{n}$ using the same method as used to show \eqref{eq:noname5}, whose proof will be omitted for brevity.

We show that $\frac{1}{n} \sum_{i=1}^n |\what s_{n i}^2 - s_{ni}^2| = \oP(1)$.
It is clear that $\frac{1}{n} \sum_{i=1}^n \ti ^2 \I{\yii \in [\uk, \ok]} - \E[\ti ^2 \I{\yii \in [\uk, \ok]}] = \OP(n^{-1/2})$.
Next, we establish that
$$
\frac{1}{n} \sum_{i=1}^n \ti ^2 (\ib_{\tc i}' \what{U}_{\kap \tc} \what{D}_{\kap \tc i}o_{\kap i})^2 \I{i \in \nn} - \E[\ti ^2 (\ib_{\tc i}' U_{\kap \tc} D_{\kap \tc i}o_{\kap i})^2 \I{i \in \nn}] = \oP(1).
$$
Observe that
\begin{equation*}
\max_{i\in\nn}|\ib_{\tc i}' \what{U}_{\kap \tc} \what{D}_{\kap \tc i}o_{\kap i}-\ib_{\tc i}' {U}_{\kap \tc} \what{D}_{\kap \tc i}o_{\kap i}| \le \|\what{U}_{\kap \tc} - U_{\kap \tc}\|_F \max_{i\in\nn} \|\ib_{\tc i}\| \|\what{D}_{\kap \tc i}\| \|o_{\kap i}\| \leP C^\tc \frac{\chi_k^{-2} \tc^{-2\tc+1/2}\kap^{2\tc + 2}}{\sqrt n} \tc\kap,
\end{equation*}
which follows from the established fact that $\|\what{U}_{\kap \tc} - U_{\kap \tc}\|_F^2 \leP C^\tc \frac{\chi_k^{-4} \tc^{-4\tc+1}\kap^{4\tc + 4}}{n}$, $\|\ib_{\tc i}\| \lesssim \tc$, $\|\what{D}_{\kap \tc i}\|\lesssim C^\tc$, and $\|o_{\kap i}\|\lesssim \kap$.
Additionally, we have for each $i\in\nn$,
\begin{align*}
& |\ib_{\tc i}' U_{\kap \tc} \what{D}_{\kap \tc i}o_{\kap i}-\ib_{\tc i}' {U}_{\kap \tc} D_{\kap \tc i}o_{\kap i}| \\
\lesssim &\ \tc \kap \|U_{\kap \tc}\|_F \|\what{D}_{\kap \tc i} - {D}_{\kap \tc i}\| \leP C^\tc \kap^{2\tc} \tc^{-2\tc + 3/2} \chi_{\kap}^{-1/2} \sup_{1\le j\le \tc} |\what{f}_{\kap j i} - f_{\kap j i}|,
\end{align*}
which follows from \eqref{eq:noname6}.
Also note that
$
|\ib_{\tc i}' {U}_{\kap \tc} D_{\kap \tc i}o_{\kap i}|  \lesssim C^\tc \kap^{2\tc} \tc^{-2\tc + 3/2} \chi_{\kap}^{-1/2}
$
for all $i \in \nn$ with probability $1$.
Putting these together, we have
\begin{align*}
& \frac{1}{n} \sum_{i\in\nn}|(\ib_{\tc i}' \what{U}_{\kap \tc} \what{D}_{\kap \tc i}o_{\kap i})^2-(\ib_{\tc i}' {U}_{\kap \tc} D_{\kap \tc i}o_{\kap i})^2| \\
\lesssim &\ \frac{1}{n} \sum_{i\in\nn} |\ib_{\tc i}' \what{U}_{\kap \tc} \what{D}_{\kap \tc i}o_{\kap i}-\ib_{\tc i}' {U}_{\kap \tc} D_{\kap \tc i}o_{\kap i}|^2 + 2|\ib_{\tc i}' \what{U}_{\kap \tc} \what{D}_{\kap \tc i}o_{\kap i} - \ib_{\tc i}' {U}_{\kap \tc} D_{\kap \tc i}o_{\kap i}| |\ib_{\tc i}' {U}_{\kap \tc} D_{\kap \tc i}o_{\kap i}| \\
\leP &\ \frac{1}{n} \sum_{i\in\nn} |\ib_{\tc i}' \what{U}_{\kap \tc} \what{D}_{\kap \tc i}o_{\kap i}-\ib_{\tc i}' {U}_{\kap \tc} \what{D}_{\kap \tc i}o_{\kap i}|^2 + |\ib_{\tc i}' U_{\kap \tc} \what{D}_{\kap \tc i}o_{\kap i}-\ib_{\tc i}' {U}_{\kap \tc} D_{\kap \tc i}o_{\kap i}|^2 \\
& \ \ + \max_{i\in\nn}|\ib_{\tc i}' {U}_{\kap \tc} D_{\kap \tc i}o_{\kap i}| \cdot \frac{1}{n} \sum_{i\in\nn} \left( |\ib_{\tc i}' \what{U}_{\kap \tc} \what{D}_{\kap \tc i}o_{\kap i}-\ib_{\tc i}' {U}_{\kap \tc} \what{D}_{\kap \tc i}o_{\kap i}| + |\ib_{\tc i}' U_{\kap \tc} \what{D}_{\kap \tc i}o_{\kap i}-\ib_{\tc i}' {U}_{\kap \tc} D_{\kap \tc i}o_{\kap i}|\right) \\
\leP &\ C^\tc \frac{\chi_\kap^{-4}  \tc^{-4\tc+1}\kap^{4\tc + 4}}{n}(\tc\kap)^2 + C^\tc  \frac{\chi_\kap^{-1}\kap^{4\tc}\tc^{-4\tc + 3} }{n} \kap\\
& \qquad + C^\tc \kap^{2\tc} \tc^{-2\tc + 3/2} \chi_{\kap}^{-1/2}\left( \frac{\chi_\kap^{-2}  \tc^{-2\tc+1/2}\kap^{2\tc + 2}}{\sqrt n} \tc\kap+ \frac{\chi_\kap^{-1/2}\kap^{2\tc}\tc^{-2\tc + 3/2}}{\sqrt n}\kap^{1/2}\right)\\
& = \oP(1),
\end{align*}
by Assumption~\ref{asm:selection-tuning}.
Here, to move from the second to the third line, we used
\begin{align*}
     |\ib_{\tc i}' \what{U}_{\kap \tc} \what{D}_{\kap \tc i}o_{\kap i}-\ib_{\tc i}' {U}_{\kap \tc} D_{\kap \tc i}o_{\kap i}|^2&      \lesssim |\ib_{\tc i}' \what{U}_{\kap \tc} \what{D}_{\kap \tc i}o_{\kap i}-\ib_{\tc i}' {U}_{\kap \tc} \what{D}_{\kap \tc i}o_{\kap i}|^2 + |\ib_{\tc i}' U_{\kap \tc} \what{D}_{\kap \tc i}o_{\kap i}-\ib_{\tc i}' {U}_{\kap \tc} D_{\kap \tc i}o_{\kap i}|^2    
\end{align*}
and
\begin{align*}
    &|\ib_{\tc i}' \what{U}_{\kap \tc} \what{D}_{\kap \tc i}o_{\kap i} - \ib_{\tc i}' {U}_{\kap \tc} D_{\kap \tc i}o_{\kap i}| |\ib_{\tc i}' {U}_{\kap \tc} D_{\kap \tc i}o_{\kap i}| \\
    \lesssim & \ \max_{i\in\nn}|\ib_{\tc i}' {U}_{\kap \tc} D_{\kap \tc i}o_{\kap i}| \cdot \left( |\ib_{\tc i}' \what{U}_{\kap \tc} \what{D}_{\kap \tc i}o_{\kap i}-\ib_{\tc i}' {U}_{\kap \tc} \what{D}_{\kap \tc i}o_{\kap i}|+ |\ib_{\tc i}' U_{\kap \tc} \what{D}_{\kap \tc i}o_{\kap i}-\ib_{\tc i}' {U}_{\kap \tc} D_{\kap \tc i}o_{\kap i}|\right).
\end{align*}
We also used Lemma~\ref{lem:average-consistency-fkj} to bound the following terms to control sums involving $\sup_{1\le j\le \tc} |\what{f}_{\kap j i} - f_{\kap j i}|$: 
\begin{align*}
    \frac{1}{n} \sum_{i \in \nn} \sup_{1\le j\le \tc} |\what{f}_{\kap j i} - f_{\kap j i}|^2 &\le \frac{1}{n} \sum_{i\in\nn} \sum_{j=1}^{\tc} |\what{f}_{\kap j i} - f_{\kap j i}|^2 \leP C^\tc \kap,\\
    \frac{1}{n} \sum_{i\in\nn} \sup_{1\le j\le \tc} |\what{f}_{\kap j i} - f_{\kap j i}| &\le \left(  \frac{1}{n} \sum_{i\in\nn} \sum_{j=1}^{\tc} |\what{f}_{\kap j i} - f_{\kap j i}|^2 \right)^{1/2}\leP C^\tc \kap^{1/2}.
\end{align*}
Finally, by Hoeffding's inequality (\citeay{wainwrightHighDimensionalStatisticsNonAsymptotic2019}), we get
\begin{align*}
&\frac{1}{n} \sum_{i=1}^n \left\{ \ti ^2 (\ib_{\tc i}' U_{\kap \tc} D_{\kap \tc i}o_{\kap i})^2 \I{i \in \nn} - \E[\ti ^2 (\ib_{\tc i}' U_{\kap \tc} D_{\kap \tc i}o_{\kap i})^2 \I{i \in \nn}]\right\}\\
= &\ \OP \left( C^{\tc}\frac{\kap^{2\tc}\tc^{-2\tc+3/2}\chi_\kap^{-1/2} }{\sqrt n}\right) = \oP(1),   
\end{align*}
which follows from the fact that 
$$
0\le \ti ^2 (\ib_{\tc i}' U_{\kap \tc} D_{\kap \tc i}o_{\kap i})^2 \I{\yis{0} \in \S} \lesssim C^{\tc}\kap^{2\tc}\tc^{-2\tc+3/2}\chi_\kap^{-1/2}
$$
with probability 1.
We conclude that $|\what \sigma_n^2 - \sigma_n^2| \convp 0$ and $\what \sigma_n/ \sigma_n = 1 + \oP(1)$.
This implies
\begin{align*}
\mathcal{T}_n (1+\oP(1)) &= \sqrt{n}\frac{\what \mu_{n}}{ \sigma_n} \\
& = \frac{G_n[s_{ni}]}{ \sigma_n} + \frac{\sqrt{n} b_{\tc}}{ \sigma_n} + {\oP(1)}\\
&= \xi  +  \frac{\sqrt{n} b_{\tc}}{ \sigma_n} + \oP(1),
\end{align*}
where $\xi \sim N(0,1)$. 
By the anti-concentration inequality applied to $\xi + \frac{\sqrt n b_\tc}{\sigma_n} \sim N(\frac{\sqrt{n} b_{\tc}}{ \sigma_n},1)$, we get 
$$
\sup_{c \in \mathbb{R}_+}\left|\P\left( \left|\xi + \frac{\sqrt{n} b_{\tc}}{ \sigma_n}+\oP(1)\right| > c(1+\oP(1))\right) - \P\left( \left| N\left( \frac{\sqrt{n} b_{\tc}}{ \sigma_n},1\right)\right| > c\right)\right| \to 0
$$
as $n\to\infty$ uniformly over $\P \in \mathfrak{P}_0$.
Using $|b_{\tc}| \le \maxb_{\tc} := \beta \delta^{\tc} \E[\ti \I{\yii \in [\uk, \ok]}]$ and the fact that $\cv{\alpha}(b)$ is increasing in $|b|$, we obtain
\begin{align*}
    \sup_{\P \in \mathfrak{P}_0}\P( |\mathcal{T}_n| > \cv{\alpha}(\sqrt{n} \maxb_{\tc}/ \what \sigma_n) ) \le \alpha + o(1)
\end{align*}
as $n \to \infty$.


\end{document}